\documentclass[]{elsarticle}

\usepackage[english]{babel}
\usepackage[utf8]{inputenc}
\usepackage{amsmath}
\usepackage{algorithm}
\usepackage{algpseudocode}
\usepackage{hyperref}
\usepackage{cleveref}
\usepackage{booktabs}
\usepackage[size=tiny]{todonotes}
\usepackage{soul}
\usepackage{comment}
\excludecomment{comment}
\usepackage{subcaption}
\usepackage{geometry}
\usepackage{amssymb}
\usepackage{algpseudocode}
\usepackage{algorithm}
\usepackage{transparent}
\usepackage[nolist]{acronym}
\usepackage{tabularx} 
\usepackage{relsize}
\usepackage{tikz}
\usetikzlibrary{shapes}

\newcommand{\mysquare}{\raisebox{0pt}{\tikz{\node[draw,scale=0.65,regular polygon, regular polygon sides=4,line width=0.1em,fill=none](){};}}}
\newcommand{\mytriangle}{\tikz{\node[draw,isosceles triangle,isosceles triangle stretches,shape border rotate=90,line width= 0.1em,minimum width=0.2cm,minimum height=0.2cm,inner sep=0pt] at (0,0) {};}}

\usepackage{color}

\makeatletter
\def\ps@pprintTitle{%
 \let\@oddhead\@empty
 \let\@evenhead\@empty
 \def\@oddfoot{}%
 \let\@evenfoot\@oddfoot}
\makeatother

\usepackage{natbib}
\usepackage{graphicx}
\usepackage{mathtools}
\usepackage{caption}
\usepackage{listings}
\usepackage{xcolor}
\usepackage{sectsty}
\usepackage{pgfplots}
\usepackage{bm} 
\usepackage{gensymb}
\usepackage{multirow}
\definecolor{royalblue}{RGB}{65, 105, 225}
\definecolor{mygreen}{RGB}{0,155,0}
\definecolor{Cerulean}{RGB}{0,123,167}

\usepackage{ulem}
\colorlet{Reviewer1}{black}

\colorlet{Reviewer2}{black}

\colorlet{Reviewer3}{black}

\colorlet{Tomislav}{black}
\usepackage{lineno}

\newcommand{\DeltaT}{\ensuremath{\Delta t}}

\renewcommand{\vec}[1]{\ensuremath{\mathbf{#1}}}

\newcommand{\x}{\ensuremath{\mathbf{x}}}

\newcommand{\Nsigma}[1]{\ensuremath{\mathbf{n}_{\Sigma}}}

\graphicspath{{figures/}}

\begin{document}

\begin{frontmatter}


\begin{acronym}[]
  \acro{AABB}{Axis-Aligned Bounding Box}
  \acro{ALE}{Arbitrary Lagrangian-Eulerian}
  \acro{AMR}{Adaptive Mesh Refinement}
  \acro{BDS}{Backward Differencing Scheme}
  \acro{BE}{Backward Euler}
  \acro{BGL}{Boost Geometry Library}
  \acro{BT}{Barycentric Triangulation}
  \acro{BCPT}{Barycentric Convex Polygon Triangulation}
  \acro{CAD}{Computer Aided Design}
  \acro{CCI}{Cell / Cell Intersection}
  \acro{CCU}{Cell-wise Conservative Unsplit}
  \acro{CCNR}{Cell Cutting Normal Reconstruction}
  \acro{CDS}{Central Differencing Scheme}
  \acro{CG}{Computer Graphics}
  \acro{CGAL}{Computational Geometry Algorithms Library}
  \acro{CIAM}{Calcul d'Interface Affine par Morceaux}
  \acro{CLCIR}{Conservative Level Contour Interface Reconstruction} 
  \acro{CBIR}{Cubic B\'{e}zier Interface Reconstruction} 
  \acro{CVTNA}{Centroid Vertex Triangle Normal Averaging} 
  \acro{CFD}{Computational Fluid Dynamics}
  \acro{CFL}{Courant-Friedrichs-Lewy}
  \acro{CPT}{Cell-Point Taylor}
  \acro{CPU}{Central Processing Unit}
  \acro{CSG}{Computational Solid Geometry}
  \acro{CV}{Control Volume}
  \acro{DG}{Discontinuous Galerking Method}
  \acro{DR}{Donating Region}
  \acro{DRACS}{Donating Region Approximated by Cubic Splines}
  \acro{DDR}{Defined Donating Region}
  \acro{DGNR}{Distance-Gradient Normal Reconstruction}
  \acro{DNS}{Direct Numerical Simulations}
  \acro{EGC}{Exact Geometric Computation}
  \acro{EI-LE}{Eulerian Implicit - Lagrangian Explicit}
  \acro{EILE-3D}{Eulerian Implicit - Lagrangian Explicit 3D}
  \acro{EILE-3DS}{Eulerian Implicit - Lagrangian Explicit 3D Decomposition Simplified}
  \acro{ELVIRA}{Efficient Least squares Volume of fluid Interface Reconstruction Algorithm}
  \acro{EMFPA}{Edge-Matched Flux Polygon Advection}
  \acro{EMFPA-SIR}{Edge-Matched Flux Polygon Advection and Spline Interface Reconstruction}
  \acro{FDM}{Finite Difference Method}
  \acro{FEM}{Finite Element Method}
  \acro{FMFPA-3D}{Face-Matched Flux Polyhedron Advection}
  \acro{FNB}{Face in Narrow Band test}
  \acro{FT}{Flux Triangulation}
  \acro{FV}{Finite Volume}
  \acro{FVM}{Finite Volume method}
  \acro{GPCA}{Geometrical Predictor-Corrector Advection}
  \acro{HTML}{HyperText Markup Language}
  \acro{HPC}{High Performance Computing}
  \acro{HyLEM}{Hybrid Lagrangian–Eulerian Method for Multiphase flow}
  \acro{IO}{Input / Output}
  \acro{IDW}{Inversed Distance Weighted}
  \acro{ISA}{iso-advector scheme}
  \acro{IDWGG}{Inversed Distance Weighted Gauss Gradient}
  \acro{LENT}{Level Set / Front Tracking}
  \acro{LE}{Lagrangian tracking / Eulerian remapping}
  \acro{LEFT}{hybrid level set / front tracking}
  \acro{LFRM}{Local Front Reconstruction Method}
  \acro{LVIRA}{Least squares Volume of fluid Interface Reconstruction Algorithm}
  \acro{LLSG}{Linear Least Squares Gradient}
  \acro{LS}{Least Squares}
  \acro{LSF}{Least Squares Fit}
  \acro{IDWLSG}{Inverse Distance Weighted Least Squares Gradient}
  \acro{LCRM}{Level Contour Reconstruction Method}
  \acro{LSG}{Least Squares Gradient}
  \acro{LSP}{Liskov Substitution Principle}
  \acro{MCE}{Mean Cosine Error}
  \acro{MoF}{Moment of Fluid}
  \acro{MS}{Mosso-Swartz}
  \acro{NIFPA}{Non-Intersecting Flux Polyhedron Advection}
  \acro{NS}{Navier-Stokes}
  \acro{NP}{Non-deterministic Polynomial}
  \acro{OOD}{Object Oriented Design}
  \acro{OD}{Owkes-Desjardins scheme}
  \acro{ODE}{ordinary differential equation}
  \acro{OD-S}{Owkes-Desjardins Sub-resolution}
  \acro{OT}{Oriented Triangulation}
  \acro{OCPT}{Oriented Convex Polygon Triangulation}
  \acro{EPT}{Edge-based Polygon Triangulation}
  \acro{PDE}{Partial Differential Equation}
  \acro{PAM}{Polygonal Area Mapping Method}
  \acro{iPAM}{improved Polygonal Area Mapping Method}
  \acro{PIR}{Patterned Interfacce Reconstruction}
  \acro{PCFSC}{Piecewise Constant Flux Surface Calculation}
  \acro{PLIC}{Piecewise Linear Interface Calculation}
  \acro{RTS}{Run-Time Selection}
  \acro{RKA}{Rider-Kothe Algorithm}
  \acro{RK}{Runge-Kutta}
  \acro{RTT}{Reynolds Transport Theorem}
  \acro{SCL}{Space Conservation Law}
  \acro{SMCI}{Surface Mesh / Cell Intersection}
  \acro{SFINAE}{Substitution Failure Is Not An Error}
  \acro{SIR}{Spline Interface Reconstruction}
  \acro{SLIC}{Simple Line Interface Calculation}
  \acro{SRP}{Single Responsibility Principle}
  \acro{STL}{Standard Template Library}
  \acro{TBDS}{Taylor-Backward Differencing Scheme}
  \acro{TBES}{Taylor-Euler Backward Scheme}
  \acro{THINC/QQ}{Tangent of Hyperbola Interface Capturing with Quadratic surface representation and Gaussian Quadrature}
  \acro{UML}{Unified Modeling Language}
  \acro{UFVFC}{Unsplit Face-Vertex Flux Calculation}
  \acro{VOF}{Volume of Fluid}
  \acro{YSR}{Youngs' / Swartz Reconstruction algorithm}
\end{acronym}                     

\title{Numerical wetting simulations using the plicRDF-isoAdvector unstructured Volume-of-Fluid (VOF) method}

\author[]{Muhammad Hassan Asghar}
\ead{asghar@mma.tu-darmstadt.de}

\author[]{Mathis Fricke}
\ead{fricke@mma.tu-darmstadt.de}

\author[]{Dieter Bothe}
\ead{bothe@mma.tu-darmstadt.de}

\author[]{{Tomislav Mari\'{c}}\corref{corr}}
\cortext[corr]{Corresponding author}
\ead{maric@mma.tu-darmstadt.de}

\address{Mathematical Modeling and Analysis Group, TU Darmstadt}

\begin{abstract}
\textbf{This is the preprint version of the published manuscript \url{https://doi.org/10.1016/j.camwa.2024.12.015}: please cite the published manuscript when referring to the contents of this document.}

Numerical simulation of wetting and dewetting of geometrically complex surfaces benefits from the boundary-fitted unstructured Finite Volume method because it discretizes boundary conditions on geometrically complex domain boundaries with second-order accuracy and simplifies the simulation workflow. \textcolor{Reviewer3}{The plicRDF-isoAdvector method, an unstructured geometric Volume-of-Fluid (VOF) method,} reconstructs the Piecewise Linear Interface Calculation (PLIC) interface by \underline{r}econstructing signed \underline{d}istance \underline{f}unctions (RDF). \textcolor{Reviewer3}{This method is chosen} to investigate wetting processes because of its volume conservation property and high computational efficiency. The present work verifies and validates the plicRDF-isoAdvector method for wetting problems, employing five different case studies.\ The first study investigates the accuracy of the interface advection near walls. The method is further investigated for the spreading of droplets on a flat and a spherical surface, respectively, for which excellent agreement with the reference solutions is obtained. Furthermore, a validation study using a droplet spreading test case is carried out. The uncompensated Young stress is introduced in the contact angle boundary condition, which significantly improves the validation of the numerical method. Furthermore, a 2D capillary rise is considered, and a numerical comparison based on results from previous work is performed.\ A suite with all case studies, input data, and Jupyter Notebooks used in this study are publicly available to facilitate further research and comparison with other numerical codes.
\end{abstract}

\end{frontmatter}


\section{Introduction}
\label{sec:Introduction} 
The wetting of a solid surface by a liquid is encountered in many natural and technical processes, including the spreading of paint, ink, lubricant, dye, or pesticides. In many typical applications, the solid surface is not flat or homogeneous but shows geometrically complex structures, chemical heterogeneity or porosity.\ In particular, it has been demonstrated extensively \cite{Marengo2022} that features like surface structure and roughness can be used as a tool to significantly enhance the process performance. 
To describe and predict wetting processes on complex surfaces, it is necessary to develop simulation tools that can handle geometrically complex boundaries and dynamics of the contact line. The unstructured Volume of Fluid (VOF) method (cf.\ \citep{maric2020unstructured} for a recent review) is widely used for simulating two-phase flows in geometrically complex technical systems, as it effectively handles geometrical complexity of the solution domain using polyhedral finite volume discretization.

%

The unstructured VOF method can be classified into two categories regarding the underlying approach for the advection of volume fractions, i.e., the discretized version of the phase indicator function - algebraic and geometric VOF methods. Algebraic VOF methods solve a linear algebraic system for the advection of the volume fraction field.\ A well-known OpenFOAM~\citep{openfoam1} solver that uses the algebraic VOF method is interFoam \cite{deshpande2012evaluating}.\ Although computationally very efficient, algebraic VOF methods may lead to inaccurate results \citep{deshpande2012evaluating,roenby2016computational} caused by artificial diffusion of the interface. On the other hand, geometric VOF methods reconstruct the fluid interface to approximate the phase-specific volumes fluxed through each face of the cell (see \citep{maric2020unstructured} for a recent review). An efficient geometric approach is underlying the plicRDF-isoAdvector unstructured geometric VOF method \citep{roenby2016computational,scheufler2019accurate} developed by \citet{roenby2016computational}.

%
The plicRDF-isoAdvector method is based on geometric approximations both in the interface reconstruction step and in the interface advection step (cf.\ \cref{sec:numerical-method}).\ It achieves second-order convergence of the geometrical VOF advection error in the $L_1$-norm on unstructured meshes for time steps restricted to CFL numbers below $0.2$\ ~\citep{scheufler2019accurate}.\ \citet{gamet2020validation} have validated the plicRDF-isoAdvector method for rising bubbles and benchmarked it against interFoam~\citep{deshpande2012evaluating}, Basilisk~\citep{popinet2015quadtree} and results from the Finite Element Methods available in the existing literature (TP2D~\citep{osher1988fronts,turek1999efficient}, FreeLIFE~\citep{parolini2005finite}, and MooNMD~\citep{john2004moonmd}).\ In particular, the plicRDF-isoAdvector method performed better than interFoam in capturing the helical trajectory and shape of the bubble for surface tension dominant flow. \citet{siriano2022numerical} have tested the numerical method for rising bubbles with high-density ratios.\ \citet{lippert2022benchmark} have proposed a set of simulation benchmarks for surface-tension-driven incompressible two-phase flows. They have studied the occurrence of spurious currents at the interface and benchmarked the plicRDF-isoAdvector method against interFoam~\citep{deshpande2012evaluating}, and Basilisk~\citep{popinet2015quadtree}. They have also investigated the different curvature models available in the TwoPhaseFlow OpenFOAM project~\citep{TwoPhaseFlow} and showed that, in general, Basilisk and plicRDF-isoAdvector with RDF curvature model produce the smallest error. \citet{giefer2023impact} studied the effects of wettability on interfacial propagation in microcapillary tubes using the interFlow~\citep{TwoPhaseFlow} solver, which employs the plicRDF-isoAdvector method. They categorized different propagation regimes based on the balance of viscous forces to interfacial forces. These studies show that the plicRDF-isoAdvector method is an efficient and accurate unstructured geometrical VOF method for two-phase flow problems. Here we verify, validate, and further extend the plicRDF-isoAdvector method for wetting problems.

%

We investigate five canonical wetting case studies. For all five case studies, we have made the input data, the primary data, the secondary data, and the post-processing utilities publicly available online~\citep{B01code,asgharinput,asgharcodeocean}.\ The post-processing, based on Jupyter notebooks~\citep{jupyter}, simplifies the verification/validation of wetting processes, not just for OpenFOAM but for any other simulation software, provided that the files storing the secondary data (error norms) are formatted as described in the documentation ~\citep{asgharnotebooks,asgharcodeocean}.

The structure of the paper is the following: the first case study is the near-wall interface advection verification test proposed by \citet{fricke2020contact}. It investigates the numerical contact angle evolution when the interface is advected using a known divergence-free velocity field. The results are presented in \cref{sec:Interface-advection-test}. 

Next, droplet spreading on a flat surface is considered - a classical wetting validation case study. \citet{dupont2010numerical} have studied the droplet shape with stationary geometrical relations.\ The droplet spreading with and without the influence of gravity is studied in \cref{sec:Droplet-spreading-on-a-flat-surface}. The dynamics of droplet spreading are validated by experimental results of \citet{lavi2004}, and comparison with other numerical methods are presented in \cref{sec:partial_wetting_dynamics}.

Subsequently, droplet spreading on a spherical surface \citep{patel2017coupled} is considered for testing the accuracy of the plicRDF-isoAdvector method on unstructured near-wall refined meshes for a geometrically more complex surface. The results are presented in \cref{sec:Droplet-spreading-on-a-spherical-surface}. 

Lastly, in line with \citet{grunding2020comparative}, we study the transient capillary rise based on full continuum mechanical simulations.\ Here, contact line singularity as first described by Huh and Scriven \citep{huh1971hydrodynamic} is regularized by means of the Navier slip condition. \citet{grunding2020comparative} show that both the dimensionless group identified in \citep{Quere1999,fries2009dimensionless}, and the Navier slip length have a major impact on the rise dynamics.\ The present study compares the simulation results for the capillary rise dynamics with the data from \cite{grunding2020comparative}. The results are presented in \cref{sec:2D-capillary-rise}. 
%


This study uses the ESI OpenFOAM version (git tag OpenFOAM-v2312)~\citep{OpenFOAMv2312}.\ The solver interFlow from the TwoPhaseFlow OpenFOAM project~\citep{TwoPhaseFlow} (branch of2312) is used for the plicRDF-isoAdvector method. A Python library, PyFoam~\cite{pyFoam}, version 2021.6, has been used for setting up parameter studies. For initialization of the droplet, the volume fraction field is computed using the Surface-Mesh/Cell Approximation Algorithm (SMCA) \citep{TOLLE2022108249} and exact implicit surfaces.\ An OpenFOAM submodule, cfMesh~\citep{cfMesh}, is employed for discretizing the domain using unstructured meshes, and meshing details of individual studies are provided in respective sections. An open-source software, FreeCAD~\citep{freecad}, version 0.18.4, is used to create .stl files. 
%

%

\section{Numerical method}
\label{sec:numerical-method}
\begin{figure}[H]
    \centering
    \includegraphics[width=\textwidth]{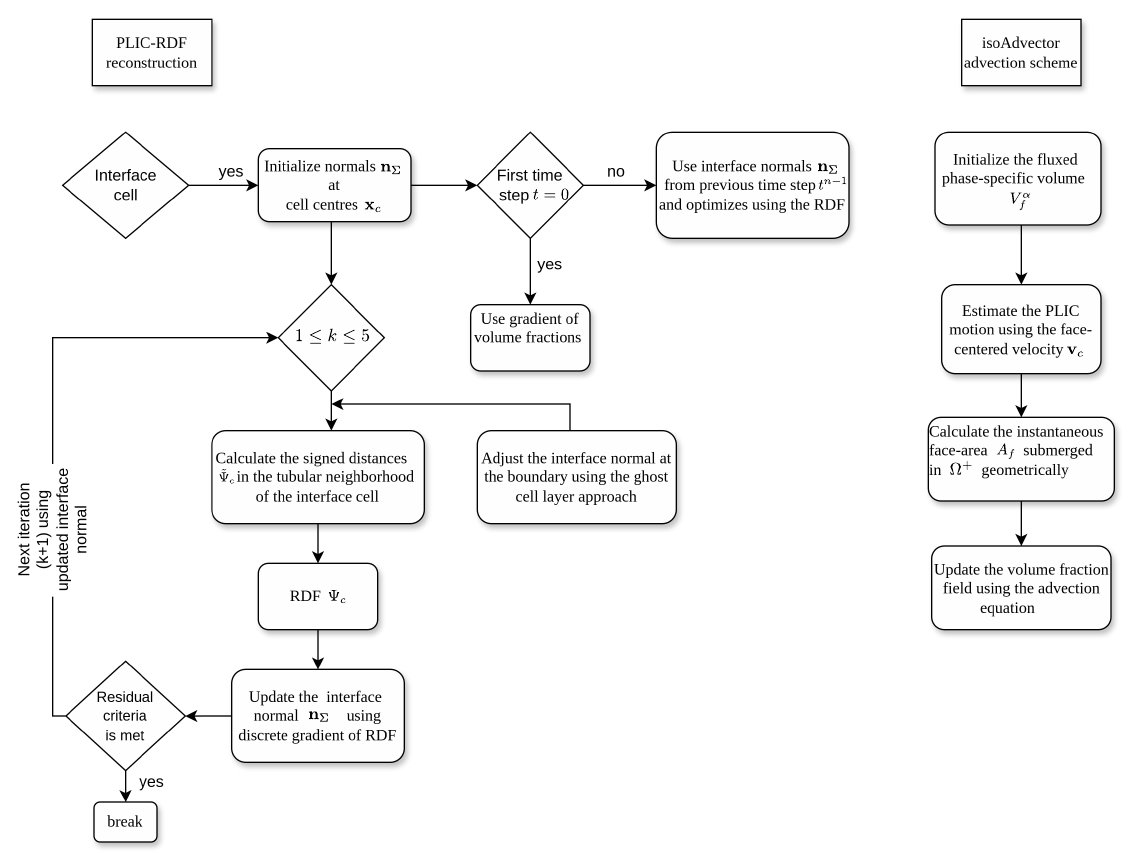}
    \caption{Flow diagram of the solution algorithm of the plicRDF-isoAdvector method. The plicRDF reconstruction scheme reconstructs the signed distance functions RDF and iteratively improves the orientation of the interface normals. The isoAdvector scheme then updates the phase-specific volume fluxes through each face of the control volume $\Omega$ by geometrical calculations of the face-area $A_f$, submerged in phase $\Omega^+$.}
    \label{app:flow-chart}
\end{figure}

In this section, we provide an overview of the plicRDF-isoAdvector numerical method~\citep{scheufler2019accurate,TwoPhaseFlow} (see \Cref{app:flow-chart} for an overview of the overall solution algorithm of the plicRDF-isoAdvector method).

\subsection{ Volume-of-Fluid method}
\begin{figure}[tb]
	\centering
	\captionsetup{position=top}
	\def\svgwidth{0.5\textwidth}
	{\footnotesize
		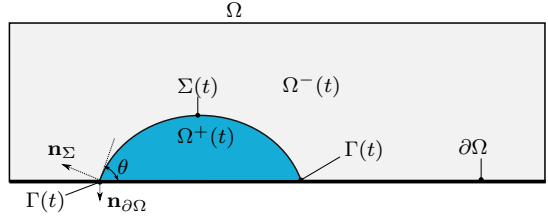
	}
	\vspace{0.5em}
	\caption{Schematic diagram of a multiphase domain $\Omega$ with contact line $\Gamma(t)$. The interface $\Sigma(t)$ with normal $\mathbf{n}_\Sigma$ intersects the domain solid boundary $\partial\Omega$ at the contact angle $\theta$.}
	\label{fig:schematic-diagram-of-multiphase-domain}
\end{figure}

Consider a flow domain $\Omega$ as illustrated in \cref{fig:schematic-diagram-of-multiphase-domain}, composed of two sub-domains occupied with different incompressible fluids, denoted by $\Omega^+(t)$ and $\Omega^-(t)$. 
The phase-indicator function
\begin{equation}
\chi(t,\textbf{x}) =\begin{cases}
1 & \textbf{x} \in \Omega^+(t) \\
0 & \textbf{x} \notin \Omega^+(t)
\end{cases}
\label{eq:phase-indicator}
\end{equation}
distinguishes the sub-domains $\Omega^{\pm}$. Evidently, the sub-domain $\Omega^+$ is then 
\begin{equation}
\Omega^+(t) := \{\textbf{x} \in\Omega: \chi(t, \textbf{x}) =1\}.
\label{eq:phase-description-using-phase-indicator-function}
\end{equation}
The volume fraction $\alpha_c(t)$ of the phase $\Omega^+(t)$ inside a fixed control volume $\Omega_c$ at time $t$ is defined as
\begin{equation}
\alpha_c(t) :=\frac{1}{|\Omega_c|} \int_{\Omega_c} \chi(t,\textbf{x}) \,dV\,.
\label{eq:def-of-volume-fraction}
\end{equation}
The value of the volume fraction inside a cell indicates whether phase $\Omega^+(t)$ is present inside the cell. Indeed, it holds that
\begin{equation}
\begin{split}
\alpha_c  = 1  & \Longleftrightarrow \text{cell is inside } \Omega^+ (t), \\
\alpha_c  \in (0,1)  & \Longleftrightarrow  \text{cell intersects the interface (``interface cell")}, \\
\alpha_c  = 0 & \Longleftrightarrow\text{cell is inside } \Omega^- (t). 
\end{split}
\end{equation}
Within each phase $\Omega^+(t)$ or $\Omega^-(t)$, mass conservation for incompressible fluids has the form of
\begin{equation}
\nabla \cdot \mathbf{v} = 0,
\end{equation}
where $\mathbf{v}$ is the fluid velocity.\ 
In situations without phase change, the phase indicator function keeps its value along trajectories of the two-phase flow, i.e.,\ it satisfies (in a distributional sense) the transport equation
\begin{equation}
   \partial_t \chi + {\bf v} \cdot \nabla \chi = 0.
   \label{eq:chi-transport}
\end{equation}

The phase-indicator $\chi(t, \cdot)$ determines the phase-dependent local values of the physical quantities such as the single-field density $\rho$ and viscosity $\mu$ in the single-field formulation of the two-phase Navier Stokes equations. For constant densities $\rho^+$, $\rho^-$ and constant viscosities $\mu^+$, $\mu^-$, we have
\begin{equation}
\rho(t, \mathbf{x}) = \chi(t, \mathbf{x}) \rho^+ + (1-\chi(t,\mathbf{x}))\rho^-,
\end{equation}
and
\begin{equation}
\mu(t, \mathbf{x}) = \chi(t, \mathbf{x}) \mu^+ + (1-\chi(t,\mathbf{x}))\mu^-.
\end{equation}
The momentum balance reads as 
\begin{equation}
\partial_t  (\rho \mathbf{v}) + \nabla \cdot (\rho \mathbf{v}\mathbf{v}) =  -\nabla p^{'} - \nabla  (\rho\mathbf{g} \cdot \x) + \nabla \cdot \left(\mu(\nabla \mathbf{v} +\nabla \mathbf{v}^T)\right) + \mathbf{f}_\Sigma,
\label{eq:momentum-eq}
\end{equation}
where $p^{'}=p-\rho\mathbf{g}h$ is the modified pressure, $\textbf{g}$ is gravitational acceleration, and
\begin{equation}
    \mathbf{f}_\Sigma :=  \sigma\kappa\mathbf{n}_\Sigma\delta_\Sigma
    \label{eq:surf-tension-force}
\end{equation}
is the surface tension force with the surface tension $\sigma$, the mean curvature $\kappa$ , the Dirac's delta distribution $\delta_\Sigma$ at the interface $\Sigma(t)$ with normal $\mathbf{n}_\Sigma$.

The VOF methods integrate~\cref{eq:chi-transport} over a fixed control volume $\Omega_c$ within a time step $[t^n, t^{n+1}]$, followed by the application of the Reynolds transport theorem. This leads to the integral form of the volume fraction transport equation (see \citep{maric2020unstructured} for details) according to
\begin{equation}
\alpha_{c}(t^{n+1}) = \alpha_{c}(t^n) - \frac{1}{|\Omega_c|} \int_{t^n}^{t^{n+1}} \int_{\partial\Omega_c} \chi(t,\textbf{x})  \mathbf{v}\cdot\mathbf{n}~dS~dt .
\label{eq:intergral-form-of-volume-fraction-transport}
\end{equation}
The boundary $\partial\Omega_c$ of the cell $\Omega_c$, used on the r.h.s. of \cref{eq:intergral-form-of-volume-fraction-transport}, is a union of surfaces (faces) that are bounded by line segments (edges), namely
\begin{equation}
\partial\Omega_c = \cup_{f\in F_c} S_f,
\label{eq:cellboundary}
\end{equation}
where $F_c$ is the set of cell-local indices of all faces of the cell $\Omega_c$.
Using this decomposition (\cref{eq:cellboundary}) of $\partial \Omega_c$, \cref{eq:intergral-form-of-volume-fraction-transport} can be written as
\begin{equation}
\alpha_{c}(t^{n+1}) = \alpha_{c}(t^n) - \frac{1}{|\Omega_c|} \sum_{f \in F_{c}} \int_{t^n}^{t^{n+1}} \int_{S_f} \chi(t,\textbf{x})  \mathbf{v}\cdot\mathbf{n}~dS~dt .
\label{eq:volfrac-final}
\end{equation}
The double integral on the right-hand side of \cref{eq:volfrac-final} defines the amount of the phase-specific volume $|V_f^\alpha|$ fluxed over the face $S_f$ within a time interval $[t^n, t^{n+1}]$.
\Cref{eq:volfrac-final} is still an exact equation since no approximations have been made so far. It is the basis for every geometric VOF method \citep{maric2020unstructured}, that all obtain the form, 
\begin{equation}
\alpha_{c}(t^{n+1}) = \alpha_{c}(t^n) - \frac{1}{|\Omega_c|}   \sum_{f \in F_{c}} V_f^\alpha, 
\label{eq:volfraction-vfalpha}
\end{equation}
differing only in how $V_f^\alpha$ is calculated. If one could calculate $V_f^\alpha$ exactly for arbitrary $\mathbf{v}$ and $\chi$, and exactly represent the domain boundary $\bigcup_{c\in C} \partial \Omega_c$, where $C$ is the set that contains the indices of all boundary faces, \cref{eq:volfraction-vfalpha} would be an exact equation.
\subsection{The plicRDF-isoAdvector method}
\label{subsec-plicRDF-isoAdvector}
Geometric VOF methods rely on a cell-local geometrical approximation of the interface.\ Each interface reconstruction algorithm thus aims to compute the interface normal $\mathbf{n}_{\Sigma,c}$ ($\mathbf{n}_{c}$ for short) and position $\mathbf{P}_{c}$.\ At first, the interface orientation algorithm approximates $\mathbf{n}_{c}$.\ Then the interface positioning algorithm places the interface at $\mathbf{P}_{c}$. The well-known Piecewise Linear Interface Calculation (PLIC) algorithm is the common interface reconstruction algorithm. The Youngs' algorithm~\cite{youngs1982time}, the Least Squares Fit (LSF) algorithm~\cite{scardovelli2003interface}, and the plicRDF reconstruction method~\cite{scheufler2019accurate} are a few examples of the PLIC-based interface orientation algorithms.  

\begin{figure}[h!]
	\centering
	\captionsetup{position=top}
	\def\svgwidth{0.85\textwidth}
	{\footnotesize
\begingroup%
  \makeatletter%
  \providecommand\color[2][]{%
    \errmessage{(Inkscape) Color is used for the text in Inkscape, but the package 'color.sty' is not loaded}%
    \renewcommand\color[2][]{}%
  }%
  \providecommand\transparent[1]{%
    \errmessage{(Inkscape) Transparency is used (non-zero) for the text in Inkscape, but the package 'transparent.sty' is not loaded}%
    \renewcommand\transparent[1]{}%
  }%
  \providecommand\rotatebox[2]{#2}%
  \newcommand*\fsize{\dimexpr\f@size pt\relax}%
  \newcommand*\lineheight[1]{\fontsize{\fsize}{#1\fsize}\selectfont}%
  \ifx\svgwidth\undefined%
    \setlength{\unitlength}{582.58573547bp}%
    \ifx\svgscale\undefined%
      \relax%
    \else%
      \setlength{\unitlength}{\unitlength * \real{\svgscale}}%
    \fi%
  \else%
    \setlength{\unitlength}{\svgwidth}%
  \fi%
  \global\let\svgwidth\undefined%
  \global\let\svgscale\undefined%
  \makeatother%
  \begin{picture}(1,0.27991198)%
    \lineheight{1}%
    \setlength\tabcolsep{0pt}%
    \put(0,0){\includegraphics[width=\unitlength,page=1]{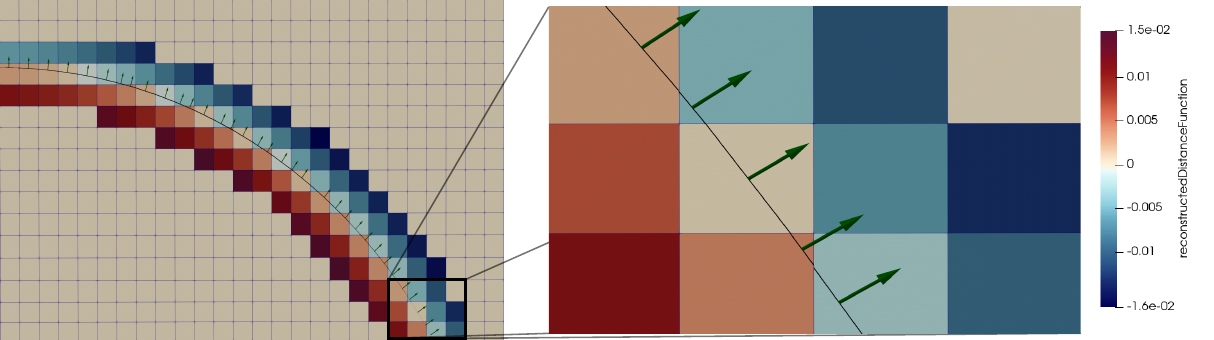}}%
    \put(0.66817867,0.09655975){\color[rgb]{0,0,0}\rotatebox{35.943178}{\makebox(0,0)[lt]{\lineheight{1.25}\smash{\begin{tabular}[t]{l}$\Tilde{\Psi}_c$\end{tabular}}}}}%
    \put(0.63935441,0.16600896){\color[rgb]{0,0,0}\makebox(0,0)[lt]{\lineheight{1.25}\smash{\begin{tabular}[t]{l}$\textbf{n}_{n}$\end{tabular}}}}%
    \put(0.58360638,0.11931201){\color[rgb]{0,0,0}\makebox(0,0)[lt]{\lineheight{1.25}\smash{\begin{tabular}[t]{l}$\textbf{P}_{n}$\end{tabular}}}}%
    \put(0,0){\includegraphics[width=\unitlength,page=2]{RDF.pdf}}%
    \put(0.73105475,0.11715768){\color[rgb]{0,0,0}\makebox(0,0)[lt]{\lineheight{1.25}\smash{\begin{tabular}[t]{l}$\textbf{x}_c$\end{tabular}}}}%
    \put(0,0){\includegraphics[width=\unitlength,page=3]{RDF.pdf}}%
  \end{picture}%
\endgroup%

	}
	\vspace{0.5em}
	\caption{The reconstructed signed distance function field for a droplet, reconstructed in the tubular neighborhood of the interface cells. The straight line inside each interface cell represents the reconstructed interface. The interface normal $\mathbf{n}_{n}$ positioned at $\mathbf{P}_{n}$ is given by the Least Squares Finite Volume gradient of the RDF $\Psi_c$ using \cref{eq:normal-update}.}
	\label{fig:plicRDF}
\end{figure}

The plicRDF reconstruction method~\cite{scheufler2019accurate}
\underline{r}econstructs the so-called signed \underline{d}istance \underline{f}unctions (RDFs) in the tubular neighborhood of the interface cells (cf.  \cref{fig:plicRDF}).
%
The normals are then approximated as a discrete gradient of the RDFs $\Psi$, i.e., 
\begin{equation}
\textbf{n}^{k+1}_{c} = \frac{\nabla_c \Psi_c^k}{\|\nabla_c \Psi^k_c \|},
\label{eq:normal-update}
\end{equation}
where $\nabla_c$ is a discrete unstructured Finite Volume least squares gradient and $k = 1, \dots, 5$ is the reconstruction algorithm iteration index used in \citep{scheufler2019accurate}. The interface positioning algorithm uses $\{\mathbf{n}^{k+1}_{n}\}_{n \in I}$, with $I$ as the cell-local index set of all interface-cells, to position the PLIC interface planes, resulting in $\{\mathbf{P}^{k+1}_{n}\}_{n \in I}$ needed for the next iteration.

\subsubsection{The isoAdvector advection scheme}
The isoAdvector numerical method~\citep{roenby2016computational} calculates the phase-specific fluxed volume $V_f^\alpha$ as
\begin{equation}
\begin{aligned}
V_{f}^\alpha & =  \int_{t^n}^{t^{n+1}} \int_{S_f} \chi(t,\textbf{x}) \mathbf{v}\cdot\mathbf{n}~dS~dt 
=  \int_{t^n}^{t^{n+1}} \mathbf{v}_f(t) \cdot \hat{\mathbf{n}}_f \int_{S_f} \chi(t,\textbf{x}) ~dS~dt 
+ O(h^2) \\
& = \int_{t^n}^{t^{n+1}}  \frac{F_f(t)}{|S_f|} \int_{S_f} \chi(t,\textbf{x}) ~dS~dt + O(h^2) \\
& =  \frac{0.5(F_f^n + F_f^{n+1})}{|S_f|} \int_{t^n}^{t^{n+1}} \int_{S_f} \chi(t,\textbf{x}) ~dS~dt + O(\Delta t^2) + O(h^2)
\label{eq:phase-specific-flux}
\end{aligned}
\end{equation}
with  velocity  $\mathbf{v}_f(t)$ as the face-centered velocity resulting in $O(h^2)$, defining $F_f \coloneqq \mathbf{v}_f \cdot \mathbf{n}_f |S_f|$, the volumetric flux across the face $S_f$.\ The vector $\mathbf{n}_f$ is the unit-normal vector of the face $S_f$. For a fixed prescribed velocity (for example, for the purpose of verifying advection), the velocity $\mathbf{v}_f$ is directly known. Otherwise, it is obtained from the solution of the Navier Stokes Equation, as detailed in \citep{Tolle2020}.  

%
The isoAdvector scheme geometrically evaluates the integral
\begin{equation}
\int_{t^n}^{t^{n+1}} \int_{S_f} \chi(t,\textbf{x}) ~dS~dt =  \int_{t^n}^{t^{n+1}} A_f(t)~dt,
\label{eq:instantaneous-sbmerged-area}
\end{equation}
where $A_f(t)$ is the instantaneous face-area submerged in $\Omega^+(t)$ at time $t$. Details on the submerged face-area integration are available in \citep{roenby2016computational}.

%
With the calculation of the phase-specific fluxed volume $V_f^\alpha$, the cell volume fraction value is updated using~\cref{eq:volfraction-vfalpha}.\ The isoAdvector scheme restores strict boundedness $\alpha_c \in [0,1]$ by redistributing over-and-undershoots in $\alpha_c$ in the upwind direction.
\subsubsection{Boundary conditions at the solid boundary}
In the following, we work in the frame of reference where the solid boundary $\partial\Omega$ is at rest. We assume the boundary of the domain to be impermeable, i.e.,
\begin{equation}
\mathbf{v}_\perp \left.\right|_{\partial\Omega} = 0,
\label{eq:permeability-condition}    
\end{equation}
where $\mathbf{v}_\perp$ is the velocity component normal to the boundary. The Navier slip boundary condition (for a flat boundary) is given as%
\begin{equation}
\left. \mathbf{v}_\parallel\right|_{\partial\Omega} = \lambda \left.\frac{\partial\mathbf{v}_\parallel}{\partial y} \right|_{\partial\Omega},
\label{eq:Navier-slip}
\end{equation}
where $\mathbf{v}_{||}$ is the velocity component tangential to the boundary, $\lambda$ is the slip length, and $\frac{\partial\mathbf{v}_\parallel}{\partial y}$ is the velocity gradient in the wall-normal direction $y$. Note that the no-slip boundary condition, i.e.,\
\begin{equation}
\mathbf{v}_{||} \left.\right|_{\partial\Omega} = 0,
\label{eq:no-slip}    
\end{equation}
is recovered from \cref{eq:Navier-slip} for $\lambda=0$.
In this study, we apply the impermeability condition \eqref{eq:permeability-condition} together with either the Navier slip (\cref{eq:Navier-slip}) or the no-slip~(\cref{eq:no-slip}) condition.\\
\\
It is important to note that for the no-slip case, the numerical method relies on the so-called ``numerical slip''~\citep{renardy2001numerical} to move the contact line.\ The numerical slip is an inherent property of the advection algorithm, which uses the face-centred velocity to transport the volume fraction field.\ The face-centred velocity at the boundary cell is not strictly zero and therefore allows for the motion of the contact line. Since the amount of numerical slip is related to the mesh size, one will typically find a mesh dependence of the contact line speed if a numerical slip is used. On the other hand, mesh convergence of the contact line speed can be reached with the Navier slip condition provided that the slip length $\lambda$ is resolved by the mesh (see Section~\ref{sec:2D-capillary-rise}).\\
Although the solutions with the numerical slip and Navier slip differ in the contact line dynamics, it is to be noted that the stationary state is mesh-independent, and the numerical simulation results converge to the reference equilibrium solution regardless of the choice of numerical/Navier slip boundary conditions, as discussed in~\cref{sec:2D-capillary-rise}.
%

%
\subsubsection{Boundary condition at the contact line}
At the contact line, i.e., the region where the interface touches the domain boundary, the contact angle $\theta$ (see~\cref{fig:schematic-diagram-of-multiphase-domain}) is defined by the geometric relation
\begin{equation}
\cos \theta = -\mathbf{n}_\Sigma \cdot \mathbf{n}_{\partial\Omega},
\label{eq:contact_angle}
\end{equation}
where $\mathbf{n}_{\partial\Omega}$ is the outer unit normal vector of the domain boundary $\partial \Omega$.\ In this paper, for simplicity, the contact angle $\theta$ is always prescribed as the equilibrium contact angle, unless stated otherwise.\\

\begin{figure}[ht]
	\centering
	\captionsetup{position=top}
	\def\svgwidth{0.5\textwidth}
	{\footnotesize
		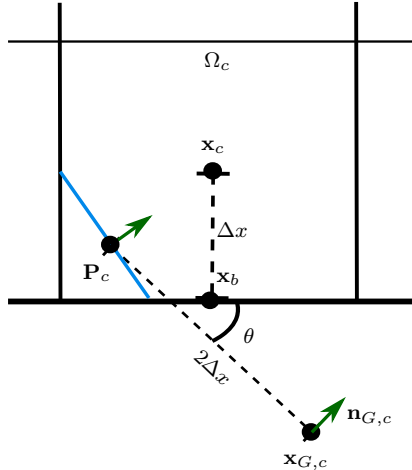
	}
	\caption{Contact angle treatment by the plicRDF-isoAdvector method. The ghost interface point $\textbf{x}_{G,c}$ with ghost interface normal $\textbf{n}_{G,c}$ satisfies the contact angle boundary condition and is used as an additional contribution to reconstructing the RDF at the centroid $\mathbf{x}_c$ of cell $\Omega_c$. }
	\label{fig:contact-angle-treatment}
\end{figure}

For the numerical treatment of boundary interface cells $\Omega_c$, \citet{scheufler2021twophaseflow} have introduced the concept of a ghost interface point $\textbf{x}_{G,c}$ placed on the opposite side of the boundary face $b$ of the cell $\Omega_c$~(\cref{fig:contact-angle-treatment}). It is the unique point at a distance of $2\Delta x$ from the PLIC centroid position $\textbf{P}_{c}$ such that the ghost interface normal $\textbf{n}_{G,c}$ becomes oriented in such a way that it satisfies the contact angle boundary condition. In order to transmit this information about the contact angle into the algorithm, the point $\textbf{x}_{G,c}$ and the normal $\textbf{n}_{G,c}$ are then used as an additional contribution to reconstruct the RDF at the centroid $\mathbf{x}_c$ of the cell $\Omega_c$.
\subsubsection{The curvature model}
The plicRDF-isoAdvector method models the surface tension force using \cref{eq:surf-tension-force}. The TwoPhaseFlow OpenFOAM project~\citep{TwoPhaseFlow} provides different models for the approximation of the mean curvature $\kappa$ (see \citep{scheufler2021twophaseflow} for details).

The present study presents the results using the \textit{parabolic fit} curvature model~\citep{scheufler2021twophaseflow}.\ For the mean curvature calculation in the interface cell $\Omega_c$, a surface given by a quadratic form is computed by fitting it to the PLIC centroids $\{\mathbf{P}_{ n}\}_{n \in N_c}$ inside all neighboring interface-cells. The curvature of the surface is then approximated by one of the quadratic surfaces.

The simulation results using \textit{parabolic fit}, \textit{RDF} and \textit{height function} curvature models with Navier slip boundary condition are also publicly available online \citep{fpnotebooks,rdfnotebooks,hfnotebooks}. The \textit{parabolic fit} curvature model has shown better performance for convergence with respect to the equilibrium state.

\section{Verification of the advection accuracy near the contact line}
\label{sec:Interface-advection-test}
\subsection{Definition of case study}
In this study, we consider the advection of the droplet interface using a divergence-free velocity field and report the accuracy of the interface advection near walls. It has been shown in \citep{fricke2019kinematic,fricke2018kinematics} that the contact line advection problem is a well-posed initial value problem if the velocity field is sufficiently regular and tangential to the domain boundary. The interface motion and the contact angle evolution can be computed from the velocity field and the initial geometry.\ Notably, the full time evolution of the contact angle can be inferred from the solution of a system of ordinary differential equations \cite{fricke2019kinematic}.\ In this study, we replicate one of the case studies presented by \citet{fricke2020contact}.\\
\begin{figure}[h!]
	\centering
	\captionsetup{position=top}
	\def\svgwidth{0.5\textwidth}
	{\footnotesize
\begingroup%
  \makeatletter%
  \providecommand\color[2][]{%
    \errmessage{(Inkscape) Color is used for the text in Inkscape, but the package 'color.sty' is not loaded}%
    \renewcommand\color[2][]{}%
  }%
  \providecommand\transparent[1]{%
    \errmessage{(Inkscape) Transparency is used (non-zero) for the text in Inkscape, but the package 'transparent.sty' is not loaded}%
    \renewcommand\transparent[1]{}%
  }%
  \providecommand\rotatebox[2]{#2}%
  \newcommand*\fsize{\dimexpr\f@size pt\relax}%
  \newcommand*\lineheight[1]{\fontsize{\fsize}{#1\fsize}\selectfont}%
  \ifx\svgwidth\undefined%
    \setlength{\unitlength}{136.03639303bp}%
    \ifx\svgscale\undefined%
      \relax%
    \else%
      \setlength{\unitlength}{\unitlength * \real{\svgscale}}%
    \fi%
  \else%
    \setlength{\unitlength}{\svgwidth}%
  \fi%
  \global\let\svgwidth\undefined%
  \global\let\svgscale\undefined%
  \makeatother%
  \begin{picture}(1,0.50515259)%
    \lineheight{1}%
    \setlength\tabcolsep{0pt}%
    \put(0,0){\includegraphics[width=\unitlength,page=1]{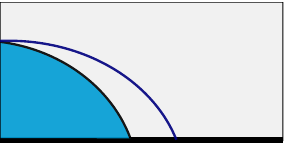}}%
    \put(0.0665325,0.2840944){\color[rgb]{0,0,0}\makebox(0,0)[lt]{\lineheight{1.25}\smash{\begin{tabular}[t]{l}$\Sigma(t_0)$\end{tabular}}}}%
    \put(0.22502156,0.3631666){\color[rgb]{0,0,0}\makebox(0,0)[lt]{\lineheight{1.25}\smash{\begin{tabular}[t]{l}$\Sigma(t)$\end{tabular}}}}%
    \put(0.42257312,0.28931552){\color[rgb]{0,0,0}\makebox(0,0)[lt]{\lineheight{1.25}\smash{\begin{tabular}[t]{l}$\dot{\mathbf{x}}(t) = \textbf{v}(t,\textbf{x}(t))$\end{tabular}}}}%
    \put(0,0){\includegraphics[width=\unitlength,page=2]{kinematic_transport_of_contact_angle.pdf}}%
    \put(0.2513726,0.21002234){\color[rgb]{0,0,0}\makebox(0,0)[lt]{\lineheight{1.25}\smash{\begin{tabular}[t]{l}$\mathbf{x}_0$\end{tabular}}}}%
    \put(0,0){\includegraphics[width=\unitlength,page=3]{kinematic_transport_of_contact_angle.pdf}}%
  \end{picture}%
\endgroup%

	}
	\vspace{0.5em}
	\caption{Kinematic transport of the contact angle.}
	\label{fig:kinematic-transport-of-the-contact-angle}
\end{figure}
\paragraph{Kinematics of contact angle transport:} In the following, we will study the time-evolution of the contact angle (see \Cref{fig:kinematic-transport-of-the-contact-angle} for a sketch of the kinematics transport) along a flow trajectory $\mathbf{x}(t)$ defined as the solution of the initial value problem
\begin{equation}
\dot{\mathbf{x}}(t;t_0,x_0) = \textbf{v}(t,\textbf{x}(t;t_0,x_0)), 
~ \textbf{x}(t_{0};t_0,x_0) = \textbf{x}_{0}.
\label{eq:trajectory-of-flow-ode}
\end{equation}
We may also write $\mathbf{x}(t)$ for short, keeping in mind that an initial position must be specified.\ It has been shown in \cite{fricke2019kinematic} that the contact line is invariant with respect to the flow generated by \eqref{eq:trajectory-of-flow-ode} provided that $\vec{v}$ is tangential, i.e.,\ if $\vec{v}_\bot = 0$ at the solid boundary. This means that a trajectory that starts at the contact line will always stay at the contact line and the function
\[ \theta(t) := \theta(t,\vec{x}(t)) \]
is well-defined. Mathematically, the time-evolution of the contact angle can be deduced from the time-evolution of the normal field $\mathbf{n}_{\Sigma}$ along $\mathbf{x}(t)$ via
\begin{align} 
\label{eq:contact-angle-numerical-relation}
\cos \theta(t,\vec{x}(t)) = - \mathbf{n}_{\Sigma}(t, \vec{x}(t)) \cdot \mathbf{n}_{\partial\Omega}(t,\vec{x}(t)).
\end{align}
The solution of the ordinary differential equation (ODE)  
\begin{equation}
\dot{\bm{\nu}}(t) = - \nabla \mathbf{v}(t,\mathbf{x}(t))^{\sf T} \cdot \bm{\nu}(t), 
~\bm{\nu}(t_{0}) = \mathbf{n}_{\Sigma}(t_0,x_0),
\label{eq:evolution-of-interface-normal-ode-I}
\end{equation} 
and normalization of $\bm{\nu}$ according to
\begin{equation} 
\textbf{n}_{\Sigma}(t, \textbf{x}(t)) = \frac{\bm{\nu}(t)}{\|\bm{\nu}(t)\|}
\label{eq:evolution-of-interface-normal-ode-II}
\end{equation}
provides $\textbf{n}_{\Sigma}(t, \textbf{x}(t))$.\\
\\
For this case study, we follow one of the examples in \cite{fricke2020contact}.\ The interface $\Sigma$ is advected using a velocity field called ``vortex-in-a-box" given by
\begin{equation}
\textbf{v}(t, x_1, x_2) = v_0  \cos \left(\frac{\pi t}{\tau}\right) (-\sin(\pi x_1)\cos(\pi x_2), \cos(\pi x_1)\sin(\pi x_2))
\label{eq:velocity-vortex}.
\end{equation}
The periodicity of the field in time allows for comparing the droplet's initial shape at $t=t_0$ and the shape after the time period $\tau$. If the advection problem is solved exactly, the droplet shape at $t=0$ will coincide with the shape at $t=\tau$. Otherwise, the difference in the volume fractions fields can be used to quantify the error. Moreover, the full time-evolution of the transported contact angle is obtained from the solution of the ODE system~(\cref{eq:trajectory-of-flow-ode,eq:evolution-of-interface-normal-ode-I,eq:evolution-of-interface-normal-ode-II}).
%

\subsubsection{Computational setup}
We consider a 2D computational domain with a droplet $\Omega^+(t)$ of a dimensionless radius of $R_0=0.2$ initialized on a flat surface $\partial \Omega$ 
%
%
at $(0.4,-0.1)$ position (cf. \cref{fig:schematic-diagram-of-multiphase-domain}), resulting in an initial contact angle
\begin{equation}
\theta_0 = \cos^{-1} \left(\frac{0.1}{0.2}\right) = 60^{\circ}. 
\end{equation}
In this study, we choose $v_0=0.1$, $\tau=0.2$, and ensure that the Courant number satisfies $ Co= U \frac{\Delta t}{\Delta x} < 0.01$. We quantify the geometrical shape error as
\begin{equation}
E_1 = \sum_{c} |\alpha_{c}(\tau) - \alpha_{c}(0)|V_c,
\label{eq:L1-error}
\end{equation}
and the maximum error in the transported contact angle as 
\begin{equation}
E_{\infty} = \text{max}_{t\in [0,T]} | \theta_{\text{num}}(t) - \theta_{\text{ref}}(t) |,
\label{eq:E_inf}
\end{equation}
where $T$ is the maximum simulation time.

\paragraph{Remark:} For practical reasons, a no-slip boundary condition at the bottom boundary and a zero-gradient boundary condition for the volume fractions are formally specified for the solver. However, the solution is not affected since this study is only an initial value problem.

\subsubsection{Simulation results}
\begin{figure}[h!]
	\centering
	\includegraphics[width=.5\textwidth]{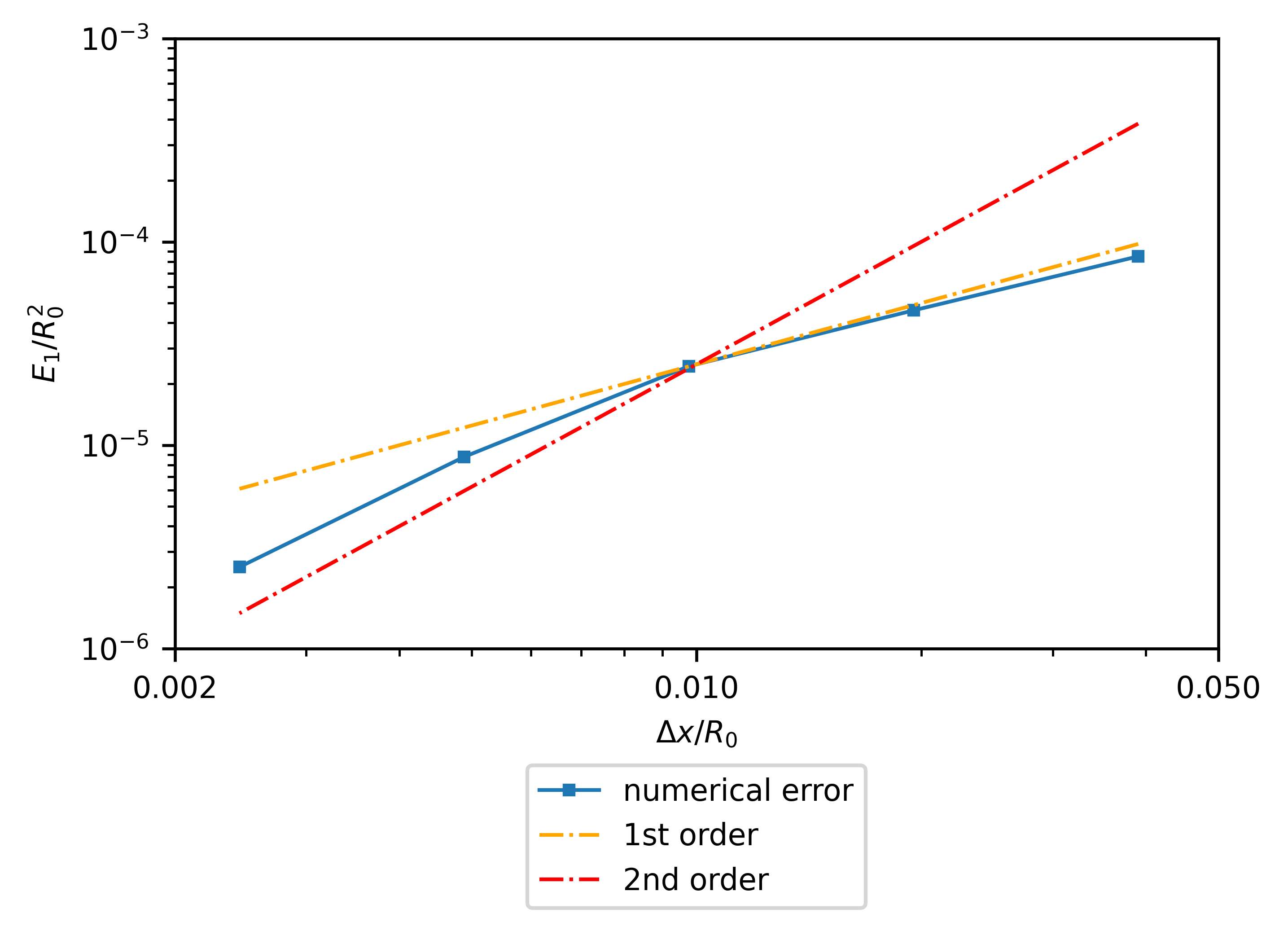}\hfill
	\caption{Convergence of $E_1$ errors for the velocity ﬁeld (\ref{eq:velocity-vortex}).}
	\label{fig:convergence-with-L1-norm}
\end{figure}

We discretized the domain using uniform meshes. 
\Cref{fig:convergence-with-L1-norm} shows the convergence of $E_1$ errors for the velocity field given by~\cref{eq:velocity-vortex}. As the mesh refinement increases, the order of convergence for $E_1$ errors also increases. A critical mesh size beyond which a transition in the $E_1$ errors convergence is observed as presented in \cref{fig:convergence-with-L1-norm}.\\

\begin{figure}[h!]
	\centering
	\includegraphics[width=.5\textwidth]{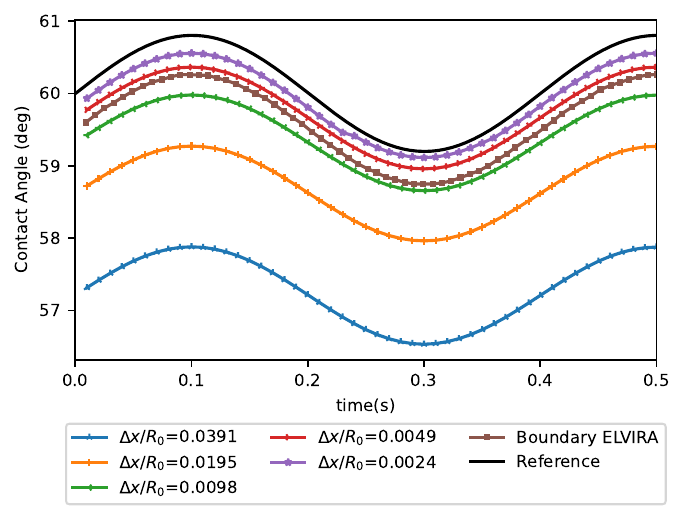}\hfill
	 \includegraphics[width=.49\textwidth, height=0.285\textheight]{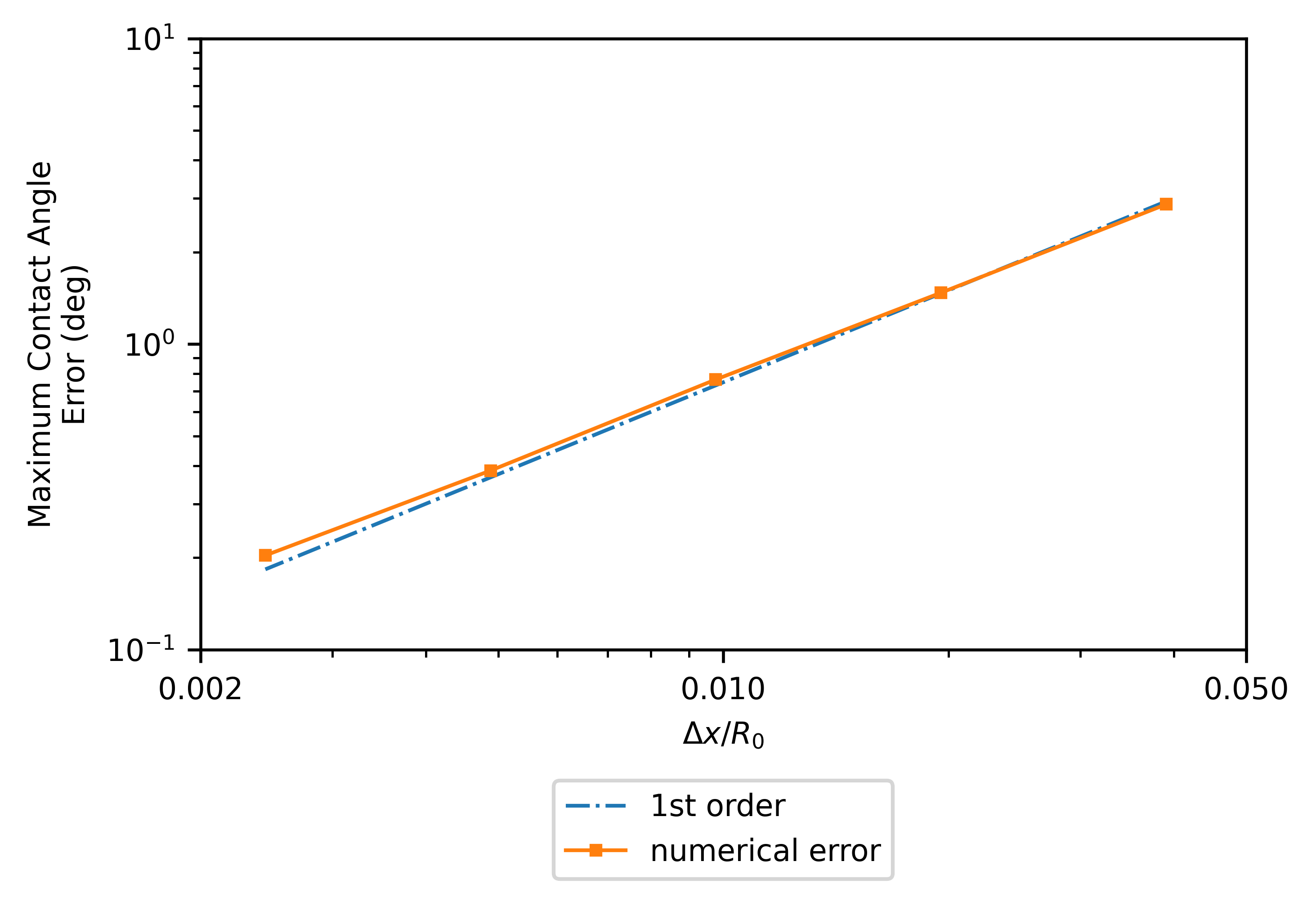}
	\caption{Left: Numerical contact angle evolution for the velocity field (\cref{eq:velocity-vortex}) with a uniform mesh. Additionally, the contact angle evolution results from the Boundary ELVIRA method~\cite{fricke2020contact} is shown for the highest available mesh resolution, i.e., $\Delta x / R_0 = 0.005$. Right: Maximum error in the transported contact angle (\cref{eq:E_inf}).}
	\label{fig:advection-results-for-uniform-mesh}
\end{figure}

\Cref{fig:advection-results-for-uniform-mesh} show the simulation results of the numerical evolution of the transported contact angle.\ The solutions of the ODE (\cref{eq:contact-angle-numerical-relation,eq:trajectory-of-flow-ode,eq:evolution-of-interface-normal-ode-I,eq:evolution-of-interface-normal-ode-II}) provide the reference value of the instantaneous transported contact angle $\theta_{\text{ref}}(t)$. It is noted that for the same mesh resolution, i.e., $\Delta x / R_0= 0.005$, the absolute error reported in \cite{fricke2020contact} is greater than that produced by the isoAdvector advection method. However, the isoAdvector method exhibits asymmetry in absolute error at the peaks and troughs of the contact angle evolution curve (cf. \cref{fig:advection-results-for-uniform-mesh}). The numerical solution converges to the reference solution and delivers near first-order convergent results for $E_\infty$ (see \cref{fig:advection-results-for-uniform-mesh}). These findings align with those obtained using the Boundary ELVIRA method \cite{fricke2020contact}. 

\section{Droplet spreading on a flat surface}
\label{sec:Droplet-spreading-on-a-flat-surface}
\subsection{Definition of the case study}
In this study, we investigate the spreading of a droplet on a flat surface \citep{dupont2010numerical}. The focus is on the effect of the static contact angle boundary condition and the Bond number, $Bo = \frac{\rho_l g R_0}{\sigma}$, on the equilibrium shape of the droplet.\ The set of constant physical parameters listed in \cref{table:phyical-parameters-of-the-droplet} with varying values of the gravitational acceleration $g$ are used to achieve different values for the Bond number (see \cref{table:Bond-numbers}). For a droplet that spreads with $Bo\ll1$, the surface tension forces dominate, and the droplet at equilibrium maintains a spherical cap shape and satisfies the contact angle boundary condition.\ On the other hand, for $Bo\gg1$, the gravitational forces dominate, and the droplet forms a puddle, whose height is directly proportional to the capillary length, $l_{Ca}=\sqrt{\frac{\sigma}{\rho_l g}}$. The droplet's volume $V$ and the equilibrium contact angle $\theta_e$ are used to derive the geometrical relations for the equilibrium shape of the droplet \citep{dupont2010numerical, fricke2020geometry,patel2017coupled}.\ Furthermore, we have also studied the mesh convergence of the spreading droplets.

\begin{figure}[h!]
	\centering
	\def\svgwidth{0.5\textwidth}
	{\footnotesize
		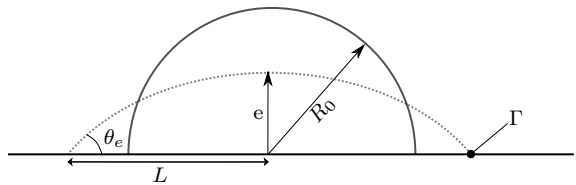
	}
	\caption{Schematic diagram of initial configuration (--) of a droplet with initial radius $R_0$ and final equilibrium shape ($\vcenter{\hbox{...}}$) attained by spreading on a flat surface with an equilibrium contact angle $\theta_e$.}
	\label{fig:initial-config-spreading-over-flat-surface}
\end{figure}

\Cref{fig:initial-config-spreading-over-flat-surface} illustrates the schematic diagram of a semi-spherical droplet initialized on a flat surface with an initial radius $R_0$. Note that $\theta_0 = 90^\circ$ is an (arbitrary) choice for the initial contact angle. The droplet spreads and attains an equilibrium state at $\theta_e$, having height $e$ and wetted radius $L$. The droplet \textit{wets} the surface if the initial contact angle is larger than the equilibrium contact angle. Contrary to this behavior, we observe \textit{dewetting} if the initial contact angle is smaller than the equilibrium contact angle.

We have considered droplets of water-glycerol (75\% glycerol) and pure water. The viscosity of water-glycerol is larger than that of pure water by a factor of 30, while the surface tension is slightly smaller. The physical properties of both liquids are presented in~\cref{table:phyical-parameters-of-the-droplet}. 

\begin{table}[htbp]
	\centering
	\begin{tabular}{c c c c} 
		\toprule
		Fluid & $\rho$  & $\nu$ & $\sigma$  \\ 
		& $kg~m^{-3}$ & $m^2 s^{-1}$ & $N m^{-1}$ \\[0.5ex] 
		\midrule
		water & $10^3$ & $10^{-6}$ & $0.072$ \\
		water-glycerol & $1194.9$ & $2.51e^{-5}$ & $0.0635$ \\[1ex] 
		\bottomrule
	\end{tabular}
	\caption{Physical parameters of the liquid droplet.}
	\label{table:phyical-parameters-of-the-droplet}
\end{table}

A three-dimensional computational domain, discretized using unstructured uniform hexahedral mesh elements, (see \cref{table:computationalParamForDropletSpreading} for domain parameters) is simulated. The droplet is initialized at the center of the domain's \underline{b}ottom boundary
\begin{equation}
\partial \Omega_b = \{(x,y,0): 0 \leq x \leq 5 , 0 \leq y \leq 5\}.
\end{equation}

\begin{table}[h!]
	\centering
	\begin{tabular}{c c c} 
		\toprule
		Parameter & Value & Unit \\ [0.5ex] 
		\midrule
		Droplet initial radius, $R_0$& $1$ & mm \\
		Droplet initial position, $(x_0, y_0, z_0)$& $(2.5, 2.5, 0)$ & mm \\
		Domain size & $(5, 5, 4)$ & mm \\
		$(n_x, n_y, n_z)$ & $(100, 100, 80)$ & cells \\[1ex] 
		\bottomrule
	\end{tabular}
	\caption{Computational and geometrical parameters for the droplet spreading on a flat surface with an equilibrium contact angle $\theta_e$.}
	\label{table:computationalParamForDropletSpreading}
\end{table}
The bottom boundary has a no-slip boundary condition for the velocity.\ The time step $\DeltaT$ is restricted to CFL number below $0.01$.

\subsubsection{Geometrical relations for a droplet at equilibrium}
%
Droplet spreading with a very small Bond number attains a spherical cap shape at the equilibrium. The wetted radius $L$ and the height $e$ of the spherical cap are given by the geometrical relations
\begin{equation}
\label{eq:equilibrium-radius}
\frac{L}{V^{\frac{1}{3}}} = g(\theta) \coloneqq \sin{\theta} \left(\frac{\pi(1-\cos{\theta})^2  (2+\cos{\theta})}{3}\right)^{-\frac{1}{3}},
\end{equation}
\begin{equation}
\label{eq:equilibrium-height}
e = L\tan\left(\frac{\theta}{2}\right).
\end{equation}
The intersection of a spherical cap and the horizontal flat surface produces a circular contact line $\Gamma$ (see~\cref{fig:initial-config-spreading-over-flat-surface}), whose area is referred to as wetted area
\begin{equation}
\label{eq:wetted-area}
A = \pi L^2.
\end{equation}
%
For a droplet spreading with a large Bond number ($Bo\gg1$), the puddle height $e$ is given by
\begin{equation}
\label{eq:height_at_higher_Bo}
e = 2\sqrt{\frac{\sigma}{\rho_l g}}\sin\left(\frac{\theta}{2}\right).
\end{equation}
The estimate for the wetted area $A$ is obtained by adding up the wetted area of each boundary face $f$, which is calculated as follows:
\begin{equation}
\label{eq:Numerical-solution-of-droplet-spreading-over-a-flat-surface}
\int_{\partial\Omega} \alpha(\vec{x})\,ds\  = \sum_{f \in F_{c}} \alpha_{f} \|\vec{S_{f}}\|_{2} + O(h^2).
\end{equation} 
Here, $\alpha_f$ represents the volume fraction value of the boundary face, which is obtained using OpenFOAM functionalities in the function object and $\vec{S_{f}}$ is the face-area normal vector of the face $f$.
\subsection{Spreading of a droplet with a very small Bond number}
\subsubsection{Convergence study}
\label{subsec:convergence-study}
\begin{figure}[h!]
	\centering
	\begin{subfigure}[b]{0.49\textwidth}
		\centering
		\includegraphics[width=\textwidth, height=0.75\textwidth]{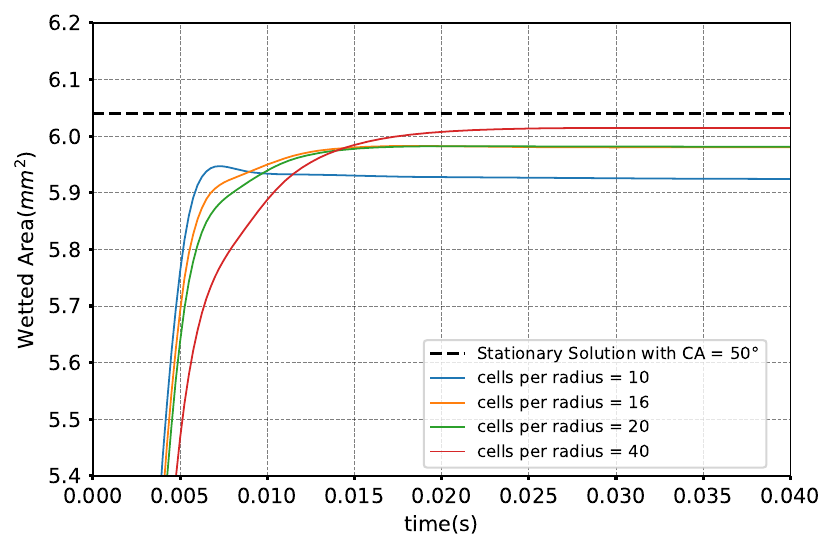}
		\caption{}
		\label{fig:mesh-convergence-study-water_glycerol_50}
	\end{subfigure}
	\hfill
	\begin{subfigure}[b]{0.49\textwidth}
		\centering
		\includegraphics[width=\textwidth, height=0.75\textwidth]{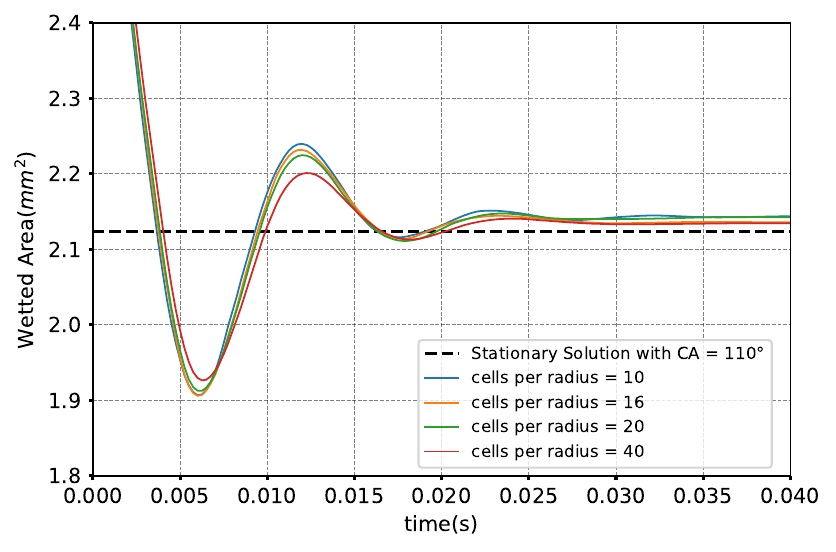}
		\caption{}
		\label{fig:mesh-convergence-study-water_glycerol_110}
	\end{subfigure}
	\centering
	\begin{subfigure}[b]{0.49\textwidth}
		\centering
		\includegraphics[width=\textwidth, height=0.75\textwidth]{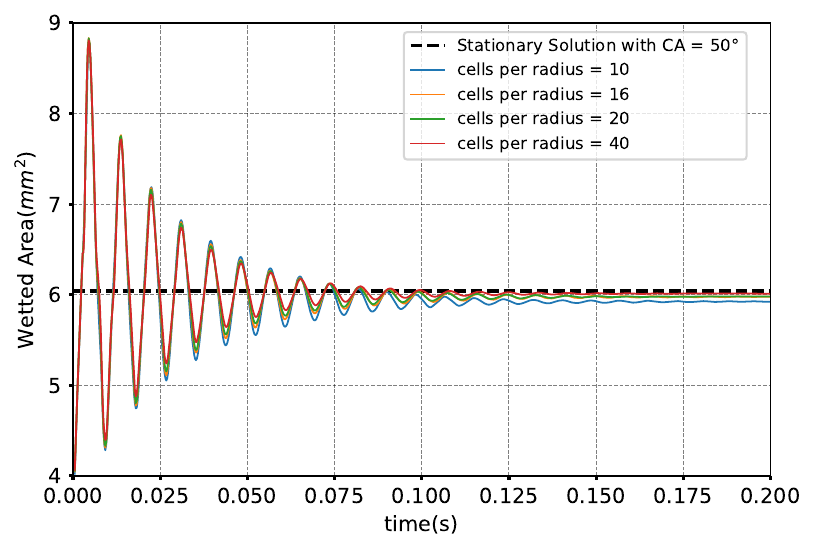}
		\caption{}
		\label{fig:mesh-convergence-study-water_50}
	\end{subfigure}
	\hfill
	\begin{subfigure}[b]{0.49\textwidth}
		\centering
		\includegraphics[width=\textwidth, height=0.75\textwidth]{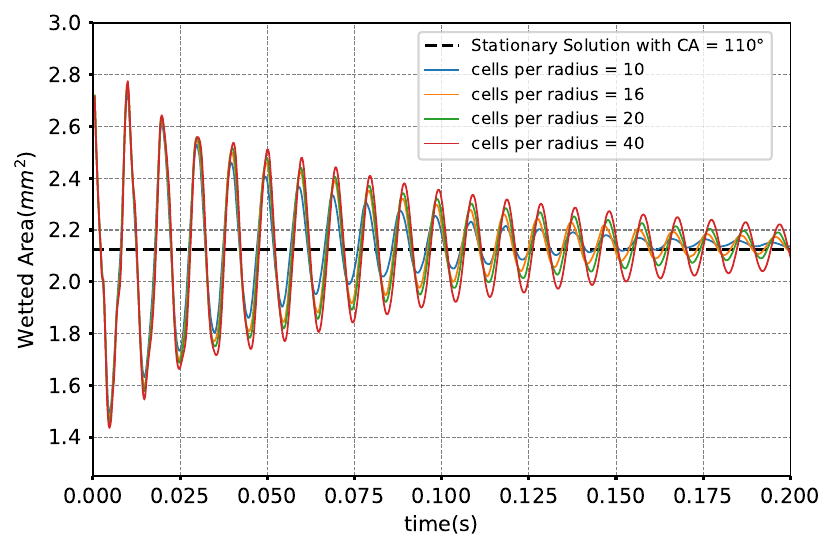}
		\caption{}
		\label{fig:mesh-convergence-study-water_110}
	\end{subfigure}
	\caption{Convergence study of the wetted area of the spreading droplet on a flat horizontal surface.\ The top images (\cref{fig:mesh-convergence-study-water_glycerol_50,,fig:mesh-convergence-study-water_glycerol_110}) and bottom images (\cref{fig:mesh-convergence-study-water_50,,fig:mesh-convergence-study-water_110}) show the droplet spreading simulation results for water-glycerol and water droplets, respectively, with a mesh size of $10, 16, 20$, and $40$ cells per radius.\ The initial contact angle is $\theta_0=90$°. The equilibrium contact angles are $\theta_e=50$° (left, wetting) and $110$° (right, dewetting).\ The stationary solution (\cref{eq:wetted-area}) is represented by the dashed horizontal (black) line.}
	\label{fig:mesh-convergence-study-for droplet-spreading}
\end{figure}

A mesh convergence study for the wetted area is conducted with four levels of mesh refinement with 10,\ 16,\ 20,\ and 40 cells per droplet radius.\ The results shown in \cref{fig:mesh-convergence-study-water_glycerol_50}, illustrate a convergent behavior of the water-glycerol droplet with respect to the stationary state at $\theta_e=50$°.\ For a coarser mesh (cells per radius=$10$), an overshoot in the wetted area curve is observed because of the numerical dissipation being small on a coarse mesh. With numerical slip, any overshoot would disappear with mesh refinement (as observed in \cref{fig:mesh-convergence-study-water_glycerol_50}), as the contact line would not be able to move in the limit $\Delta x \rightarrow 0$ ($\Delta x$ is the mesh cell size). Similar convergent behavior with respect to the stationary state is observed for the water-glycerol droplet with $\theta_e=110$° (see~\cref{fig:mesh-convergence-study-water_glycerol_110}).\ In contrast to this behavior, water droplets exhibit oscillatory spreading behavior (see ~\cref{fig:mesh-convergence-study-water_50,fig:mesh-convergence-study-water_110}).
%
%


\subsubsection{Geometrical characteristics of the droplet}
\begin{figure}[h!]
	\centering
	\begin{subfigure}[b]{0.4\textwidth}
		\centering
		\includegraphics[width=\textwidth, height=0.5\textwidth]{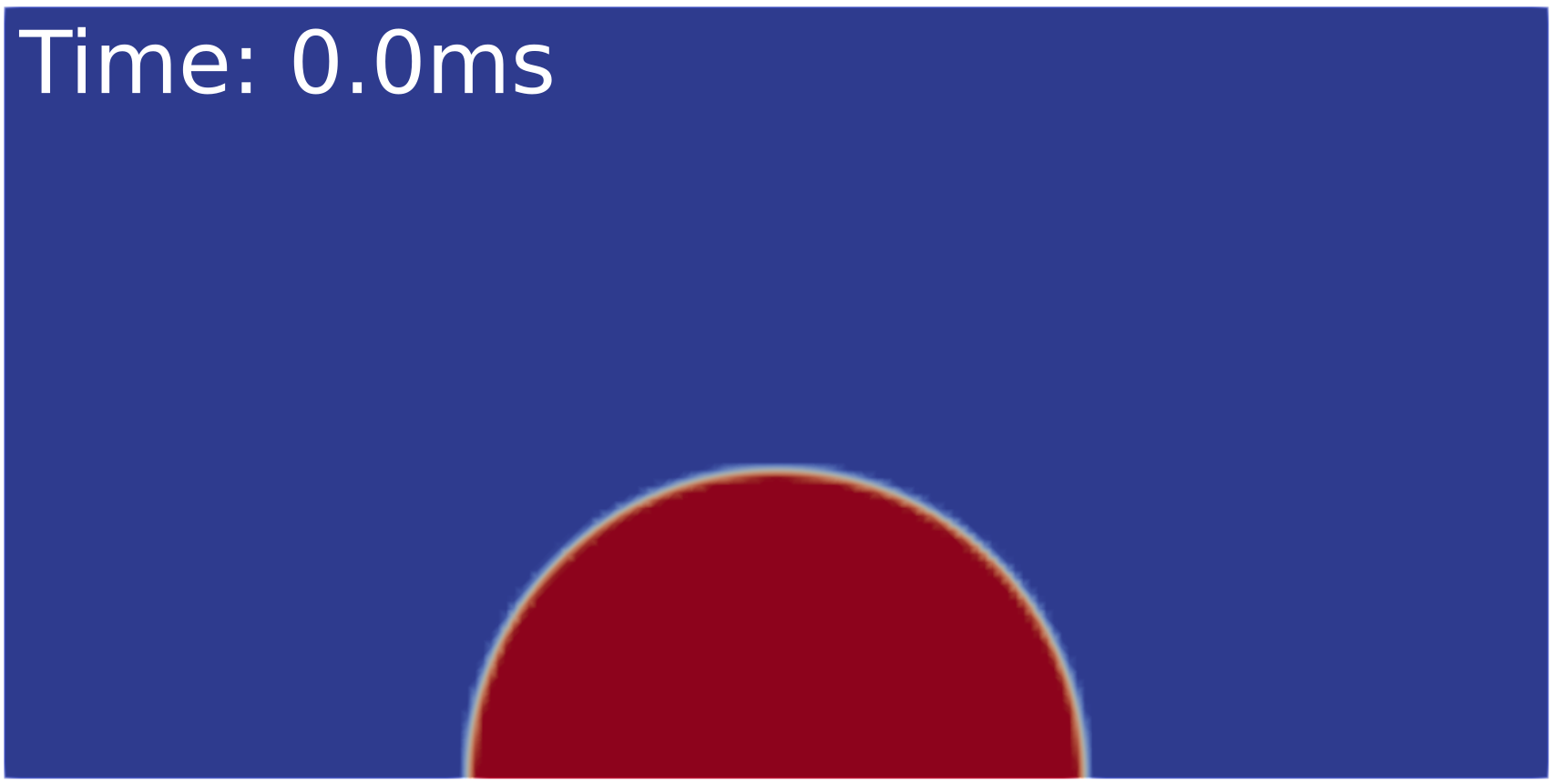}
		\caption{}
		\label{fig:transient-shapes-of-droplet-spreading-on-a-flat-surface-0_0}
	\end{subfigure}
	\hfill
	\begin{subfigure}[b]{0.4\textwidth}
		\centering
		\includegraphics[width=\textwidth, height=0.5\textwidth]{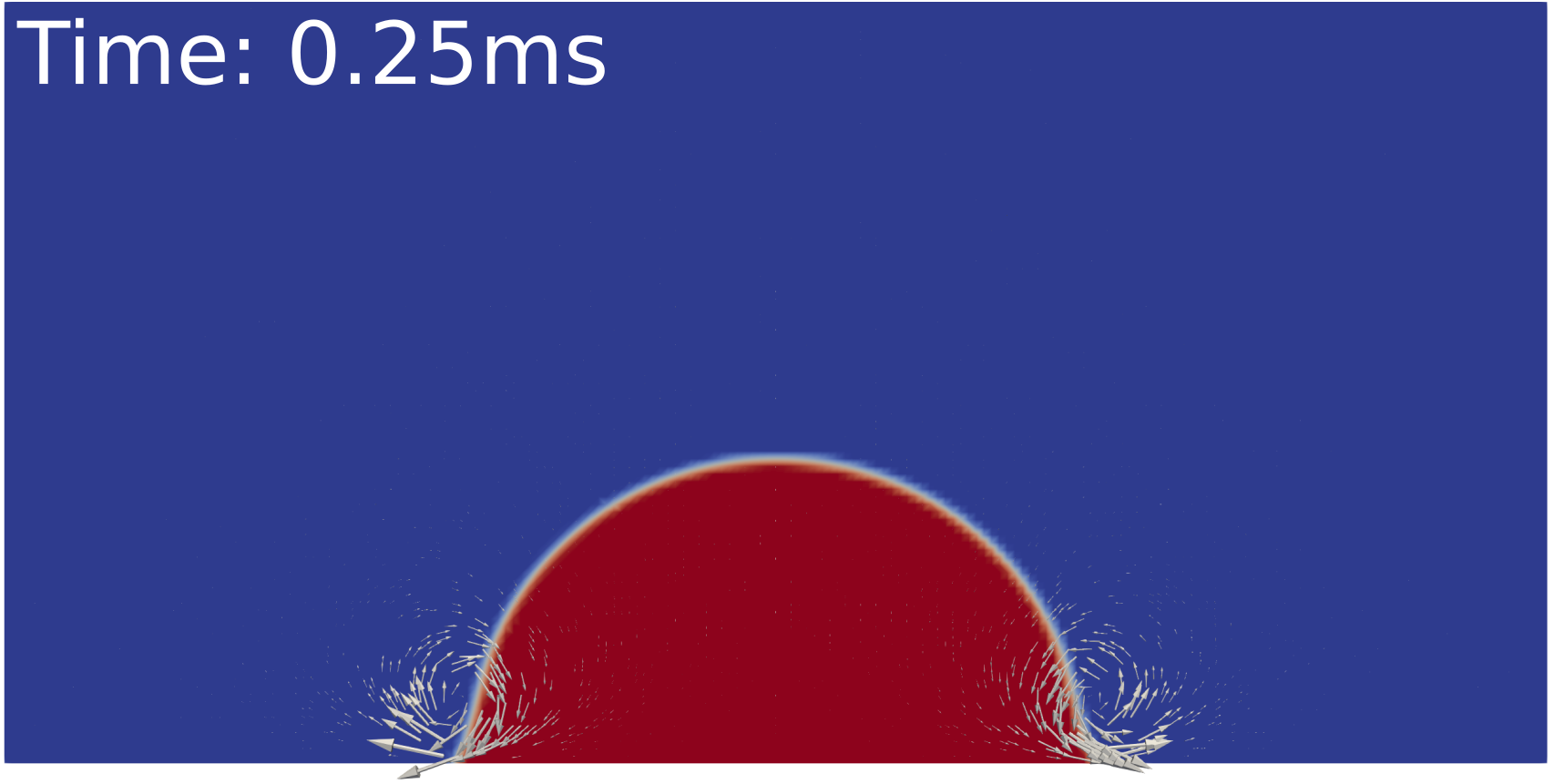}
		\caption{}
		\label{fig:transient-shapes-of-droplet-spreading-on-a-flat-surface-0_00025}
	\end{subfigure}
	\centering
	\begin{subfigure}[b]{0.4\textwidth}
		\centering
		\includegraphics[width=\textwidth, height=0.5\textwidth]{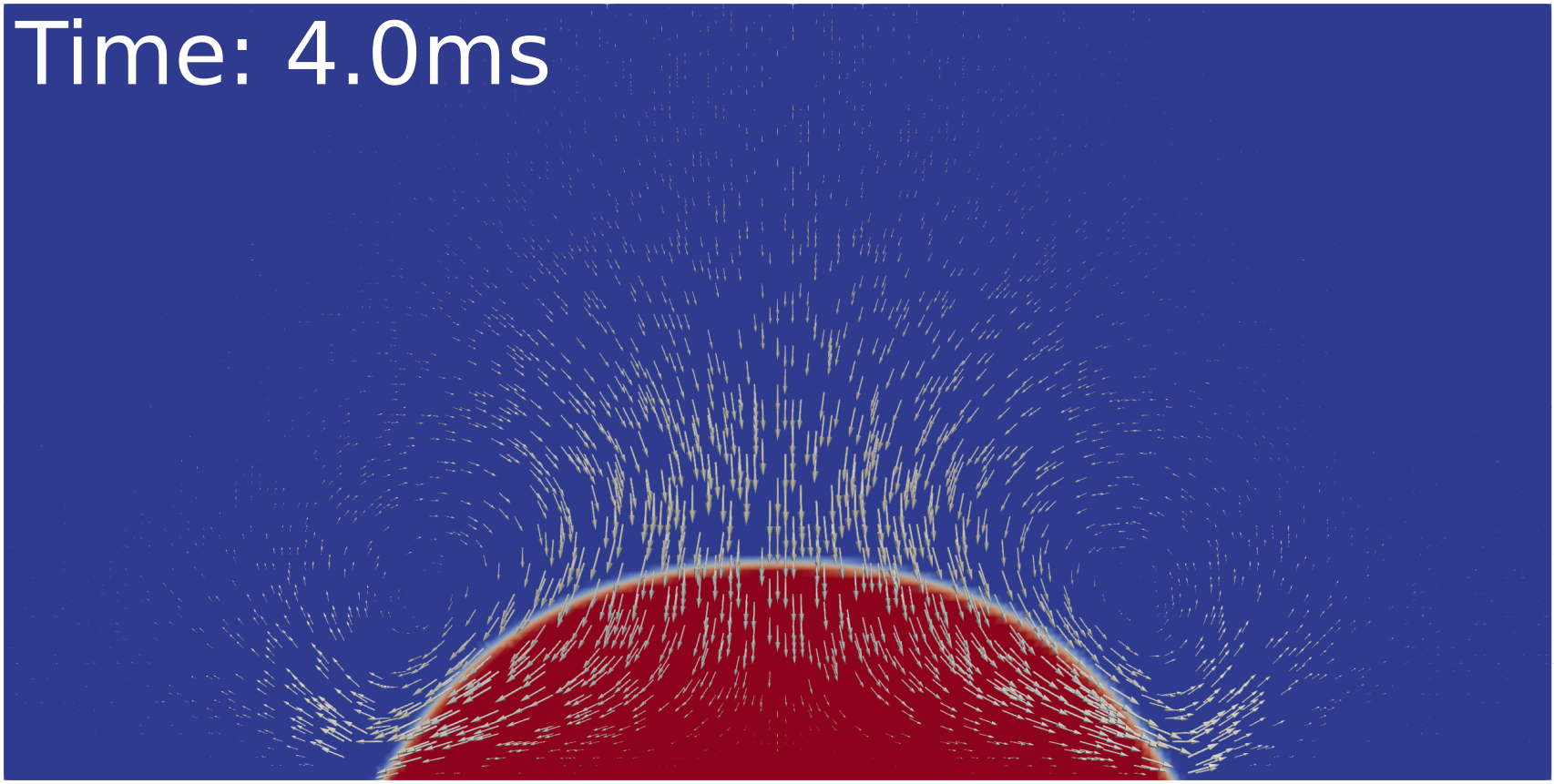}
		\caption{}
		\label{fig:transient-shapes-of-droplet-spreading-on-a-flat-surface-0_004}
	\end{subfigure}
	\hfill
	\begin{subfigure}[b]{0.4\textwidth}
		\centering
		\includegraphics[width=\textwidth, height=0.5\textwidth]{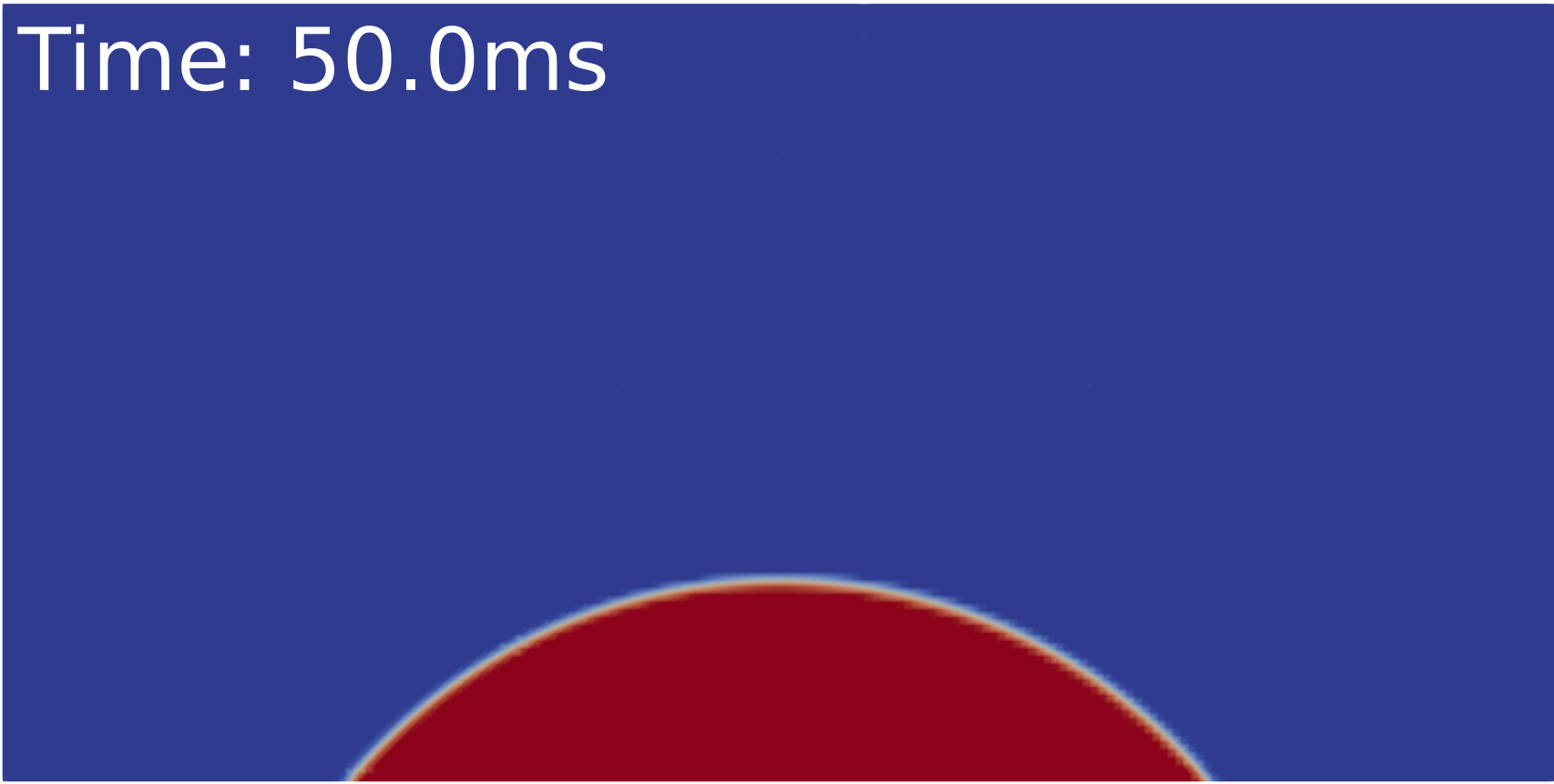}
		\caption{}
		\label{fig:transient-shapes-of-droplet-spreading-on-a-flat-surface-0_05}
	\end{subfigure}
	\caption{Simulation results of droplet spreading on a horizontal flat surface at different instants of time. The initial and equilibrium contact angles are $\theta_0=90$° and $\theta_e=50$°, respectively. The arrows represent the velocity field.}
	\label{fig:transient-shapes-of-droplet-spreading-on-a-flat-surface}
\end{figure}

\begin{figure}
    \centering
    \includegraphics[width=0.8\textwidth, height=0.225\textwidth]{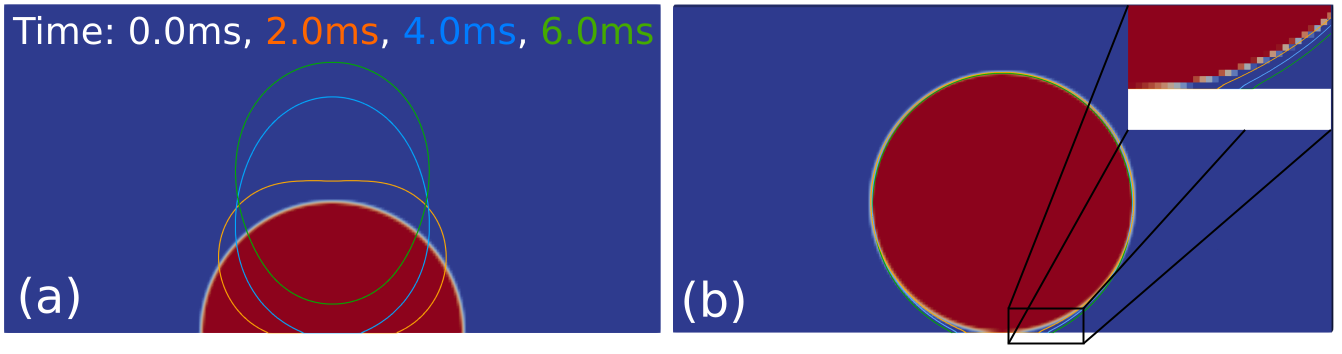}
    \caption{Droplet contours, with the contour's color representing different instants of time. The equilibrium contact angle is $\theta_e=170^\circ$.\ (a) represents the simulation results with the initial droplet shape of a semi-sphere.\ The droplet is detached from the surface during spreading due to the high potential energy stored in the droplet.\ (b) represents the simulation results with a droplet initialized as a full sphere and the initial contact angle ($\theta_0 \approx 180^\circ$) is close to the equilibrium contact angle. As the initial potential energy is relatively low, no detachment of the droplet is observed in (b). Similar behavior is observed for studies with $\theta_e > 150^\circ$.}
    \label{fig:detached-droplet}
\end{figure}

\begin{figure}[h!]
    \centering
    \includegraphics[width=0.55\textwidth, height=0.45\textwidth]{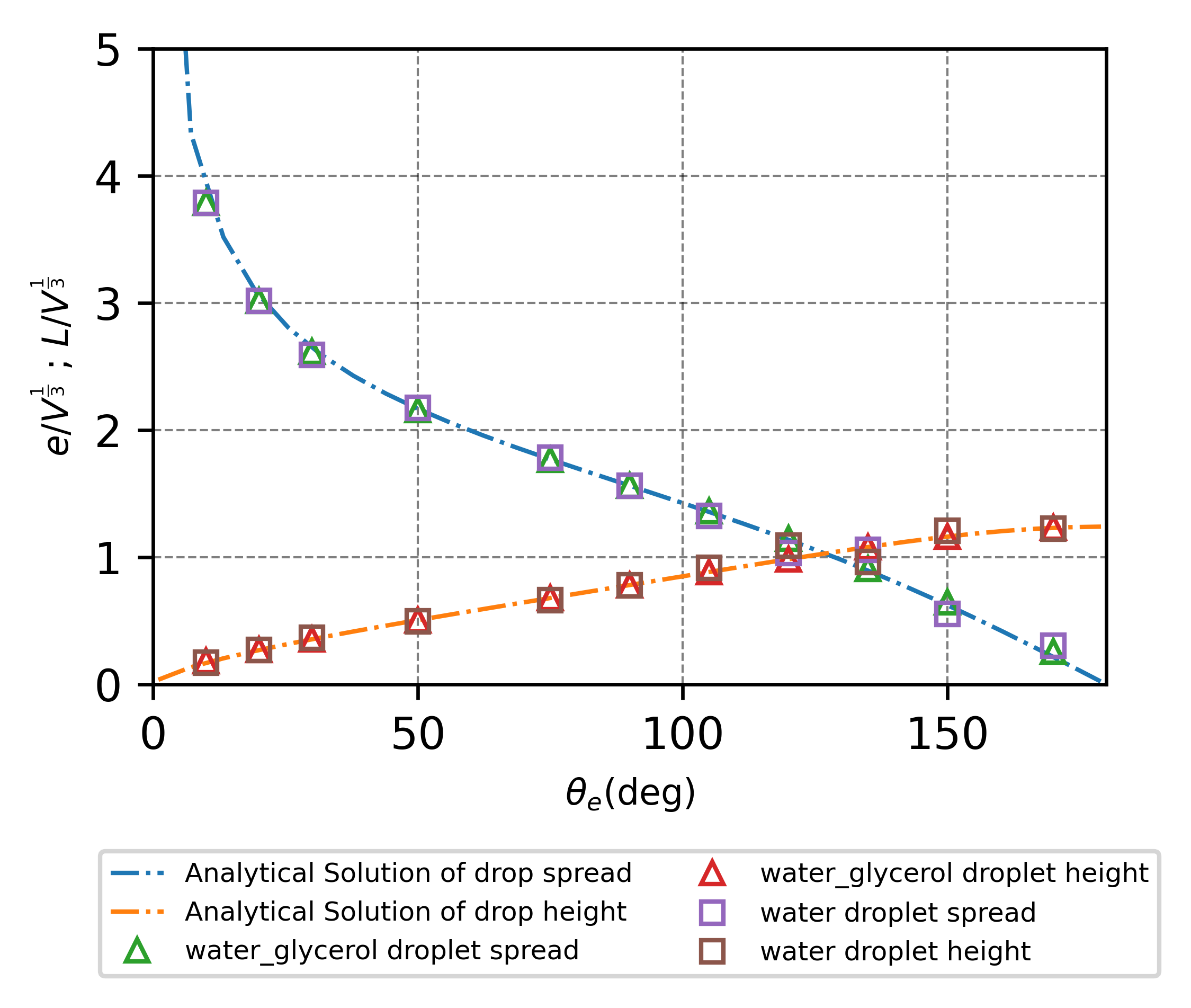}
    \caption{Droplet spreading on a horizontal flat surface:\ Geometrical characteristics of the droplet against the equilibrium contact angle $\theta_e$. \textbf{\textcolor{violet}{\protect\mysquare}}  wetted radius for water, \textbf{\textcolor{brown}{\protect\mysquare}} equilibrium height for water, \textbf{\textcolor{teal}{\protect\mytriangle}} wetted radius for water-glycerol and \textbf{\textcolor{red}{\protect\mytriangle}} equilibrium height for water-glycerol. The stationary solution for wetted radius (\textbf{\textcolor{royalblue}{--$\cdot$--}}) and equilibrium droplet height (\textbf{\textcolor{orange}{--$\cdot$--}}) is given by,~\cref{eq:equilibrium-radius,,eq:equilibrium-height}, respectively.}
    \label{fig:droplet-chractersitics-against-equilibrium-contact-angle} 
\end{figure}

\begin{figure}[H]
	\centering
	\begin{subfigure}[b]{0.4\textwidth}
		\centering
		\includegraphics[width=\textwidth, height=0.5\textwidth]{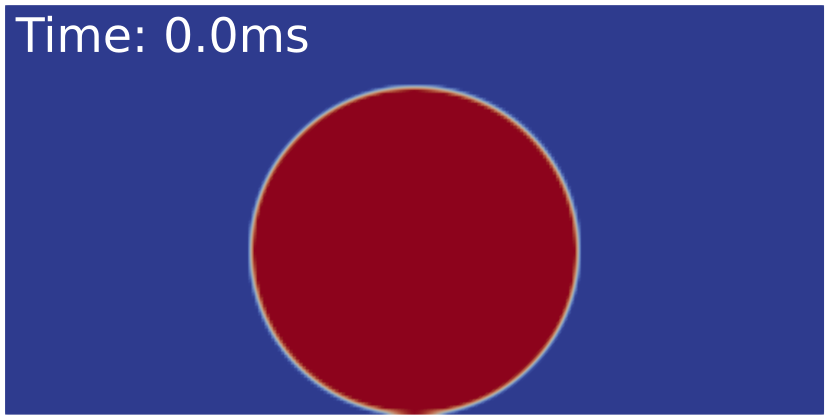}
		\caption{}
		\label{fig:full-initial-sphere-spreading-t0-1}
	\end{subfigure}
	\hfill
	\begin{subfigure}[b]{0.4\textwidth}
		\centering
		\includegraphics[width=\textwidth, height=0.5\textwidth]{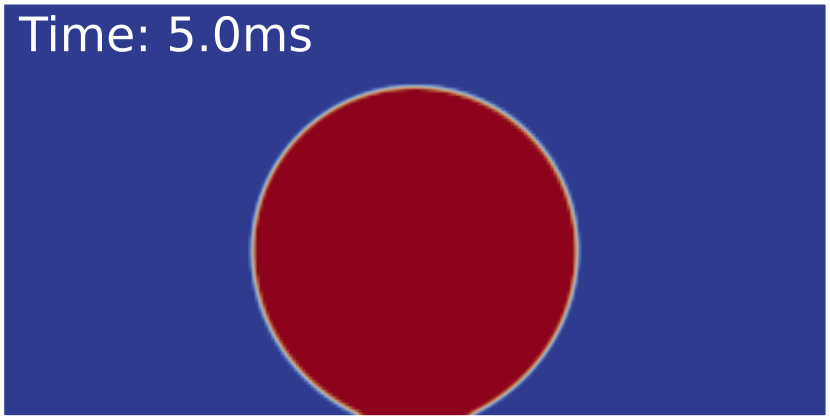}
		\caption{}
		\label{fig:full-initial-sphere-spreading-t1-1}
	\end{subfigure}
	\centering
	\begin{subfigure}[b]{0.4\textwidth}
		\centering
		\includegraphics[width=\textwidth, height=0.5\textwidth]{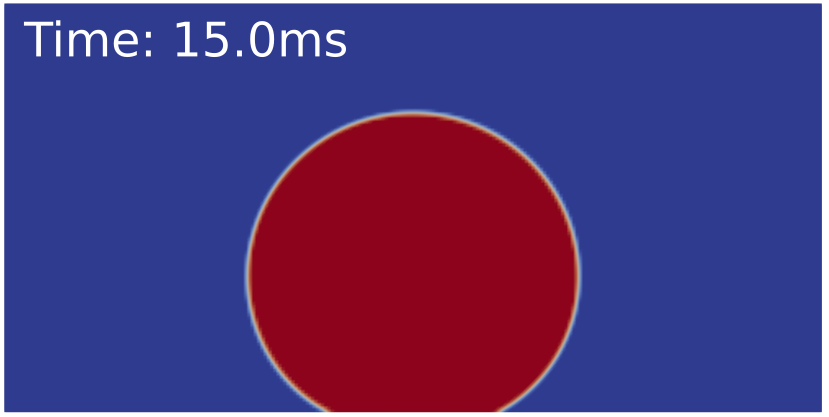}
		\caption{}
		\label{fig:full-initial-sphere-spreading-t2-1}
	\end{subfigure}
	\hfill
	\begin{subfigure}[b]{0.4\textwidth}
		\centering
		\includegraphics[width=\textwidth, height=0.5\textwidth]{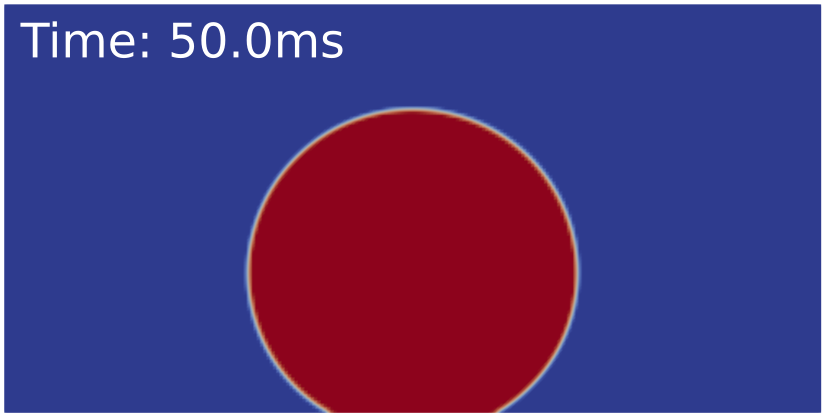}
		\caption{}
		\label{fig:full-initial-sphere-spreading-t3-1}
	\end{subfigure}
	\caption{Simulation results of droplet spreading on a horizontal flat surface at different instants of time. The initial and equilibrium contact angles are $\theta_0\approx180$° and $\theta_e=150$°, respectively.}
	\label{fig:full-sphere-spreading}
\end{figure}

\Cref{fig:transient-shapes-of-droplet-spreading-on-a-flat-surface} illustrates the different shapes of the droplet during its spreading process. In the beginning, due to the difference between the initial and equilibrium contact angles, the droplet is far from its equilibrium state, leading to a rapid movement of the contact line without any significant change in the droplet's global shape, as depicted in \cref{fig:transient-shapes-of-droplet-spreading-on-a-flat-surface-0_00025}. As the spreading continues, the droplet's apex velocity increases in a downward direction (\cref{fig:transient-shapes-of-droplet-spreading-on-a-flat-surface-0_004}), and the droplet's overall shape starts to change until it reaches its equilibrium state (\cref{fig:transient-shapes-of-droplet-spreading-on-a-flat-surface-0_05}). Similar spreading behavior is reported in the literature (\citep{afkhami2009mesh,khatavkar2007capillary,villanueva2006some}).
The comparison of dimensionless geometrical quantities at equilibrium with reference solutions for a range of contact angles is shown in \cref{fig:droplet-chractersitics-against-equilibrium-contact-angle}.\ The simulation results are in very good agreement with the reference solution for both hydrophilic and hydrophobic cases.\ We note that for the highly hydrophobic case (e.g., for $\theta_e=150^\circ$), the stationary state is not reached in a reasonable time if the droplet is initialized as a semisphere. This can be understood in terms of the initial potential energy, which is very high in the case of a semisphere (initial droplet).\ The quick release of this potential energy may even cause a droplet detachment from the surface, as illustrated in~\cref{fig:detached-droplet}.
However, if the droplet is initialized closer to the equilibrium state of the hydrophobic spreading (as a sphere  with $\theta_0\approx 180^\circ$), convergence to the reference stationary state is observed, as illustrated in~\cref{fig:full-sphere-spreading}.
\begin{figure}[h!]
	\centering
	\includegraphics[width=0.6\textwidth, height=0.15\textwidth]{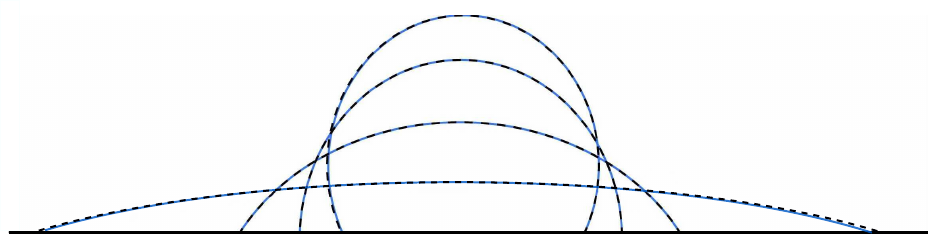}
	\caption{Droplet spreading on a horizontal flat surface: Equilibrium drop shapes for equilibrium contact angle \\ $\theta_e$ = 10°, 50°, 90°, 110°: numerical (\textcolor{royalblue}{---}) and \textcolor{black}{theoretical (\texttt{-{}-})}.}
	\label{fig:equilibrium-shape-of-droplet-spreading-over-flat-surfaces}
\end{figure}

As shown in \cref{fig:equilibrium-shape-of-droplet-spreading-over-flat-surfaces}, the equilibrium shape of the water-glycerol droplet, represented by $\alpha=0.5$ contours in ParaView~\citep{ParaView}, is compared to the reference spherical cap for a specific contact angle. The comparison illustrates that the equilibrium shape of the droplet has a very good qualitative match with the reference shape. 
\subsection{Spreading of a droplet with varying Bond numbers}
\begin{figure}[h!]
	\centering
	\begin{subfigure}[b]{0.49\textwidth}
		\centering
		\captionsetup{position=top}
		\def\svgwidth{\textwidth}
		{\footnotesize
			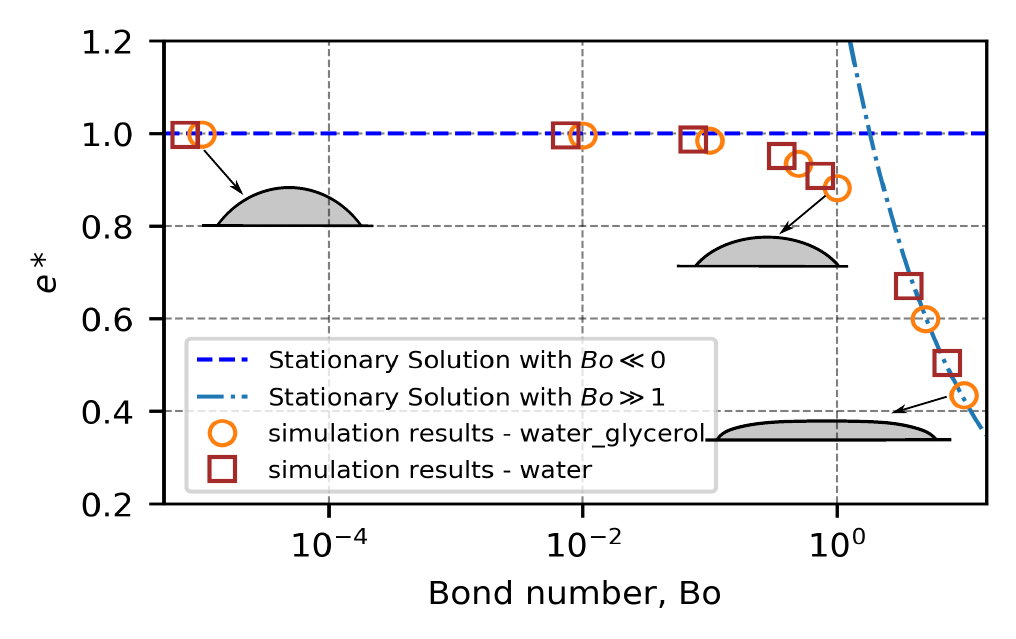
		}
		\label{fig:Normalized-droplet-height-on-a-flat-surface-for-varying-Bo-I}
	\end{subfigure}
	\hfill
	\begin{subfigure}[b]{0.49\textwidth}
		\centering
		\captionsetup{position=top}
		\def\svgwidth{\textwidth}
		{\footnotesize
			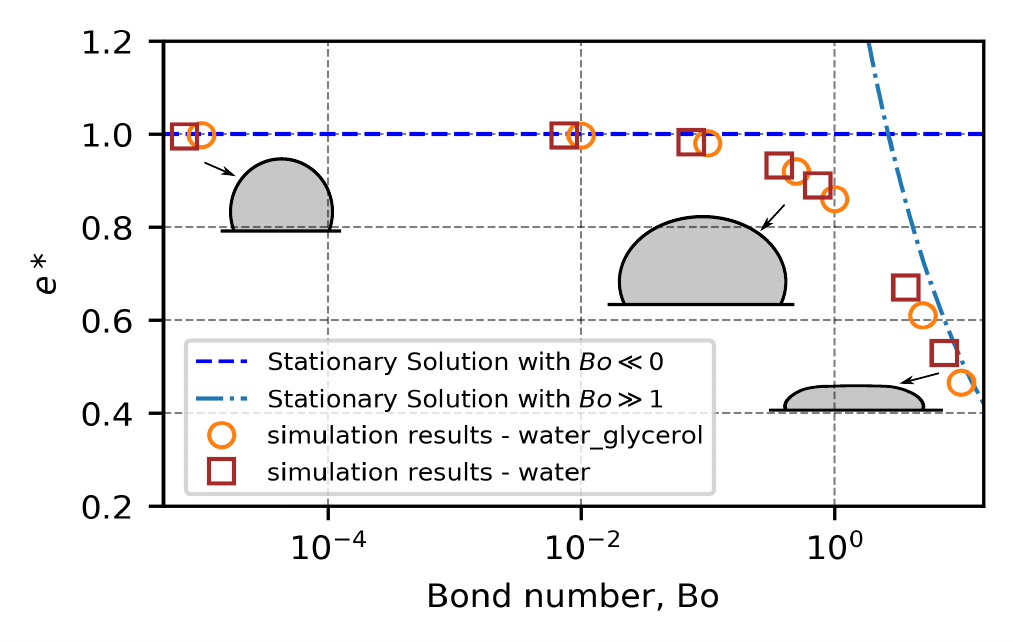
		}
		\label{fig:Normalized-droplet-height-on-a-flat-surface-for-varying-Bo-II}
	\end{subfigure}
	\caption{Normalized droplet height $e^*$ ($e^* = e/e_0$, where $e_0$ is the equilibrium height at $Bo\ll1$) on a flat surface with an equilibrium contact angle $\theta_e$;  left: $\theta_e=50$°, right: $\theta_e=110$°.}
	\label{fig:Normalized-droplet-height-on-a-flat-surface-for-varying-Bo}
\end{figure}

We now consider a droplet spreading for a range of the Bond number (see~\cref{table:Bond-numbers}). Here we have two spreading behaviors - the surface tension-dominant spreading ($Bo\ll1$) and the gravitational-dominant spreading ($Bo\gg1$), with the transition of behavior to be observed at $Bo\approx1$.\ \Cref{fig:Normalized-droplet-height-on-a-flat-surface-for-varying-Bo} shows the non-dimensional equilibrium droplet height comparison with the reference solutions in the limiting cases $Bo \rightarrow \infty$ and $Bo \rightarrow0$.
The simulation results for the spreading of the water and water-glycerol droplets are in excellent agreement with the reference solution for hydrophobic and hydrophilic cases. 

\begin{table}[h!]
	\centering
	\begin{tabular}{c c c c c c c c} 
		\toprule
		Bo - water-glycerol  & 1e-05   &   1e-02 & 1e-01   & 5e-01   & 1       & 5& 10\\ [1ex] 
		Bo - water           & 7.4e-06 & 7.4e-03 & 7.4e-02 & 3.7e-01 & 7.4e-01 & 3.7 & 7.4\\ [1ex]
		\bottomrule
	\end{tabular}
	\caption{Range of the Bond numbers for droplets spreading on a flat surface under the influence of gravity.}
	\label{table:Bond-numbers}
\end{table}

In summary, the simulation results using the plicRDF-isoAdvector method are in excellent agreement with the reference solution in terms of mesh convergence study, droplet shape comparison, and the incorporation of gravity effects on droplet spreading. It is noted, however, that for contact angles greater than 150 degrees, the simulation results show a significant dependence on the initial conditions. Handling the limiting equilibrium contact angle cases poses challenges for the numerical methods, as reported in ~\citep{dupont2010numerical,patel2017coupled}. In the present study, the method is tested for limiting contact angles, e.g., for $10^\circ$ and $170^\circ$. The results obtained show the ability of the plicRDF-isoAdvector method to handle these limiting cases very well. It can be observed that for $\theta_e = 10^\circ$, there is a small difference in equilibrium radius and stationary solution (cf. \cref{fig:droplet-chractersitics-against-equilibrium-contact-angle,fig:equilibrium-shape-of-droplet-spreading-over-flat-surfaces}). In this case, the droplet region in the vicinity of the contact line is almost flat, making a film-like region. This requires higher local mesh resolution, which, due to computational limitations, is not within the scope of this work.

\section{Validation of a droplet spreading dynamics}
\label{sec:partial_wetting_dynamics}
In this study, we test the dynamics of droplet spreading, compare the simulation results with other numerical methods, and use experimental results of \cite{lavi2004} to validate the plicRDF-isoAdvector method. Lavi and Marmur \cite{lavi2004} performed experiments with different fluids and found the temporal evolution of the wetted area to be well described by the exponential power law
\begin{equation}
    \label{eq:power-law}
    \frac{A}{A_e} = 1 - \exp{(-\frac{K}{A_e}\tau^n)},
\end{equation}
where $A_e$ is the wetted area at equilibrium, and $\tau \coloneqq t \sigma V^{1/3} / \mu$. In this study, the fluid \textit{squalane} on a substrate DTS is chosen for validation purposes. The respective physical properties and parameters are presented in~\cref{table:sq-parameters}. 
\begin{table}[h!]
	\centering
	\begin{tabular}{c c c c c} 
		\toprule
		$\mu~\mathrm{(Pa~s)}$   & $\rho~\mathrm{(kg/m^3)}$   &   $\sigma~\mathrm{(N/m)}$ & $K$     & $n$\\ [1ex] 
		32e-3            & 809                     & 34e-3          & 0.471   & 0.699 \\ [1ex]
		\bottomrule
	\end{tabular}
	\caption{The physical properties and parameters for \cref{eq:power-law} for squalane liquid on a DTS substrate~\cite{lavi2004}.}
	\label{table:sq-parameters}
\end{table}
\subsection{Definition of the case study}
The same simulation setup as from the previous study (see \cref{sec:Droplet-spreading-on-a-flat-surface} and \cref{fig:initial-config-spreading-over-flat-surface}) with the same boundary conditions are used here. The droplet is initialized such that the droplet centre is at a distance of $0.95R_0$ from the wall. Contrary to the use of a constant contact angle (CCA) as in the previous case study, we now employ the Cox-Voinov relation~\cite{cox1986dynamics}
\begin{equation}
\label{eq:cox-Voinov}
    \textcolor{Reviewer3}{\theta_d^3 = \theta_m^3 + \beta \mathrm{Ca},}
\end{equation}
\textcolor{Reviewer3}{as the contact angle boundary condition,} which describes the relationship between the dynamic contact angle (DCA) $\theta_d$, the contact line capillary number $\mathrm{Ca = \mu V_\Gamma /\sigma }$, and the microscopic contact angle $\theta_m$. \textcolor{Reviewer3}{Here, $\beta = 9 \ln(\frac{l}{\lambda})$ and} $l$ and $\lambda$ are the macroscopic and microscopic length scales, respectively. 
\subsubsection{Contact line velocity}
\begin{figure}
    \centering
    \includegraphics[width=0.4\textwidth]{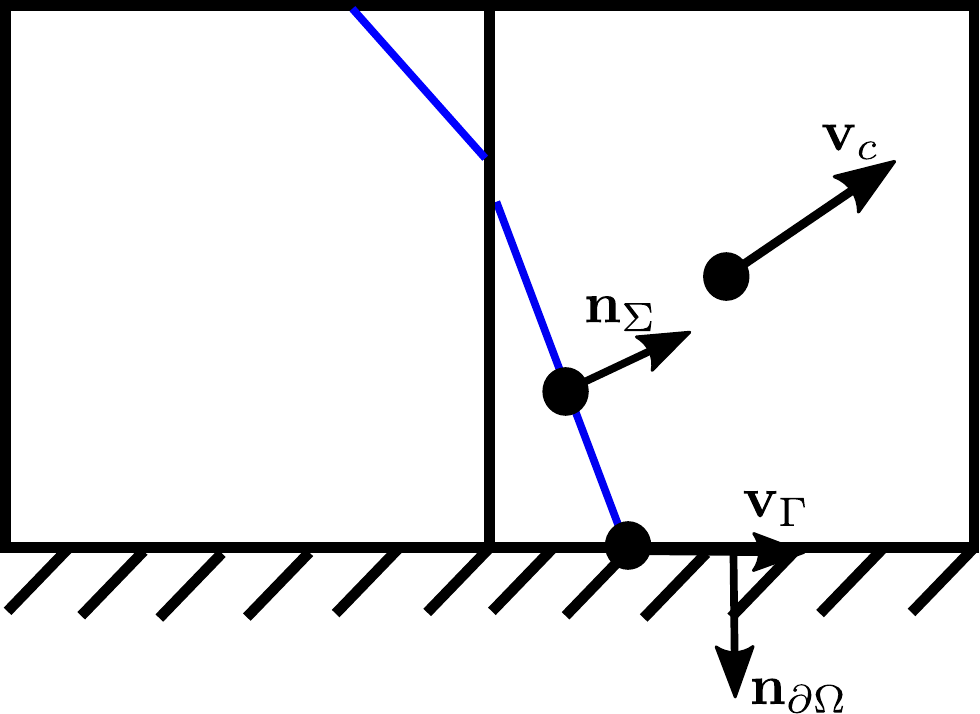}
    \caption{Interface cells adjacent to the boundary $\partial \Omega$. The cell on the right qualifies as the contact line cell as the interface $\Sigma$ intersects the domain boundary. The contact line $\Gamma$ moves with the velocity $\textbf{v}_\Gamma$.}
    \label{fig:CL-cell}
\end{figure}

Firstly, the approximation of the contact line velocity $\textbf{v}_\Gamma$ requires the accurate detection of the contact line cell (cf. \cref{algo:CL}), as not every interface cell adjacent to the boundary has a contact line, as shown in~\cref{fig:CL-cell}. \Cref{algo:CL} shows the approach employed in this study to detect the contact line cells \footnote{https://github.com/CRC-1194/b01-wetting-benchmark/blob/master/src/boundaryConditions/coxVoinov/coxVoinov.C}.

\floatname{algorithm}{Algorithm}
\begin{algorithm}
\caption{Determine the presence of a Contact Line }
\begin{algorithmic}
    \Function{hasContactLine}{$facei$}
        \State $celli \gets \text{owner of } face~i$
        \State $normal \gets \text{interface normal in } cell ~i$
        \State $centre \gets \text{interface centre in } cell~i$
        \State $vertices \gets \text{Get vertices of } face~i$
    
        \For{$i \gets 0$ \textbf{to} $\text{size}(vertices) - 1$}
            \State $dist1 \gets \text{signed distance at } vertices[i]$
            \State $dist2 \gets \text{signed distance at } vertices[i + 1]$
            \If{$dist1 \cdot dist2 < 0$}
                \State \Return \textbf{true} \Comment{A change in sign indicates a contact line}
            \EndIf
        \EndFor
        \State \Return \textbf{false} \Comment{No contact line detected}
    \EndFunction
\end{algorithmic}
\label{algo:CL}
\end{algorithm}

The relative velocity $\mathbf{v}_f$ at the boundary face $f$, w.r.t. the cell-centred velocity $\mathbf{v}_c$  is used to define $V_\Gamma$ (cf. \cref{fig:CL-cell}) as 
\begin{equation}
    \label{eq:CL-vel}
    V_\Gamma = \frac{\mathbf{n}_\Sigma - (\mathbf{n}_{\partial \Omega}\cdot \mathbf{n}_\Sigma)\mathbf{n}_{\partial \Omega}}{|\mathbf{n}_\Sigma - (\mathbf{n}_{\partial \Omega}\cdot \mathbf{n}_\Sigma)\mathbf{n}_{\partial \Omega}|} \cdot \mathbf{v}_f.
\end{equation}

\begin{figure}[h!]
	\begin{subfigure}[b]{0.49\textwidth}
            \centering
            \includegraphics[width=0.987\textwidth]{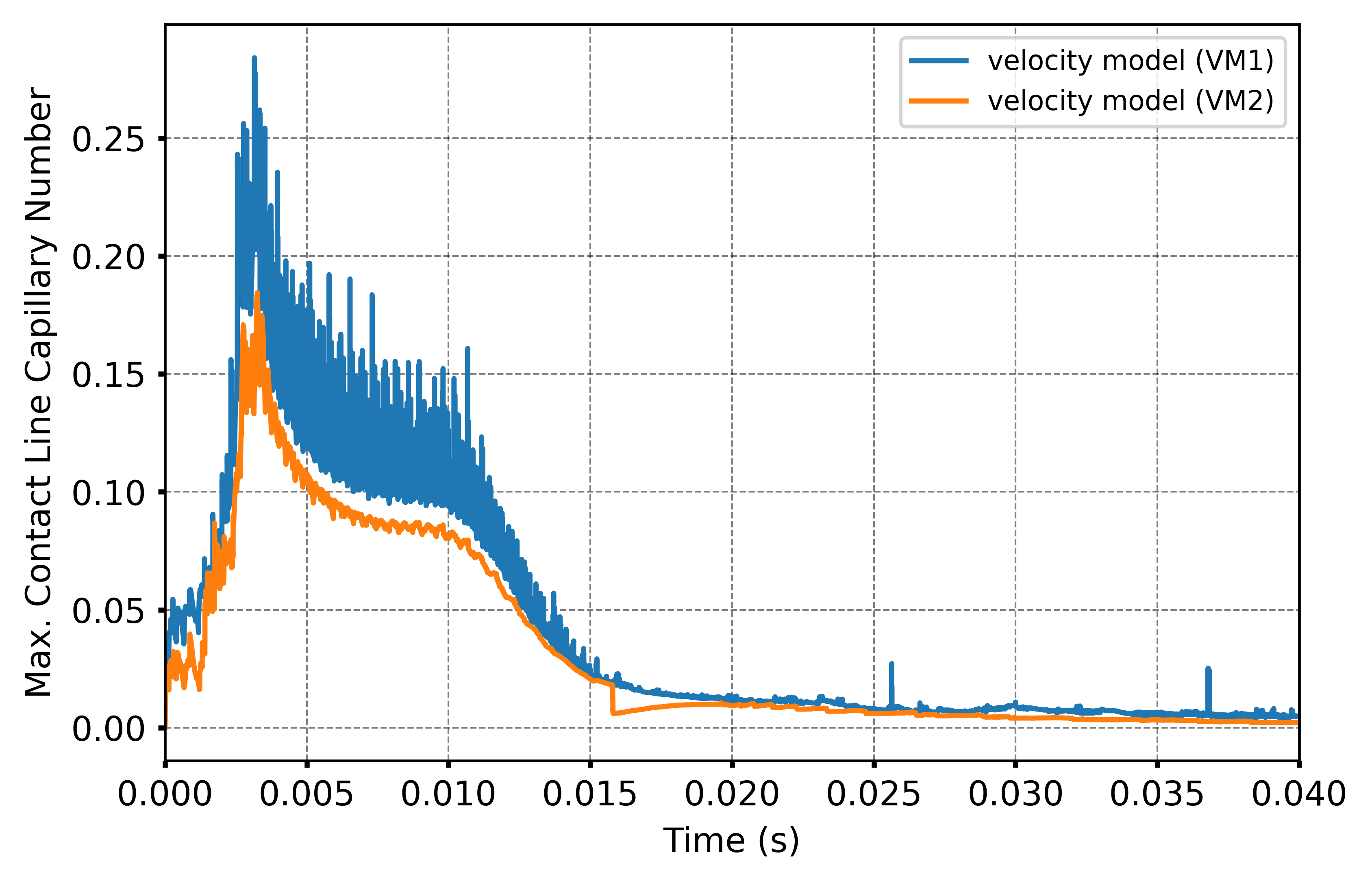}
            \caption{}
            \label{fig:sq-Ca-study}
	\end{subfigure}
	\hfill
	\begin{subfigure}[b]{0.47\textwidth}
            \centering
            \includegraphics[width=0.988\textwidth]{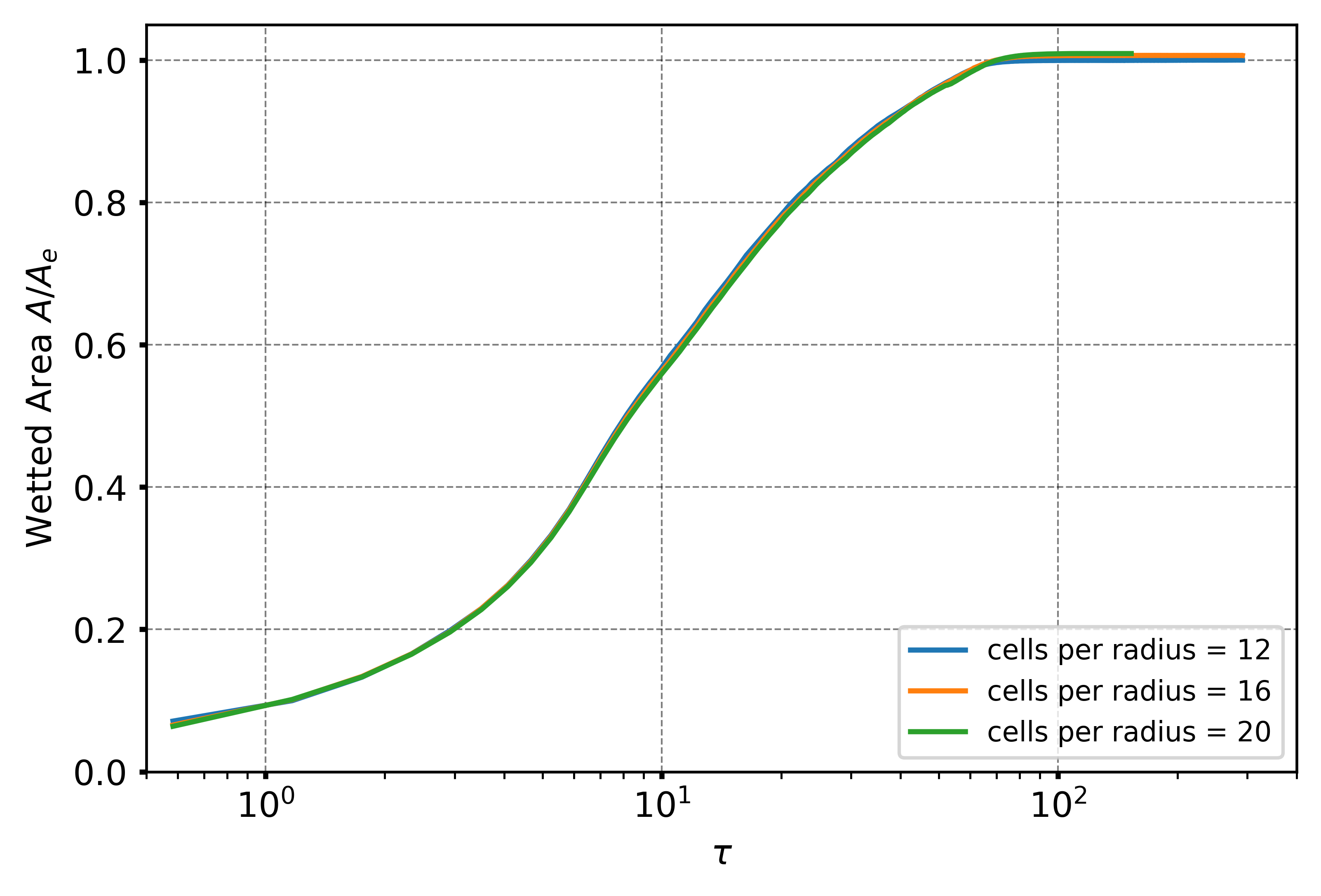}
            \caption{}
            \label{fig:sq-conv}
	\end{subfigure}
	\caption{(a): Comparison of the maximum contact line Capillary number for velocity models VM1 and VM2; which employs cell-centred velocity and interpolated velocity at the interface to compute $\textbf{v}_f$ (cf. \cref{eq:CL-vel}), respectively. (b): Convergence study for droplet spreading dynamics for squalane from \cite{lavi2004}.}
	\label{fig:sq-Conv-Ca-study}
\end{figure}

Two velocity models to obtain $\mathbf{v}_f$ are employed here; (i) VM1 by using cell-centred velocity $\textbf{v}_c$, (ii) VM2 by interpolating $\textbf{v}_c$ to the interface centre (see \cite{B01code} for computation details). \textcolor{Reviewer3}{The VM2 model is consistent with the isoAdvection method in calculating the velocity used for the advection of the interface. The isoAdvection method uses the inverse weighted interpolated velocity at the interface centre $\textbf{P}_\Sigma$ (see \cite{isoAdvection} for computation details) rather than using the cell-centred velocity $\textbf{v}_c$ to advect the interface.}  \Cref{fig:sq-Ca-study} shows that the VM1 model gives a systematically higher capillary number, which is relatively larger as compared to the one obtained using the VM2 model. The strong oscillations in $\text{Ca}$ are observed when using the VM1 model, which, close to the equilibrium stage, can cause strong oscillations in the droplet shape. To obtain further results in this study, we have used the VM2 model in the DCA boundary condition. 

Furthermore, for numerical simulations, $\theta_m = \theta_e$ in \cref{eq:cox-Voinov} is often assumed. However, this study employs the so-called uncompensated Young stress \textcolor{Reviewer3}{to model the microscopic contact angle in \cref{eq:cox-Voinov} as}
\begin{equation}
\label{eq:young-stress}
  \cos \theta_m =  \cos \theta_e -\Tilde{\zeta} ~ Ca,
\end{equation}
where $\Tilde{\zeta} = \zeta/\mu$ \textcolor{Reviewer3}{is the dimensionless contact line friction coefficient.} \textcolor{Reviewer3}{It is known from the literature that the value of $\Tilde{\zeta}$ depends, among other things, on the wettability of the surface. A more detailed discussion of the role of this parameter can be found in \cite{DuvivierExp,BlakeForcedWetting}.}
\subsubsection{Convergence study}
\Cref{fig:sq-conv} shows the convergence study for the spreading dynamics of a squalane droplet \cite{lavi2004}. The dynamics show \textcolor{Reviewer3}{a} convergent behavior for all mesh refinement levels. For further simulation results in this case study, 20 cells per radius refinement level is maintained. 
\subsubsection{Dynamics validation}
\begin{figure}[h!]
	\begin{subfigure}[b]{0.49\textwidth}
            \centering
            \includegraphics[width=\textwidth]{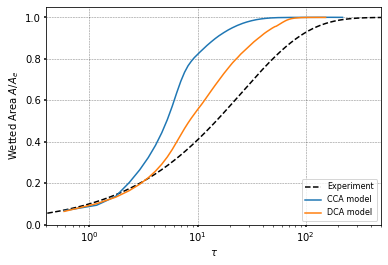}
            \caption{}
            \label{fig:sq-model-study}
	\end{subfigure}
	\hfill
	\begin{subfigure}[b]{0.49\textwidth}
            \centering
            \includegraphics[width=\textwidth]{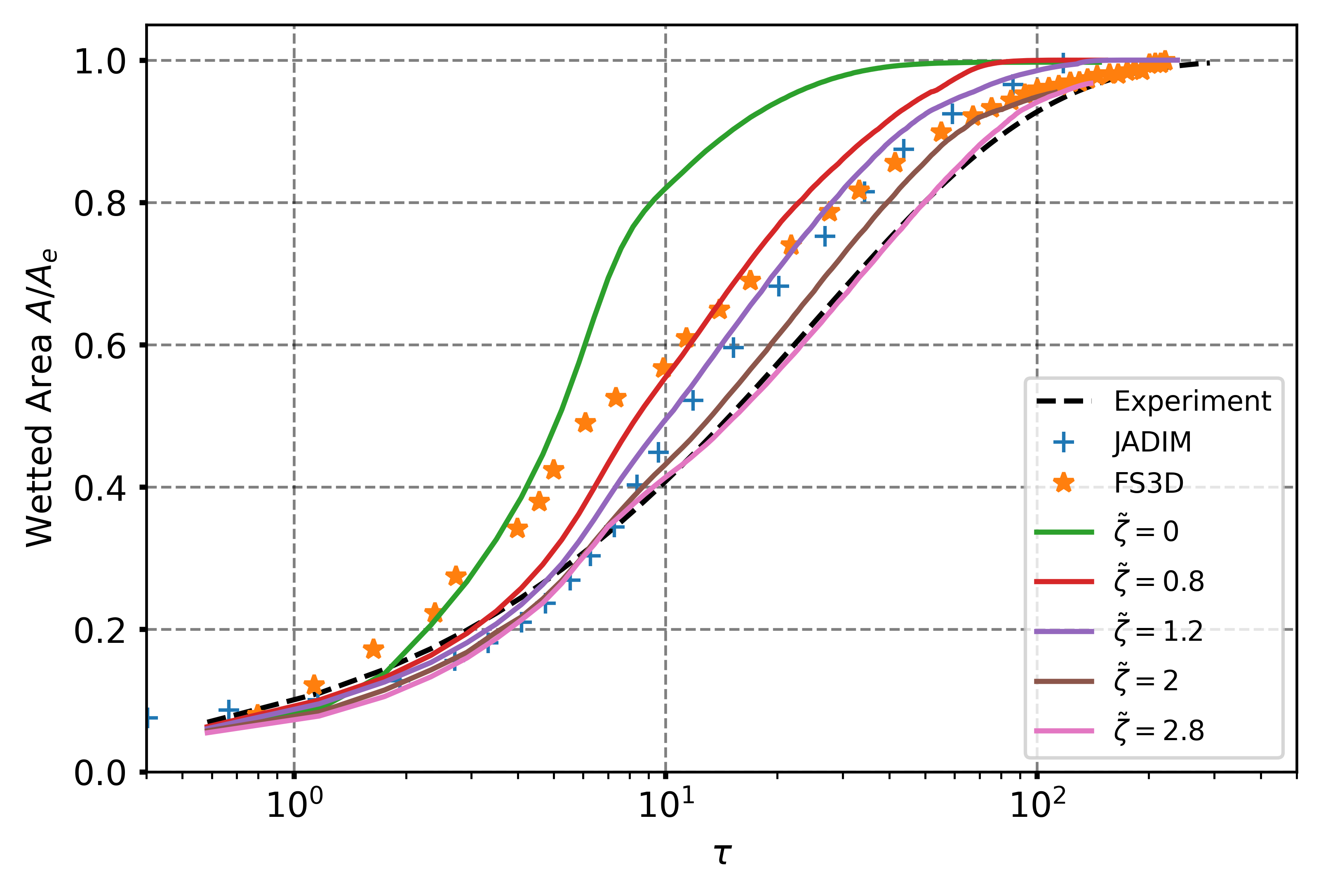}
            \caption{}
            \label{fig:sq-zeta}
	\end{subfigure}
	\caption{(a): Comparison of droplet spreading dynamics employing CCA and DCA model employing \cref{eq:young-stress} for $\theta_m$ with $\Tilde{\zeta}=0.8$. (b): Effect of dimensionless friction coefficient $\Tilde{\zeta}$ (cf. \cref{eq:young-stress}) on the dynamics of the droplet spreading and comparison with the FS3D solver~\cite{anjathesis,rieber2004numerische} and the JADIM code~\cite{JADIM,dupont2010numerical}.}
	\label{fig:sq-model-zeta-study}
\end{figure}

\Cref{fig:sq-model-study} shows the comparison of different contact angle models. The CCA model, \textcolor{Reviewer3}{i.e., for $\Tilde{\zeta}=0$},  uses the prescribed value of the contact angle, which is independent of $V_\Gamma$. This model shows faster dynamics as the difference between the initial contact angle $\theta_0$ and prescribed equilibrium contact angle $\theta_e$ is large during the early phase of the spreading. However, the DCA model with $\theta_d$ \textcolor{Reviewer3}{and $\Tilde{\zeta}>0$} shows a slower spreading rate. An increase in $\Tilde{\zeta}$ increased the contact line dissipation, hence, slows the spreading rate. \textcolor{Reviewer3}{In this study, an (arbitrary) value of $\beta=1$ is chosen, while the parameter $\Tilde{\zeta}$ is varied,  and the results are presented in \cref{fig:sq-zeta}}. The results with the plicRDF-isoAdvector method are then compared with (a) an in-house two-phase flow solver which employs the structured geometric VOF method, the Free Surface 3D (FS3D)~\cite{anjathesis,rieber2004numerische}, and (b) the JADIM code~\cite{JADIM,dupont2010numerical}. \textcolor{Reviewer3}{For the choice $\Tilde{\zeta}=2.8$,} the dynamics obtained using the plicRDF-isoAdvector method with \textcolor{Reviewer3}{~\cref{eq:cox-Voinov,eq:young-stress}} agree very well with the experimental findings of \cite{lavi2004} and perform better than the FS3D method and the JADIM code. 

The present case study evaluates the plicRDF-isoAdvector method within the scope of moving contact lines.\ By incorporating eq. \ref{eq:young-stress} in the contact angle boundary condition, this method successfully captures the dynamics of droplet spreading. However, it is observed that spurious currents affect the system's stability when approaching a steady state. These stability issues should be addressed in future work.


\section{Droplet spreading on a spherical surface} 
\label{sec:Droplet-spreading-on-a-spherical-surface}
\subsection{Definition of the case study}
This study investigates the spreading of a droplet on a complex spherical surface with a very small Bond number ($Bo\ll1$)~\citep{patel2017coupled}.\ As discussed in~\cref{sec:Droplet-spreading-on-a-flat-surface}, a droplet that spreads with $Bo\ll1$, maintains a spherical cap shape at the equilibrium.

\begin{figure}[h!]
    \centering
    \begin{subfigure}[b]{0.5\textwidth}
        \centering
	\captionsetup{position=top}
	\def\svgwidth{\textwidth}
	{\footnotesize
		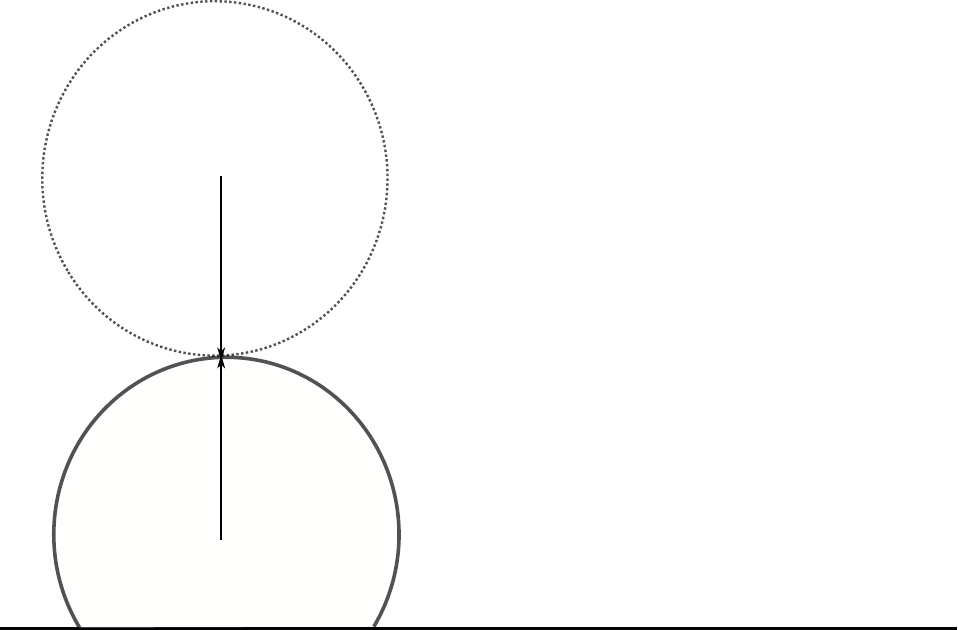
	}
	\label{fig:initial-config-spreading-over-spherical-surface}
    \end{subfigure}
    \hfill
    \begin{subfigure}[b]{0.48\textwidth}
        \centering
		\includegraphics[width=\textwidth, height=0.75\textwidth]{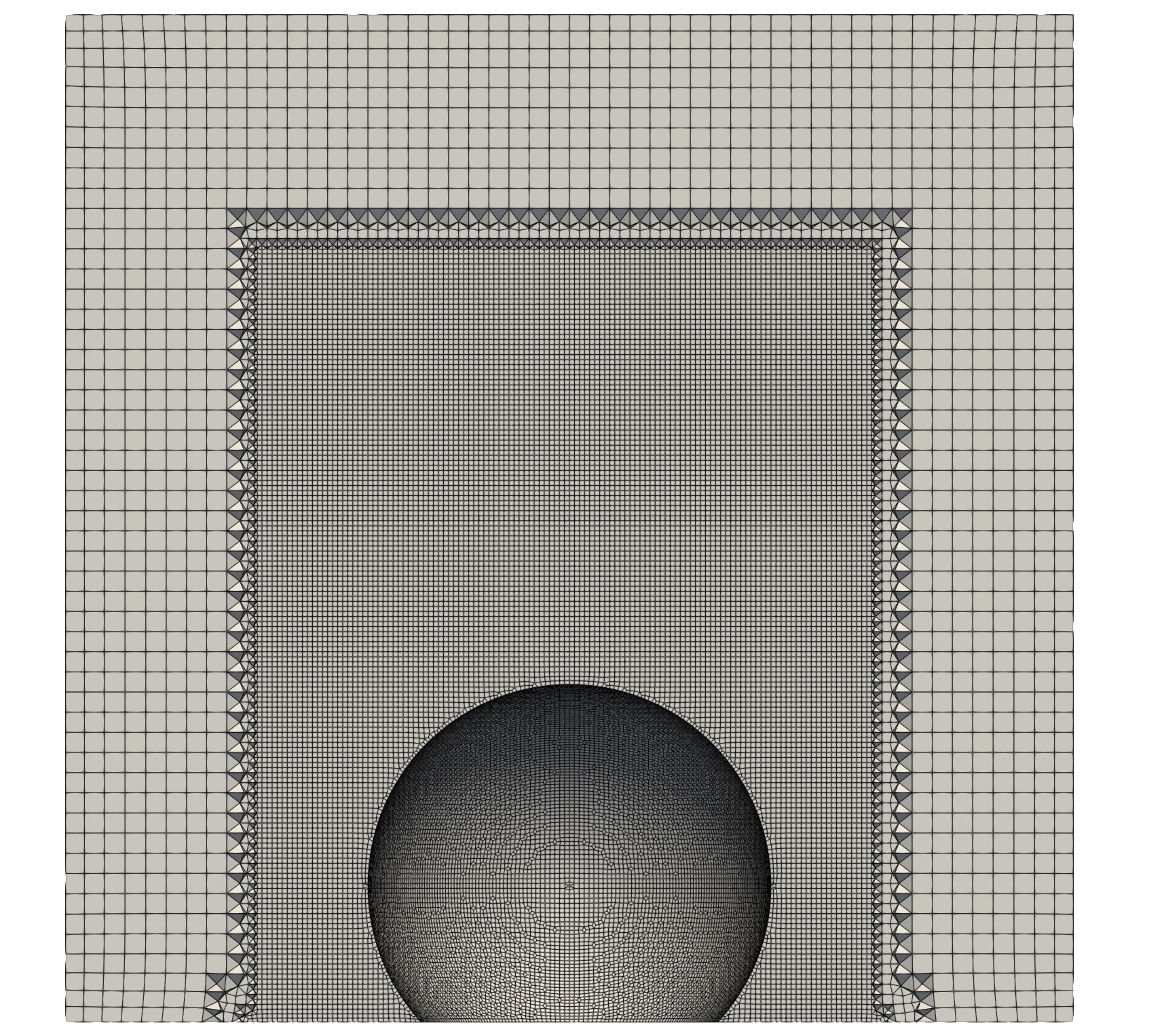}
		\label{fig:spherical-spreading-mesh}
    \end{subfigure}
    \caption{Schematic diagram of the initial configuration of a droplet ($\vcenter{\hbox{...}}$) with initial radius $R_0$\ (left) and final equilibrium shape ($\vcenter{\hbox{...}}$) (centre) spreading on a spherical surface (--) with equilibrium contact angle $\theta_e$. The equilibrium height is $e$, and the contact radius is $r$. Computational domain mesh represented on a clipped surface (right). The mesh is refined around the solid sphere, with a coarse mesh outside the region of interest.}
    \label{fig:spherical-spreading-domain}
\end{figure}

A three-dimensional computational domain is simulated, with domain parameters provided in \cref{table:computationalParamForDropletSpreadingOverSphere}. The domain is discretized using unstructured polyhedral cells, with a refined local mesh near the solid sphere and droplet.\ The refined region has a mesh density of 25 cells per radius, whereas the outer domain region has a coarser mesh with a mesh density of 10 cells per radius (see~\cref{fig:spherical-spreading-domain}).\ The droplet, with an initial radius $R_0$, is placed on the top of the spherical surface. The spherical surface has a no-slip boundary condition for the velocity (it applies numerical slip~\citep{renardy2001numerical} for the motion of the contact line).\ The droplet spreads on the spherical surface until it reaches the equilibrium, satisfying the static contact angle boundary condition. The time step $\DeltaT$ is restricted to CFL number below $0.01$ to ensure stability.
%
\begin{table}[h!]
	\centering
	\begin{tabular}{c c c} 
		\toprule
		Parameter & Value & Unit \\ [0.5ex] 
		\midrule
		Droplet initial radius, $R_0$& $1$ & mm \\
		Droplet initial position, $(x_0, y_0, z_0)$& $(2.5, 2.5, 2.672)$ & mm \\
		Spherical surface position, $(x_s, y_s, z_s)$& $(2, 2, 0.7)$ & mm \\
		Domain size & $(5, 5, 5)$ & mm \\
		Equilibrium contact angle,  $\theta_e$ & $30\text{, }50\text{, }90\text{, }110\text{, }140$& deg \\
		Gravitational acceleration, g & $0.00053$ & $\text{ms}^{-2}$ \\[1ex] 
		\bottomrule
	\end{tabular}
	\caption{Computational and geometrical parameters for the droplet spreading on a spherical surface with given equilibrium contact angle $\theta_e$.}
	\label{table:computationalParamForDropletSpreadingOverSphere}
\end{table}
\subsubsection{Geometrical relations for a droplet at equilibrium}
The conservation of the droplet's volume $V$ allows the formulation of geometrical relations for the contact radius $r$  and the droplet height $e$, which define the spherical cap at the equilibrium as
\begin{equation}
\begin{split}
\alpha + \beta & = \theta_e,  \\
r & = R \sin\beta = R_0 \sin\alpha, \\
e & = R (1+\cos\beta) - R_0(1-\cos\alpha), \\
\frac{8\pi R_0^3}{3}  & =\frac{\pi R^3}{3} (1+\cos\beta)^2 (2-\cos\beta)
+ \frac{\pi R_0^3}{3}(1+\cos\alpha)^2 (2-\cos\alpha).  
\end{split}
\label{eq:spherical_droplet_geometrical_relations}
\end{equation}
Here, the unknown parameters are $\alpha, \beta, r, e,$ and $R$. With the known droplet volume $(V_{\text{exact}} = \frac{4\pi R_0^3}{3})$ and initial guess for $\alpha$ and $\beta$, an intermediate volume $V^k(\alpha, \beta, R_0)$ is calculated by solving \cref{eq:spherical_droplet_geometrical_relations} iteratively for $k$-iterations using the bisection method. The values of $\alpha$ and  $\beta$ that minimize $|V_{\text{exact}} - V^k(\alpha, \beta, R_0)|$ are then used to approximate the contact radius $r$ and droplet height $e$.
\subsection{Post-processing}
\begin{figure}[h!]
	\centering
	\begin{subfigure}[b]{0.45\textwidth}
		\centering
		\includegraphics[width=0.6\textwidth, height=0.6\textwidth]{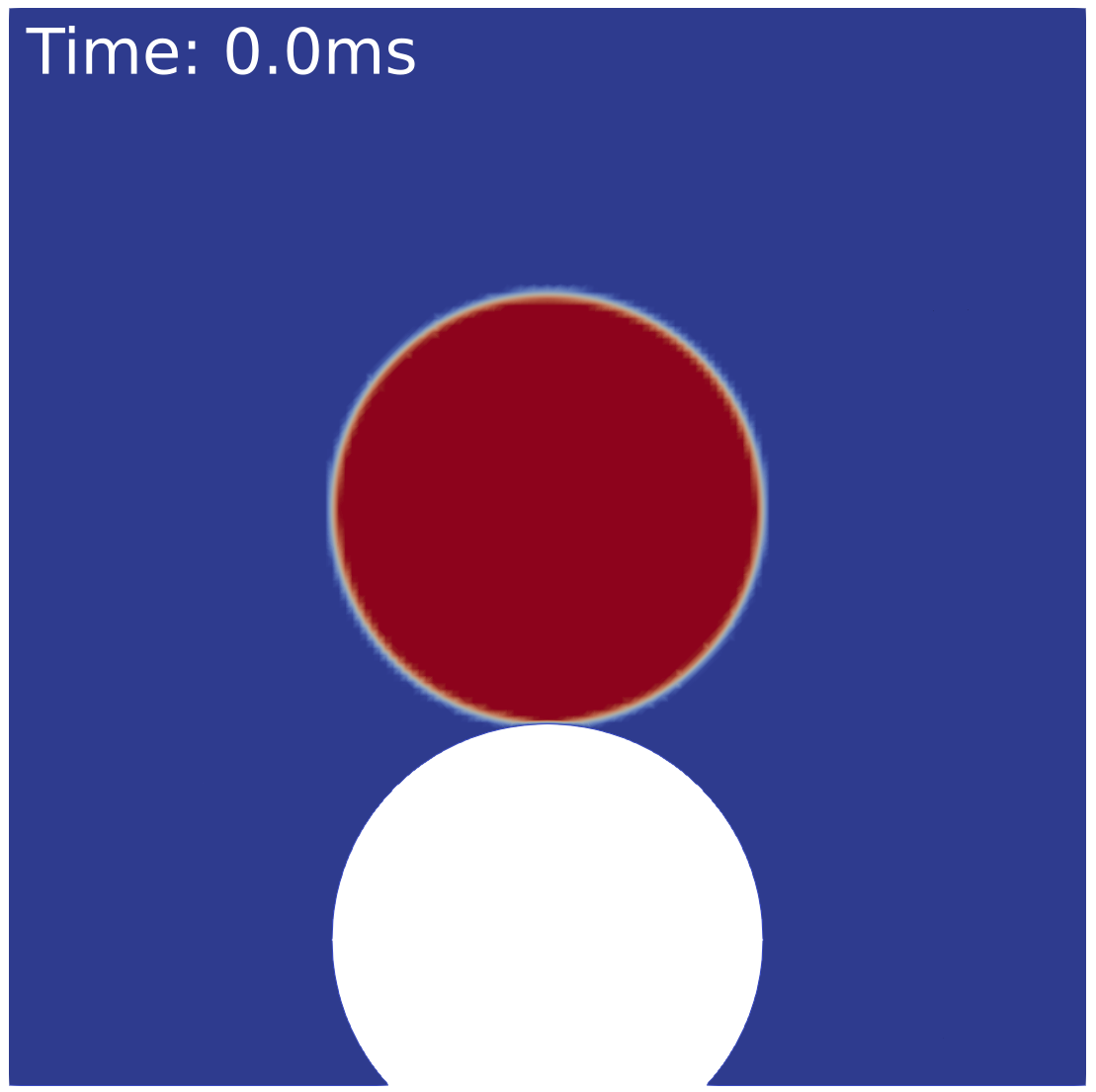}
		\caption{}
		\label{fig:transient-shapes-of-droplet-spreading-on-a-spherical-surface-0_0}
	\end{subfigure}
	\hfill
	\begin{subfigure}[b]{0.45\textwidth}
		\centering
		\includegraphics[width=0.6\textwidth, height=0.6\textwidth]{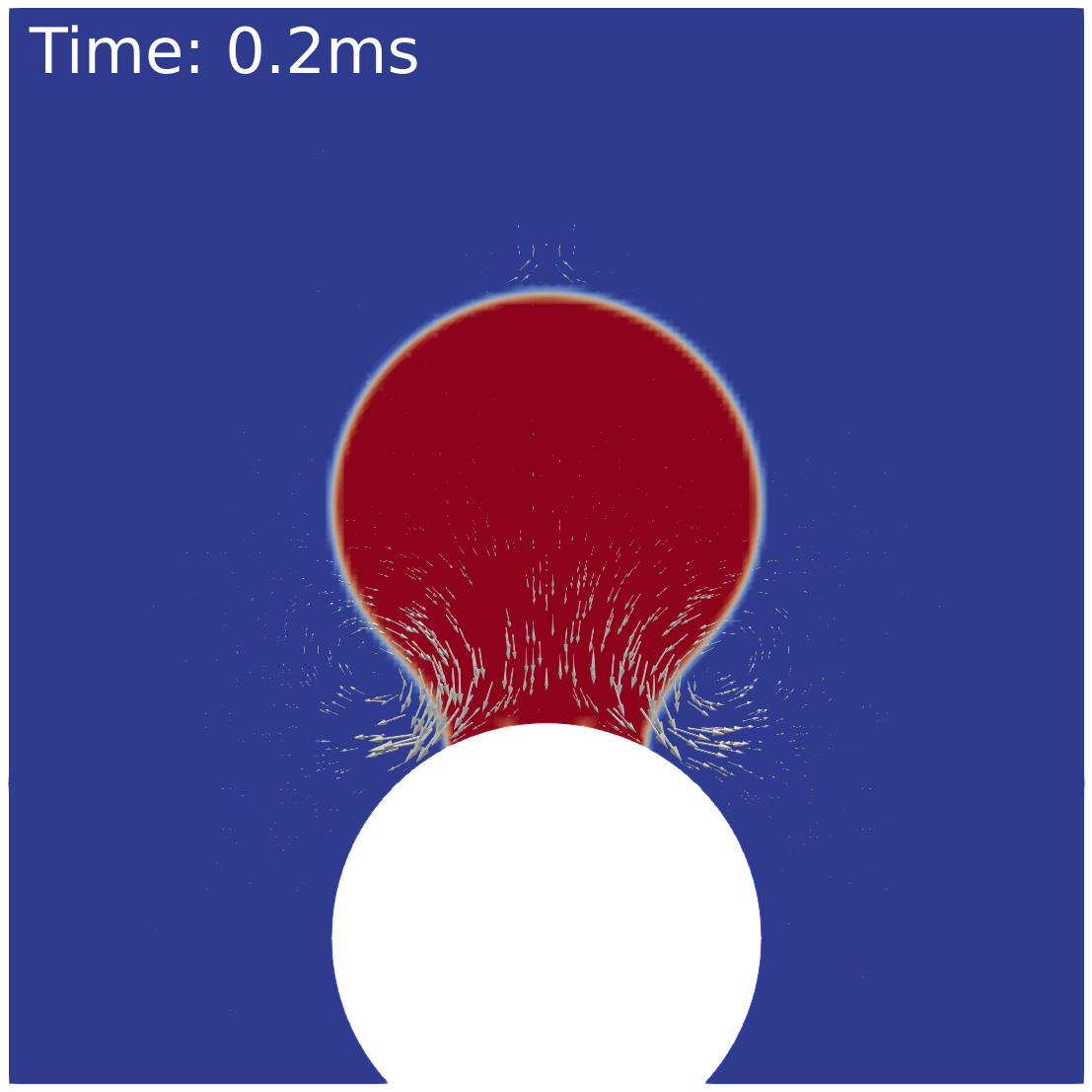}
		\caption{}
		\label{fig:transient-shapes-of-droplet-spreading-on-a-spherical-surface-0_0002}
	\end{subfigure}
	\centering
	\begin{subfigure}[b]{0.45\textwidth}
		\centering
		\includegraphics[width=0.6\textwidth, height=0.6\textwidth]{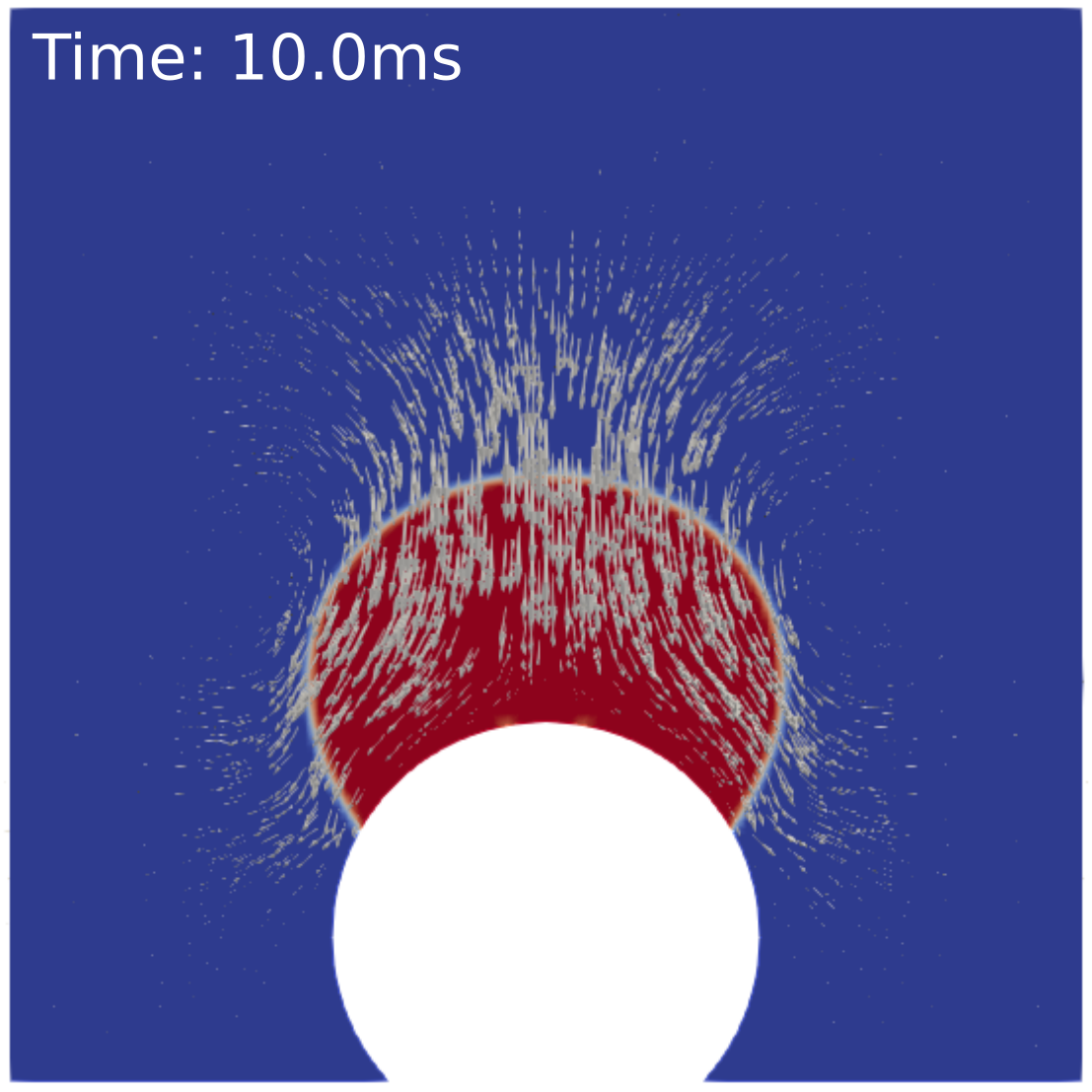}
		\caption{}
		\label{fig:transient-shapes-of-droplet-spreading-on-a-spherical-surface-0_01}
	\end{subfigure}
	\hfill
	\begin{subfigure}[b]{0.45\textwidth}
		\centering
		\includegraphics[width=0.6\textwidth, height=0.6\textwidth]{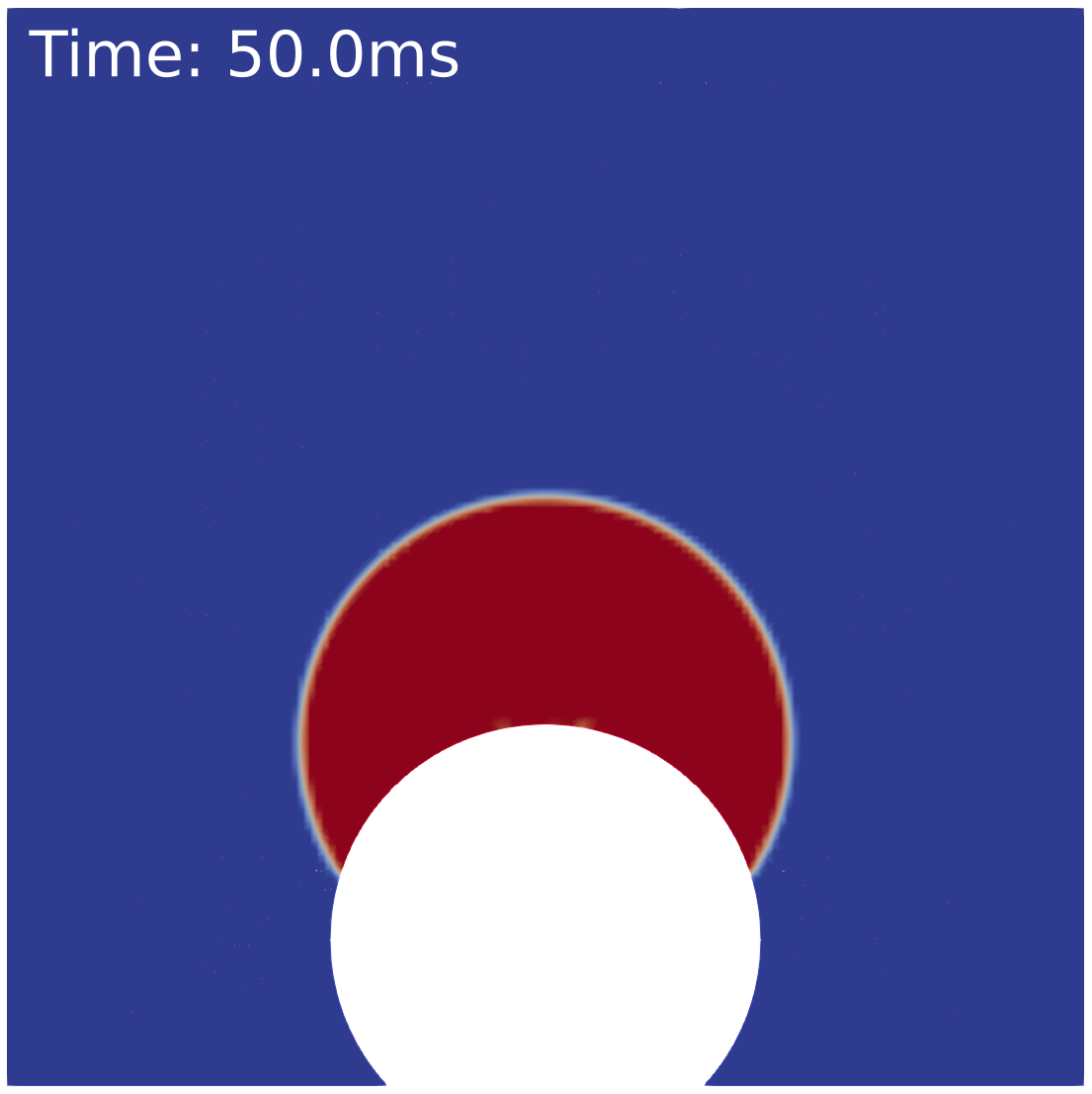}
		\caption{}
		\label{fig:transient-shapes-of-droplet-spreading-on-a-spherical-surface-0_05}
	\end{subfigure}
	\caption{Simulation results of droplet spreading on a spherical surface at different instants of time. The equilibrium contact angle is $\theta_e=50$°. The arrows represent the velocity field.}
	\label{fig:transient-shapes-of-droplet-spreading-on-a-spherical-surface}
\end{figure}

\Cref{fig:transient-shapes-of-droplet-spreading-on-a-spherical-surface} shows the droplet's shapes at different instances during spreading. The droplet's spreading dynamics are similar to those observed on a flat surface, as previously discussed in \cref{sec:Droplet-spreading-on-a-flat-surface}. At the beginning of spreading, initial rapid spreading locally at the contact line is observed, as shown in \cref{fig:transient-shapes-of-droplet-spreading-on-a-spherical-surface-0_0002}. As the simulation time progresses, the droplet's apex velocity increases in the downward direction (\cref{fig:transient-shapes-of-droplet-spreading-on-a-spherical-surface-0_01}), and the global shape of the droplet starts to change until the droplet attains the equilibrium shape (\cref{fig:transient-shapes-of-droplet-spreading-on-a-spherical-surface-0_05}). 
%


The numerical solution of the contact radius $r$ and equilibrium height $e$ involves identifying the boundary cells that contain the contact line. Identifying a contact line (cf. \cref{algo:CL}) can be challenging due to the formation of wisps, i.e., small artificial interface elements that appear in the bulk phase (as discussed by \citet{maric2018enhanced}). To address this issue, an OpenFOAM function object was developed to detect and successfully remove wisps from the contact line detection process (see \Cref{algo:wisp-detection} for details).

\floatname{algorithm}{Algorithm}
\begin{algorithm}[h!]
\caption{Detection of Wisps}
\begin{algorithmic}[1]
\Procedure{CheckWisps}{}
    \State $alphaTol$  \Comment{Volume fraction tolerance for interface cells}
    \State $wispTol$  \Comment{Volume fraction tolerance for wisp cells}
    \ForAll{$\text{patch} \in \text{patches}$}
            \State $normal,~centre \gets \text{ interface normal and centre}$
            \ForAll{$\text{facei} \in \text{patches[patch]}$}
                \State $\text{ownerCellId} \gets \text{cell index of owner cell of face~i}$
                \If{$\alpha\text{[ownerCellId]} < (1.0 - \text{alphaTol }) \text{ \& } \alpha \text{[ownerCellId]} > \text{alphaTol}$}
                    \State $\text{isAWisp} \gets \text{true}$ \Comment{All interface cells are initially marked as wisps}
                    \ForAll{$\text{NeighbourCells} \in \text{N}_{\text{facei}}$} \Comment{Loop over neighbors of face~i}
                        \State $\text{neiCellID} \gets \text{cell index of neighbor cell of face~i}$
                        \If{$\alpha\text{[neiCellID]} < (1.0 - \text{wispTol}) \text{ \& } \alpha \text{[neiCellID]} > \text{wispTol}$}
                            \State $\text{isAWisp} \gets \text{false}$
                            \State \textbf{break}
                        \EndIf
                    \EndFor
                \EndIf
            \EndFor
    \EndFor
\EndProcedure
\end{algorithmic}
\label{algo:wisp-detection}
\end{algorithm}
%

\subsection{Geometrical characteristics of the droplet}
\begin{figure}[H]
	\centering
	\includegraphics[width=0.65\textwidth, height=0.5\textwidth]{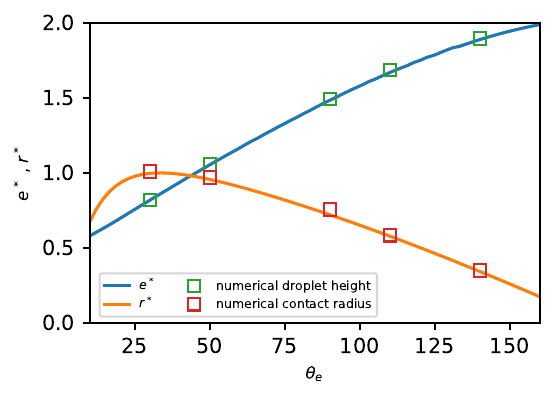}
	\caption{Dimensionless geometrical characteristics of the equilibrium droplet shape on a spherical surface: height (\textcolor{blue}{---}), contact radius (\textcolor{orange}{---}), numerical height (\textbf{\textcolor{mygreen}{\protect\mysquare}}), numerical contact radius (\textbf{\textcolor{red}{\protect\mysquare}}).}
	\label{fig:droplet-chractersitics-of-spherical-spreading-against-static-contact-angle}
\end{figure}

\begin{figure}[h!]
	\centering
	\includegraphics[width=0.4\textwidth, height=0.4\textwidth]{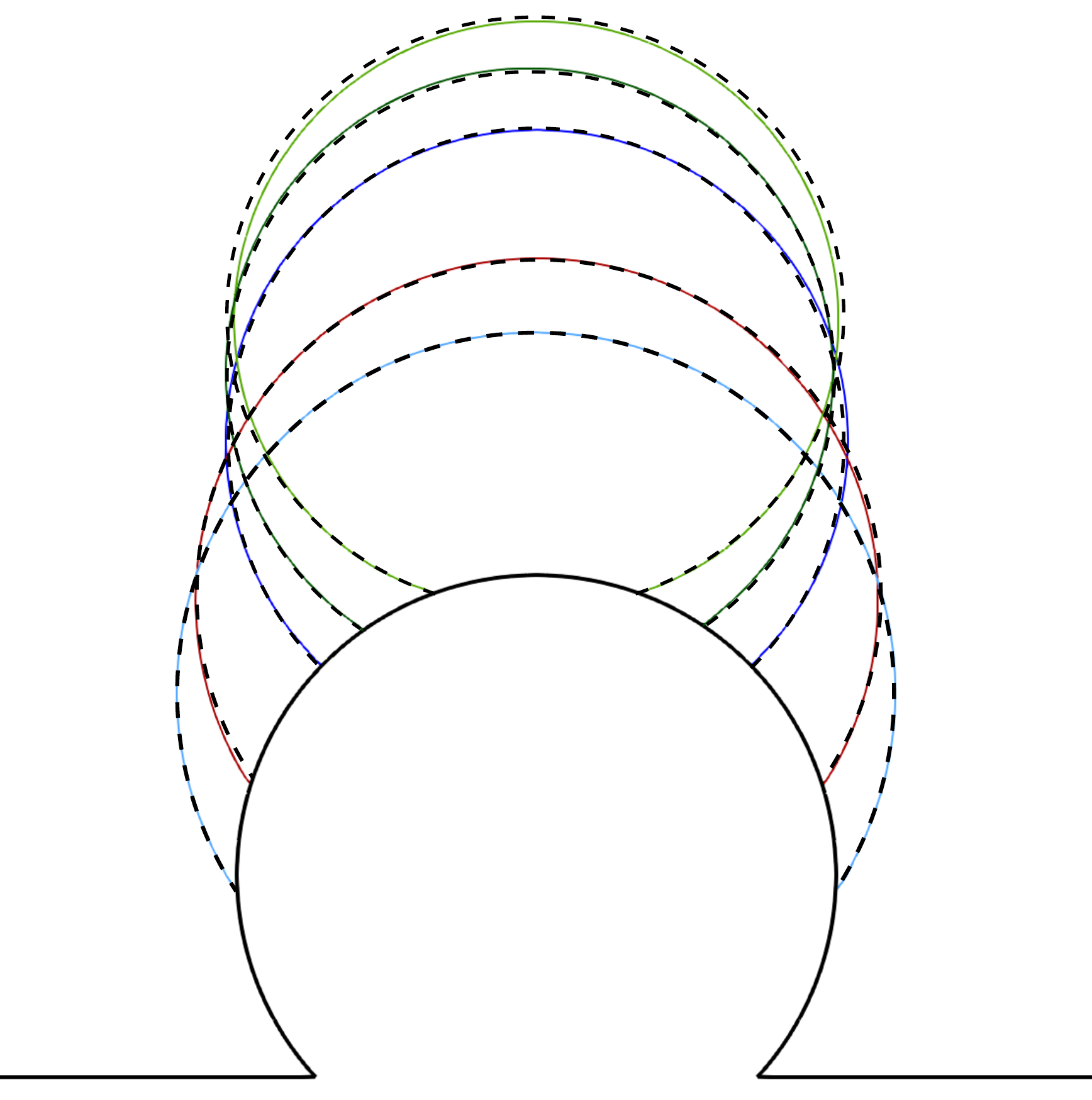}
	\caption{ Equilibrium drop shapes for equilibrium contact angle $\theta_e$ = \textcolor{cyan}{30°}, \textcolor{red}{50°}, \textcolor{blue}{90°}, \textcolor{teal}{110°}, and \textcolor{green}{140°}: geometrical reference solution (\texttt{-{}-}) (\cref{eq:spherical_droplet_geometrical_relations}), numerical (---).}
	\label{fig:sphericalSpreadingEquilibriumShapes}
\end{figure}

The simulation results for the non-dimensional equilibrium contact radius and height of a droplet spreading on a spherical surface for a range of equilibrium contact angles are presented in \cref{fig:droplet-chractersitics-of-spherical-spreading-against-static-contact-angle}. These results are compared to the reference solution from~\cref{eq:spherical_droplet_geometrical_relations} and are in excellent agreement.\ The equilibrium droplet shapes for various contact angles are illustrated in \cref{fig:sphericalSpreadingEquilibriumShapes} using $\alpha = 0.5$ contours.\ The comparison shows a good qualitative match between the reference and numerically computed droplet shapes.
The simulation results using the plicRDF-isoAdvector method have shown good agreement with the reference solution for droplet shape and geometrical characteristics. However, it should be noted that the symmetrical spreading of the droplet on the spherical surface is very sensitive to numerical noise. Simulations with small perturbations in the initial spherical shape can lead to a different equilibrium shape, with the droplet tilting to one side of the spherical surface. Even in such a case, the droplet shape remains close to a spherical cap, and the static contact angle boundary condition is still satisfied.

\section{2D capillary rise}
\label{sec:2D-capillary-rise}
\subsection{Definition of the case study}
The process of liquid flowing through narrow spaces, known as capillary action, has been studied profoundly in the literature (see, e.g., \cite{Washburn1921,Quere1999,fries2009dimensionless}).\ The process can be observed in the distribution of water from plants' roots to the rest of the body, the rising of liquids in porous media such as paper, and oil extraction from reservoirs, among others. \\
\citet{Quere1999}, and \citet{fries2009dimensionless} study the capillary rise based on a simplified model introduced by \citet{Bosanquet1923}. The latter model is an ODE resulting from empirical modeling of the forces acting on the liquid column.\ It can be shown easily \cite{Quere1999,fries2009dimensionless} that the dynamics of the capillary rise in Bosanquet's model is controlled by a single non-dimensionless group $\Omega$ defined as
\begin{equation}
\Omega = \sqrt{\frac{9\sigma\cos\theta \mu^2}{\rho^3g^2R^5}}.
\end{equation}
Moreover, \citet{Quere1999} showed that Bosanquet's model shows a regime transition at a critical value $\Omega_c = 2$. The column approaches its stationary state monotonically for $\Omega > 2$ while it shows rise height oscillations for $\Omega < 2$.\ These oscillations have been observed in experiments for low viscosity liquids~\citep{Quere1999}. Notably, \citet{grunding2020comparative} showed in their study that the Navier slip length, which is not accounted for in Bosanquet's model, can significantly influence the rise dynamics and the transient regime.\ In fact, rise height oscillations are increasingly damped out as the slip length decreases. This observation led to improved ODE modeling of capillary rise, taking into account the flow near the contact line \citep{delannoy2019dual,fricke2023bridging}.\ Moreover, it has been shown~\citep{fricke2023analytical} that the critical condition by~\citet{Quere1999} is generalized if additional channels of dissipation are active.\\
\\
In the stationary state, the height of the liquid column can be estimated by Jurin's law as
\begin{equation}
h\textsubscript{Jurin,2D} = \frac{\sigma \cos{\theta}}{R\rho g},
\label{eq:Jurins-height}
\end{equation}
where $\sigma$ is the surface tension coefficient, $R$ is the radius of the capillary, $\rho$ is the density of the liquid, $g$ is the gravitational acceleration, and $\theta$ is the contact angle.\ However, equation \eqref{eq:Jurins-height} neglects the liquid volume in the interface region, hence, overestimating the true stationary rise height. Gründing et al.\ \citep{grunding2020comparative,grunding2020enhanced} computed a corrected stationary capillary height from the liquid volume in the interface region (assuming a spherical cap shape). The corrected formula reads as
\begin{equation}
h = h\textsubscript{Jurin,2D} -\frac{R}{2\cos\theta} \left(2-\sin\theta - \frac{\sin^{-1}(\cos\theta)}{\cos\theta}\right).
\label{eq:corrected-capillary-height}
\end{equation}

In this study, we consider the two-dimensional case which corresponds to the rise of a liquid column between two planar parallel surfaces, as shown in \citep{grunding2020comparative}. We present a mesh convergence study for both the case of a no-slip (numerical slip) and a Navier slip boundary condition with a resolved slip length. As reported in \citep{grunding2020comparative}, resolving the slip length with the computational mesh is crucial for finding the mesh convergence of the contact line dynamics.\ We present the comparison of the plicRDF-isoAdvector method with other numerical methods (see~\cref{app:case-setup} for simulation setup details):

\begin{enumerate}
	\item the OpenFOAM solver interTrackFoam, an Arbitrary Lagrangian-Eulerian (ALE) method~\cite{ALE,dirkthesis}.
	\item the Free Surface 3D (FS3D), an in-house two-phase flow solver~\cite{rieber2004numerische}, employing the structured geometric VOF method (see Appendix~\ref{App:FS3D-description} for brief description).
	\item the OpenFOAM-based algebraic VOF solver, interFoam~\cite{openfoam1}.
	\item The Bounded Support Spectral Solver (BoSSS)~\cite{KummerThesis}, based on the extended discontinuous Galerkin method~\cite{kummer2017extended}.
\end{enumerate}

\subsubsection{Computational domain}
\begin{figure}[h!]
	\centering
	\def\svgwidth{0.5\textwidth}
	{\footnotesize
		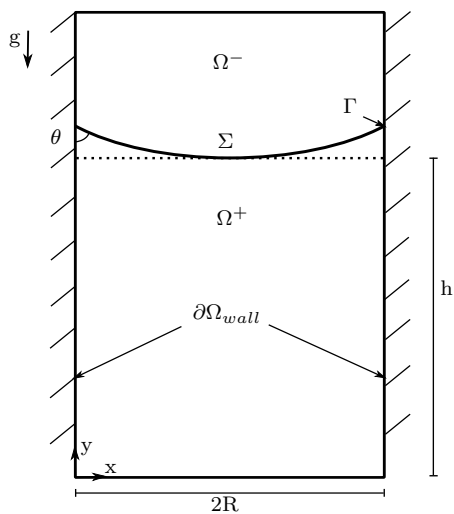
	}
	\caption{Schematic diagram of the initial configuration of the 2D capillary rise.}
	\label{fig:capillaryRiseSetup}
\end{figure}

    The initial configuration of the two-dimensional computational domain is shown in \cref{fig:capillaryRiseSetup}. The domain is discretized using the blockMesh utility of OpenFOAM, which creates a uniform Cartesian mesh in both the x and the y direction. The volume fraction field is initialized as a box $2\text{R} \times 2\text{R}$ at the bottom of the capillary. As the simulation starts, the interface evolves to satisfy the contact angle boundary condition and then rises. The set of physical parameters listed in \citep{grunding2020comparative} is designed to achieve different values for $\Omega$ while keeping the Bond number constant (see \cref{table:Omega-parameters}).
\begin{table}[h!]
	\centering
	\begin{tabular}{c c c c c c c c c} 
		\toprule
		$\Omega$ & $R$ & $\rho$& $\mu$ & $g$ & $\sigma$ & $\theta_e$ & Ca\textsubscript{max} & Bo\\ [1ex]
		\midrule  
		-  & m & $\text{kgm}^{-3}$& Pa~s& $\text{ms}^{-2}$& $\text{Nm}^{-1}$& °& - & -\\ [1ex]
		\midrule
		0.1  & 0.005 & 1663.8& 0.01& 1.04& 0.2& 30& 0.0033 & 0.217\\ [1ex]
		0.5  & 0.005 & 133.0& 0.01& 6.51& 0.1& 30& 0.015 & 0.217\\ [1ex]
		1  & 0.005 & 83.1& 0.01& 4.17& 0.04& 30& 0.029 & 0.217\\ [1ex]
		\bottomrule
	\end{tabular}
	\caption{Physical parameters for different $\Omega$. }
	\label{table:Omega-parameters}
\end{table}
\subsection{Mesh convergence study}
\begin{figure}[h!]
	\centering
	\begin{subfigure}[b]{0.49\textwidth}
		\centering
		\includegraphics[width=\textwidth, height=0.7\textwidth]{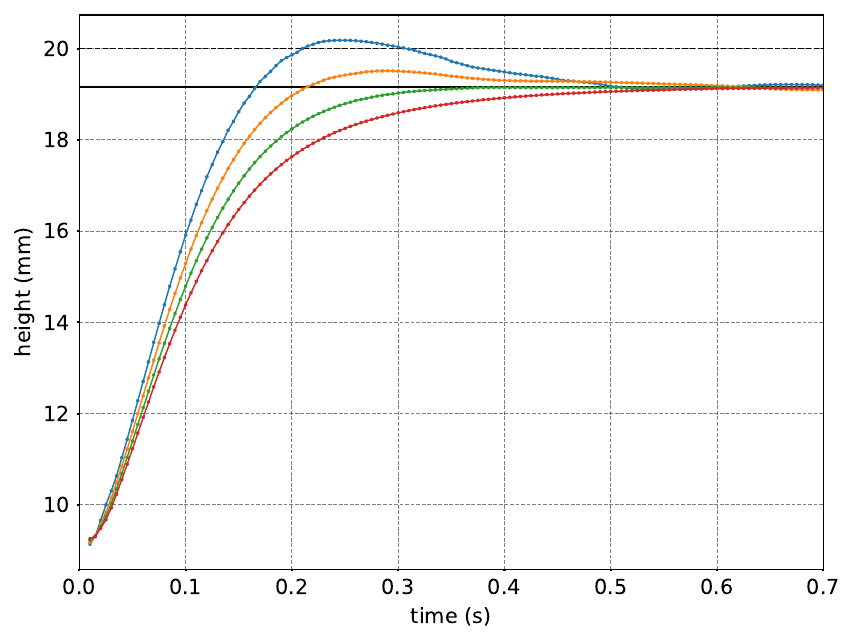}
		\caption{No-slip boundary condition.}
		\label{fig:capillary-rise-with-numerical-slip}
	\end{subfigure}
	\hfill
	\begin{subfigure}[b]{0.49\textwidth}
		\centering
		\includegraphics[width=\textwidth, height=0.7\textwidth]{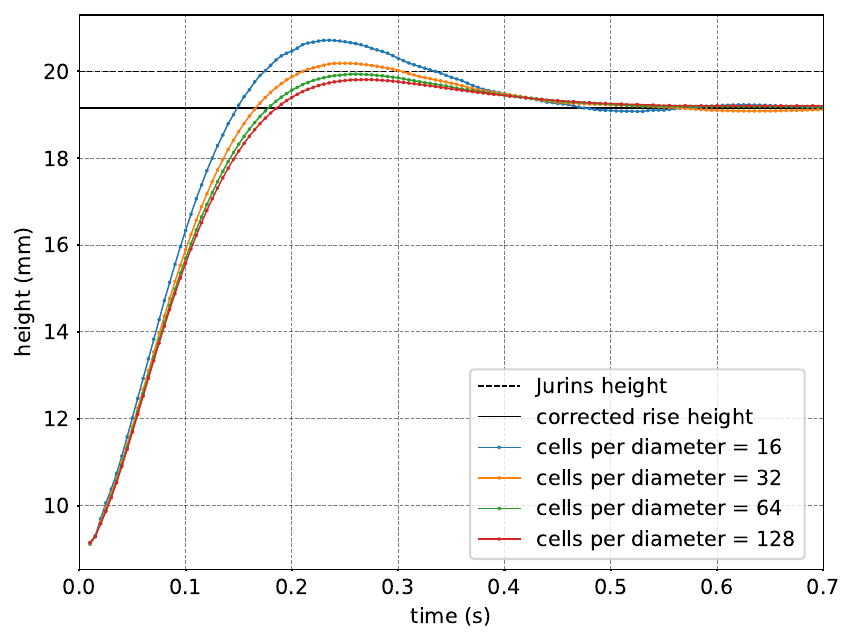}
		\caption{Navier slip with slip length $\lambda=0.1~\mathrm{mm}$.}
		\label{fig:capillary-rise-with-partial-slip}
	\end{subfigure}
	\caption{Mesh convergence study using a no-slip and a Navier slip boundary condition. Non-dimensionless number $\Omega=1$ and $\theta_e=30$°. }
	\label{fig:mesh-study-for-capillary-rise}
\end{figure}

A mesh convergence study for the 2D capillary rise with a mesh resolution of 16 to 256 cells per diameter of the capillary was conducted.\ We keep $\Omega=1$ and the slip length $\lambda = 0.1~\mathrm{mm}$ for this study.\ As $\Omega < \Omega_c$ is chosen here, we expect to see oscillations during the capillary rise from Quere's theory. 
The results in \cref{fig:capillary-rise-with-numerical-slip} are obtained using the no-slip boundary condition. However, since the method has some implicit ``numerical slip'', we still observe a contact line motion. As a consequence of the numerical slip being linked to the grid size, the results show a strong dependence on mesh resolution regarding the dynamics. In particular, the oscillations are increasingly dampened with the increase in the mesh resolution. This is expected since, for no-slip, it is well known that the viscous dissipation near the contact is non-integrably singular. As the slip length decreases, thus approaching the no-slip limit, the numerical solution starts showing the signature of the ill-posedness of the limiting problem~(Huh and Scriven paradox~\citep{huh1971hydrodynamic}). However, for Navier slip with positive slip length, pressure and viscous dissipation are integrable~\citep{Huh1977}.\ Therefore, the simulations with Navier slip (see \cref{fig:capillary-rise-with-partial-slip}) show mesh convergence. Although the solutions with the numerical slip and Navier slip differ in the rise dynamics, it is to be noted that the stationary rise height is mesh-independent for both cases and levels at the corrected stationary rise height (given by \cref{eq:corrected-capillary-height}).
\subsection{Effect of the dimensionless parameter $\Omega$ on the capillary rise}
\begin{figure}[h!]
	\centering
	\begin{subfigure}[b]{0.49\textwidth}
		\centering
		\includegraphics[width=\textwidth, height=0.7\textwidth]{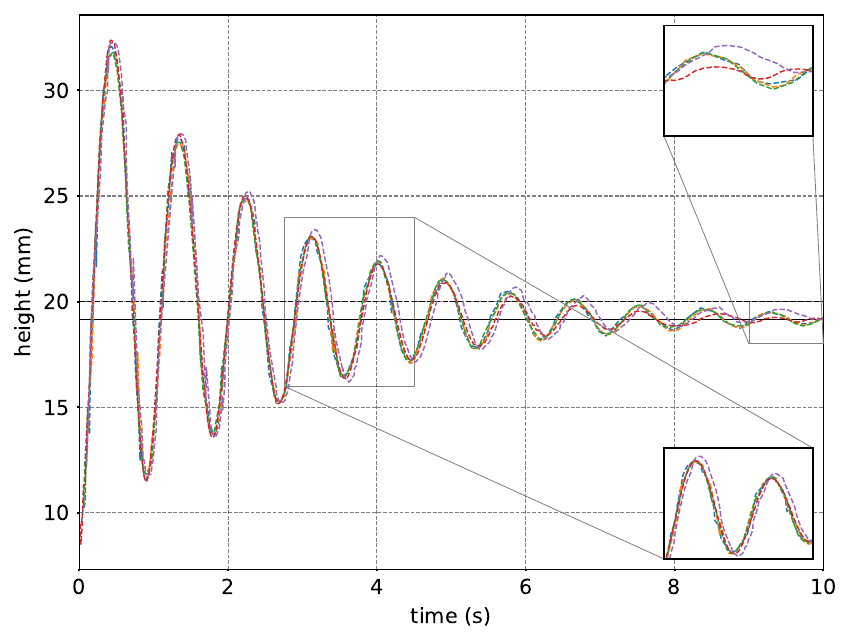}
		\caption{$\Omega=0.1$.}
		\label{fig:capillary-rise-with-Omega-0-1}
	\end{subfigure}
	\hfill
	\begin{subfigure}[b]{0.49\textwidth}
		\centering
		\includegraphics[width=\textwidth, height=0.7\textwidth]{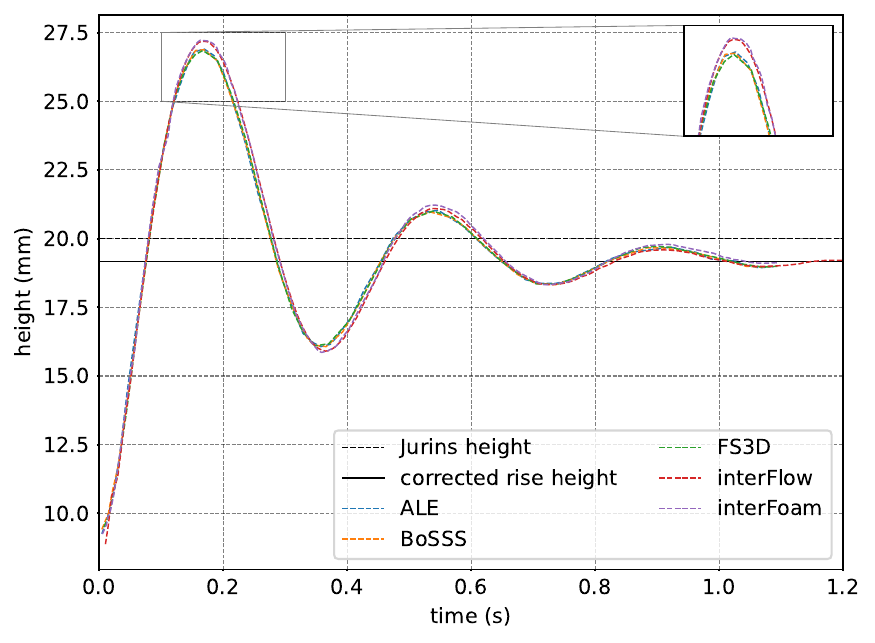}
		\caption{$\Omega=0.5$.}
		\label{fig:capillary-rise-with-Omega-0-5}
	\end{subfigure}
	\caption{Numerical result for capillary rise for the $\Omega$-study (oscillatory regime).} 
	\label{fig:omega-study-for-capillary-rise}
\end{figure}

\Cref{fig:omega-study-for-capillary-rise} shows the simulation results for $\Omega=0.1$ and $\Omega=0.5$.\ A comparison of several other numerical methods with the plicRDF-isoAdvector method is also presented. As the $\Omega$ values chosen here are below the critical value $\Omega_c=2$, we expect to see oscillations during the capillary rise from Quere's theory. The slip length is chosen to be $\lambda=R/5=1~\mathrm{mm}$. For $\Omega=0.1$ (see \cref{fig:capillary-rise-with-Omega-0-1}), we observe strong oscillations that decrease in amplitude with time, and the solution asymptotically reaches the reference (corrected) rise height value. It can be noted that the amplitude with the interFoam method, as well as the plicRDF-isoAdvector method, is higher than the oscillation's amplitude obtained with other methods. A minor phase shift for these two methods can also be observed, which, for $\Omega=0.1$, keeps on increasing with time. The plicRDF-isoAdvector method's solution dampens slightly faster than with the other numerical methods (see \cref{fig:capillary-rise-with-Omega-0-1,fig:capillary-rise-with-Omega-0-5}). Similar behavior can be observed for $\Omega=0.5$ (see \cref{fig:capillary-rise-with-Omega-0-5}), but with fewer oscillations, and the simulations results level with the stationary height according to \cref{eq:corrected-capillary-height}.
The case study has shown the ability of the plicRDF-isoAdvector method to capture the dynamics of the capillary rise in Quere’s model, and the stationary rise height obtained agrees well with the corrected capillary height~\ref{eq:corrected-capillary-height}.

\section{Summary and Outlook}
\label{sec:conclusion}

We have verified and validated the plicRDF-isoAdvector method for wetting processes through five case studies. The first study investigated interface advection, where a first- to second-order convergence transition was observed, and the method performed better than the boundary ELVIRA method \citep{fricke2020contact} in terms of absolute errors of the contact angle evolution with respect to the analytical solution. The second study verified the droplet spreading on a flat surface. The method showed excellent agreement with geometrical solutions at equilibrium, except for highly hydrophobic or hydrophilic cases. We observed that the initial state of the system significantly impacts the dynamics of the system. Systems that have an initial state closer to the equilibrium provide more stable solutions. The third case study validated the method using the experimental results of~\cite{lavi2004}. It demonstrated that employing the uncompensated Young-stress term in the contact angle boundary condition achieves very good agreement with these results and performs better than other numerical methods. The fourth study tested the method for droplet spreading on a spherical surface, showing accurate results but with some sensitivity to numerical noise. This case study provided foundational work for wetting simulations of more complex substrates and will be used to validate the interaction of droplets with complex substrates~\cite{shumaly2023deep,mitra2013} in future work. The last study was a 2D capillary rise, with mesh-convergent results for the stationary state and contact line dynamics. The dynamic oscillations of the capillary are controlled by the dimensionless parameter $\Omega$, and it is noted that the oscillations in the plicRDF-isoAdvector system dampen earlier as compared to the results of other numerical methods.\ The case setup, input data, and post-processing Jupyter Notebooks are publicly available online~\citep{B01code,asgharinput}, and this provides a valuable starting point for evaluating numerical methods for wetting processes.

As we have shown, the plicRDF-isoAdvector method can effectively simulate a broad spectrum of wetting problems.\ However, there are still challenges in simulating contact angles that are either very small (hydrophilic support) or very large (hydrophobic support). The plicRDF reconstruction indirectly considers the contact angle at the wall through a weighted sum of signed-distance values, whose discrete gradient determines the cell-centered interface orientation. Also, the influence of the contact angle in the kinematic motion of a PLIC interface is mediated by the cell-centered PLIC interface orientation. Therefore, future developments should reconsider the approach via the weighted contribution of the contact angle in the RDF-based PLIC reconstruction.\ We also recommend local adaptive mesh refinement near the interface with at least three cell layers of as uniform as possible mesh resolution surrounding the interface. Although plicRDF-isoAdvector can handle mesh grading, an interface passing through a non-uniform mesh-grading near a wall causes a loss in convergence order.\ Therefore, an unstructured geometrical VOF interface reconstruction that exactly satisfies the prescribed contact angle in wall-adjacent cells without incurring a loss of accuracy away from the wall-adjacent cell layer is necessary for contact line evolution.

\clearpage
\section*{Appendices}
\addcontentsline{toc}{section}{Appendices}
\renewcommand{\thesubsection}{\Alph{subsection}}
\label{sec:appa}
\subsection{Droplet spreading with droplet initialized as a full sphere}

\subsection{Computational setup used in \cite{grunding2020comparative} for 2D capillary rise case study}

\begin{enumerate}
    \item{The OpenFOAM solver interTrackFoam, an Arbitrary Lagrangian-Eulerian (ALE) method~\cite{ALE,dirkthesis}:}\\
        The computational domain is initialized as a box $2\text{R} \times 2\text{R}$.\ The simulation starts with a Dirichlet condition for the inlet velocity with a fixed value.\ The interface evolves into a meniscus shape with a prescribed contact angle $\theta_0$. A Laplace equation governs the mesh motion, and an anisotropic mesh diffusion coefficient is chosen for better resolution at the interface during the mesh motion. After the evolution of the interface shape to a meniscus, the simulation is restarted with a homogeneous Neumann boundary condition for the inlet velocity and a homogeneous Dirichlet boundary condition for pressure.\ The Navier slip boundary condition is applied at the capillary walls.
    \item{An in-house two-phase flow solver the Free Surface 3D (FS3D), employing the geometric VOF method:}\\
        An axisymmetric computational domain is initialized using a uniform Cartesian mesh with a symmetric boundary condition applied at the central axis of the capillary.\ The volume fraction field is initialized as a spherical interface with an equilibrium contact angle $\theta_e$. The Navier slip boundary condition is applied at the domain walls.
    \item{The OpenFOAM-based algebraic VOF solver interFoam~\cite{openfoam1}:}\\
        Similar to the FS3D setup, the symmetry boundary condition in the center of the capillary is used for simulations with interFoam. The volume fraction field is initialized as a box at the bottom of the capillary. The simulation starts with a ``closed" inlet, and the interface evolves to the meniscus shape with an equilibrium contact angle $\theta_e$. The Navier slip boundary condition is applied at the capillary wall. 
    \item{The Bounded Support Spectral Solver (BoSSS)~\cite{KummerThesis}, based on the extended discontinuous Galerkin method~\cite{kummer2017extended}:}\\
        For simulations with the BoSSS numerical method, a symmetry boundary condition is applied in the centre of the capillary.\ The domain is discretized using a uniform equidistant mesh. The interface is initialized as described for the ALE method. An \textit{inflow} and \textit{outflow} boundary condition is applied at the bottom and top of the domain, respectively. The Navier slip boundary condition is applied at the domain wall. 
        \label{app:case-setup}
\end{enumerate}
\subsection{FS3D - a geometric VOF method}
\label{App:FS3D-description}

The FS3D solver, originally developed in~\cite{rieber2004numerische,RIEBER1999455} is a geometrical Volume-of-Fluid method that solves the incompressible two-phase Navier Stokes equations on a structured Cartesian grid. Like in the plicRDF-isoAdvector method, the single-field formulation in the Continuum Surface Force (CSF) formulation by~\cite{BRACKBILL1992335} is used. The method was extended to describe moving contact lines by \cite{anjathesis,mathisThesis}. The mean curvature of the interface is computed using the height function approach by \cite{POPINET20095838}. Following the method by \cite{afkhami2008height}, the height function method is adapted in the contact line region to apply the contact angle boundary condition. Due to the Cartesian mesh structure, the method is computationally efficient but can only deal with comparably simple geometries like flat surfaces or surfaces with simple structures like stripes \cite{HARTMANN2021103582}.
\subsection{Flowchart of solution algorithm of the plicRDF-isoAdvector method}

\section*{Acknowledgements}

We acknowledge the financial support by the German Research Foundation (DFG) within the Collaborative Research Centre 1194 (Project-ID 265191195).

The use of the high-performance computing resources of the Lichtenberg High-Performance Cluster at the TU Darmstadt is gratefully acknowledged.

\clearpage

\bibliography{references}

\begin{thebibliography}{76}
\expandafter\ifx\csname natexlab\endcsname\relax\def\natexlab#1{#1}\fi
\providecommand{\url}[1]{\texttt{#1}}
\providecommand{\href}[2]{#2}
\providecommand{\path}[1]{#1}
\providecommand{\DOIprefix}{doi:}
\providecommand{\ArXivprefix}{arXiv:}
\providecommand{\URLprefix}{URL: }
\providecommand{\Pubmedprefix}{pmid:}
\providecommand{\doi}[1]{\href{http://dx.doi.org/#1}{\path{#1}}}
\providecommand{\Pubmed}[1]{\href{pmid:#1}{\path{#1}}}
\providecommand{\bibinfo}[2]{#2}
\ifx\xfnm\relax \def\xfnm[#1]{\unskip,\space#1}\fi
\bibitem[{Marengo and Coninck(2022)}]{Marengo2022}
\bibinfo{editor}{M.~Marengo}, \bibinfo{editor}{J.~D. Coninck} (Eds.),
  \bibinfo{title}{{The Surface Wettability Effect on Phase Change}},
  \bibinfo{publisher}{Springer International Publishing}, \bibinfo{year}{2022}.
  \DOIprefix\doi{10.1007/978-3-030-82992-6}.
\bibitem[{Marić et~al.(2020)Marić, Kothe, and Bothe}]{maric2020unstructured}
\bibinfo{author}{T.~Marić}, \bibinfo{author}{D.~B. Kothe},
  \bibinfo{author}{D.~Bothe},
\newblock \bibinfo{title}{{Unstructured un-split geometrical Volume-of-Fluid
  methods – A review}},
\newblock \bibinfo{journal}{Journal of Computational Physics}
  \bibinfo{volume}{420} (\bibinfo{year}{2020}) \bibinfo{pages}{109695}.
  \DOIprefix\doi{https://doi.org/10.1016/j.jcp.2020.109695}.
\bibitem[{{OpenCFD Ltd.}(2006)}]{openfoam1}
\bibinfo{author}{{OpenCFD Ltd.}}, \bibinfo{title}{{OpenFOAM: user Guide
  v2006}}, \bibinfo{year}{2006}.
  \bibinfo{note}{\url{https://openfoam.com/documentation/guides/latest/doc/},
  Last accessed on 2024-05-09}.
\bibitem[{Deshpande et~al.(2012)Deshpande, Anumolu, and
  Trujillo}]{deshpande2012evaluating}
\bibinfo{author}{S.~S. Deshpande}, \bibinfo{author}{L.~Anumolu},
  \bibinfo{author}{M.~F. Trujillo},
\newblock \bibinfo{title}{{Evaluating the performance of the two-phase flow
  solver interFoam}},
\newblock \bibinfo{journal}{Computational science \& discovery}
  \bibinfo{volume}{5} (\bibinfo{year}{2012}) \bibinfo{pages}{014016}.
  \DOIprefix\doi{10.1088/1749-4699/5/1/014016}.
\bibitem[{Roenby et~al.(2016)Roenby, Bredmose, and
  Jasak}]{roenby2016computational}
\bibinfo{author}{J.~Roenby}, \bibinfo{author}{H.~Bredmose},
  \bibinfo{author}{H.~Jasak},
\newblock \bibinfo{title}{{A computational method for sharp interface
  advection}},
\newblock \bibinfo{journal}{Royal Society open science} \bibinfo{volume}{3}
  (\bibinfo{year}{2016}) \bibinfo{pages}{160405}.
  \DOIprefix\doi{https://doi.org/10.1098/rsos.160405}.
\bibitem[{Scheufler and Roenby(2019)}]{scheufler2019accurate}
\bibinfo{author}{H.~Scheufler}, \bibinfo{author}{J.~Roenby},
\newblock \bibinfo{title}{{Accurate and efficient surface reconstruction from
  volume fraction data on general meshes}},
\newblock \bibinfo{journal}{Journal of Computational Physics}
  \bibinfo{volume}{383} (\bibinfo{year}{2019}) \bibinfo{pages}{1--23}.
  \DOIprefix\doi{https://doi.org/10.1016/j.jcp.2019.01.009}.
\bibitem[{Gamet et~al.(2020)Gamet, Scala, Roenby, Scheufler, and
  Pierson}]{gamet2020validation}
\bibinfo{author}{L.~Gamet}, \bibinfo{author}{M.~Scala},
  \bibinfo{author}{J.~Roenby}, \bibinfo{author}{H.~Scheufler},
  \bibinfo{author}{J.-L. Pierson},
\newblock \bibinfo{title}{{Validation of volume-of-fluid
  OpenFOAM{\textregistered} isoAdvector solvers using single bubble
  benchmarks}},
\newblock \bibinfo{journal}{Computers \& Fluids} \bibinfo{volume}{213}
  (\bibinfo{year}{2020}) \bibinfo{pages}{104722}.
  \DOIprefix\doi{https://doi.org/10.1016/j.compfluid.2020.104722}.
\bibitem[{Popinet(2015)}]{popinet2015quadtree}
\bibinfo{author}{S.~Popinet},
\newblock \bibinfo{title}{{A quadtree-adaptive multigrid solver for the
  Serre--Green--Naghdi equations}},
\newblock \bibinfo{journal}{Journal of Computational Physics}
  \bibinfo{volume}{302} (\bibinfo{year}{2015}) \bibinfo{pages}{336--358}.
  \DOIprefix\doi{https://doi.org/10.1016/j.jcp.2015.09.009}.
\bibitem[{Osher and Sethian(1988)}]{osher1988fronts}
\bibinfo{author}{S.~Osher}, \bibinfo{author}{J.~A. Sethian},
\newblock \bibinfo{title}{{Fronts propagating with curvature-dependent speed:
  Algorithms based on Hamilton-Jacobi formulations}},
\newblock \bibinfo{journal}{Journal of Computational Physics}
  \bibinfo{volume}{79} (\bibinfo{year}{1988}) \bibinfo{pages}{12--49}.
  \DOIprefix\doi{https://doi.org/10.1016/0021-9991(88)90002-2}.
\bibitem[{Turek(1999)}]{turek1999efficient}
\bibinfo{author}{S.~Turek}, \bibinfo{title}{{Efficient Solvers for
  Incompressible Flow Problems: An Algorithmic and Computational Approache}},
  volume~\bibinfo{volume}{6}, \bibinfo{publisher}{Springer Science \& Business
  Media}, \bibinfo{year}{1999}.
\bibitem[{Parolini and Burman(2005)}]{parolini2005finite}
\bibinfo{author}{N.~Parolini}, \bibinfo{author}{E.~Burman},
\newblock \bibinfo{title}{{A finite element level set method for viscous
  free-surface flows}},
\newblock in: \bibinfo{booktitle}{Applied and industrial mathematics in Italy},
  \bibinfo{publisher}{World Scientific}, \bibinfo{year}{2005}, pp.
  \bibinfo{pages}{416--427}.
  \DOIprefix\doi{https://doi.org/10.1142/9789812701817_0038}.
\bibitem[{John and Matthies(2004)}]{john2004moonmd}
\bibinfo{author}{V.~John}, \bibinfo{author}{G.~Matthies},
\newblock \bibinfo{title}{{MooNMD--a program package based on mapped finite
  element methods}},
\newblock \bibinfo{journal}{Computing and Visualization in Science}
  \bibinfo{volume}{6} (\bibinfo{year}{2004}) \bibinfo{pages}{163--170}.
  \DOIprefix\doi{https://doi.org/10.1007/s00791-003-0120-1}.
\bibitem[{Siriano et~al.(2022)Siriano, Balc{\'a}zar, Tassone, Rigola, and
  Caruso}]{siriano2022numerical}
\bibinfo{author}{S.~Siriano}, \bibinfo{author}{N.~Balc{\'a}zar},
  \bibinfo{author}{A.~Tassone}, \bibinfo{author}{J.~Rigola},
  \bibinfo{author}{G.~Caruso},
\newblock \bibinfo{title}{{Numerical Simulation of High-Density Ratio Bubble
  Motion with interIsoFoam}},
\newblock \bibinfo{journal}{Fluids} \bibinfo{volume}{7} (\bibinfo{year}{2022})
  \bibinfo{pages}{152}. \DOIprefix\doi{https://doi.org/10.3390/fluids7050152}.
\bibitem[{Lippert et~al.(2022)Lippert, Tolle, Dörr, and
  Maric}]{lippert2022benchmark}
\bibinfo{author}{A.~Lippert}, \bibinfo{author}{T.~Tolle},
  \bibinfo{author}{A.~Dörr}, \bibinfo{author}{T.~Maric}, \bibinfo{title}{A
  benchmark for surface-tension-driven incompressible two-phase flows},
  \bibinfo{year}{2022}. \href{http://arxiv.org/abs/2212.02904}{{\tt
  arXiv:2212.02904}}.
\bibitem[{TwoPhaseFlow-of2312(2023)}]{TwoPhaseFlow}
\bibinfo{author}{TwoPhaseFlow-of2312}, \bibinfo{year}{2023}. \URLprefix
  \url{https://github.com/tmaric/TwoPhaseFlow/tree/of2312}.
\bibitem[{Giefer et~al.(2023)Giefer, Kyrloglou, and
  Fritsching}]{giefer2023impact}
\bibinfo{author}{P.~Giefer}, \bibinfo{author}{A.~Kyrloglou},
  \bibinfo{author}{U.~Fritsching},
\newblock \bibinfo{title}{Impact of wettability on interface deformation and
  droplet breakup in microcapillaries},
\newblock \bibinfo{journal}{Physics of Fluids} \bibinfo{volume}{35}
  (\bibinfo{year}{2023}).
\bibitem[{{Asghar, M. H. and Marić, T. }(2024)}]{B01code}
\bibinfo{author}{{Asghar, M. H. and Marić, T. }},
  \bibinfo{title}{{plicRDF-isoAdvector benchmarks for wetting processes}},
  \bibinfo{howpublished}{\url{https://github.com/CRC-1194/b01-wetting-benchmark/releases/tag/release1.0}},
  \bibinfo{year}{2024}. \URLprefix
  \url{https://github.com/CRC-1194/b01-wetting-benchmark/releases/tag/release1.0},
  \bibinfo{note}{gitHub repository, Release 1.0}.
\bibitem[{Asghar et~al.(5 24)Asghar, Fricke, Bothe, and Marić}]{asgharinput}
\bibinfo{author}{M.~H. Asghar}, \bibinfo{author}{M.~Fricke},
  \bibinfo{author}{D.~Bothe}, \bibinfo{author}{T.~Marić},
  \bibinfo{title}{Validation and verification of the plicrdf-isoadvector
  unstructured volume-of-fluid (vof) method for wetting problems - input data},
  \bibinfo{year}{2024-05-24}. \URLprefix
  \url{https://tudatalib.ulb.tu-darmstadt.de/handle/tudatalib/3621.5}.
  \DOIprefix\doi{10.48328/tudatalib-982.5}.
\bibitem[{Asghar et~al.(2024)Asghar, Fricke, Bothe, and
  Maric}]{asgharcodeocean}
\bibinfo{author}{M.~H. Asghar}, \bibinfo{author}{M.~Fricke},
  \bibinfo{author}{D.~Bothe}, \bibinfo{author}{T.~Maric},
  \bibinfo{title}{Numerical wetting simulations using the plicrdf-isoadvector
  unstructured volume-of-fluid (vof) method - jupyter notebooks, csv files,
  secondary data, parameter variation file},
  \bibinfo{howpublished}{\url{https://www.codeocean.com/}},
  \bibinfo{year}{2024}. \DOIprefix\doi{10.24433/CO.1507307.v1}.
\bibitem[{Kluyver et~al.(2016)Kluyver, Ragan-Kelley, P{\'e}rez, Granger,
  Bussonnier, Frederic, Kelley, Hamrick, Grout, Corlay, Ivanov, Avila, Abdalla,
  Willing, and development team}]{jupyter}
\bibinfo{author}{T.~Kluyver}, \bibinfo{author}{B.~Ragan-Kelley},
  \bibinfo{author}{F.~P{\'e}rez}, \bibinfo{author}{B.~Granger},
  \bibinfo{author}{M.~Bussonnier}, \bibinfo{author}{J.~Frederic},
  \bibinfo{author}{K.~Kelley}, \bibinfo{author}{J.~Hamrick},
  \bibinfo{author}{J.~Grout}, \bibinfo{author}{S.~Corlay},
  \bibinfo{author}{P.~Ivanov}, \bibinfo{author}{D.~Avila},
  \bibinfo{author}{S.~Abdalla}, \bibinfo{author}{C.~Willing},
  \bibinfo{author}{J.~development team},
\newblock \bibinfo{title}{{Jupyter Notebooks - a publishing format for
  reproducible computational workflows}},
\newblock in: \bibinfo{editor}{F.~Loizides}, \bibinfo{editor}{B.~Scmidt}
  (Eds.), \bibinfo{booktitle}{{Positioning and Power in Academic Publishing:
  Players, Agents and Agendas}}, \bibinfo{publisher}{IOS Press},
  \bibinfo{address}{Netherlands}, \bibinfo{year}{2016}, pp.
  \bibinfo{pages}{87--90}. \URLprefix
  \url{https://eprints.soton.ac.uk/403913/}.
\bibitem[{Asghar et~al.(2 02)Asghar, Fricke, Bothe, and
  Marić}]{asgharnotebooks}
\bibinfo{author}{M.~H. Asghar}, \bibinfo{author}{M.~Fricke},
  \bibinfo{author}{D.~Bothe}, \bibinfo{author}{T.~Marić},
  \bibinfo{title}{Numerical wetting benchmarks - advancing the
  plicrdf-isoadvector unstructured volume-of-fluid (vof) method using the
  parabolic fit curvature model- jupyter notebooks, csv files, secondary data,
  parameter variation file}, \bibinfo{year}{2023-02-02}. \URLprefix
  \url{https://tudatalib.ulb.tu-darmstadt.de/handle/tudatalib/3622.7}.
  \DOIprefix\doi{10.48328/tudatalib-983.7}.
\bibitem[{Fricke et~al.(2020)Fricke, Mari{\'c}, and Bothe}]{fricke2020contact}
\bibinfo{author}{M.~Fricke}, \bibinfo{author}{T.~Mari{\'c}},
  \bibinfo{author}{D.~Bothe},
\newblock \bibinfo{title}{{Contact line advection using the geometrical
  Volume-of-Fluid method}},
\newblock \bibinfo{journal}{Journal of Computational Physics}
  \bibinfo{volume}{407} (\bibinfo{year}{2020}) \bibinfo{pages}{109221}.
  \DOIprefix\doi{https://doi.org/10.1016/j.jcp.2019.109221}.
\bibitem[{Dupont and Legendre(2010)}]{dupont2010numerical}
\bibinfo{author}{J.-B. Dupont}, \bibinfo{author}{D.~Legendre},
\newblock \bibinfo{title}{{Numerical simulation of static and sliding drop with
  contact angle hysteresis}},
\newblock \bibinfo{journal}{Journal of Computational Physics}
  \bibinfo{volume}{229} (\bibinfo{year}{2010}) \bibinfo{pages}{2453--2478}.
  \DOIprefix\doi{https://doi.org/10.1016/j.jcp.2009.07.034}.
\bibitem[{Lavi and Marmur(2004)}]{lavi2004}
\bibinfo{author}{B.~Lavi}, \bibinfo{author}{A.~Marmur},
\newblock \bibinfo{title}{The exponential power law: partial wetting kinetics
  and dynamic contact angles},
\newblock \bibinfo{journal}{Colloids and Surfaces A: Physicochemical and
  Engineering Aspects} \bibinfo{volume}{250} (\bibinfo{year}{2004})
  \bibinfo{pages}{409--414}. \URLprefix
  \url{https://www.sciencedirect.com/science/article/pii/S0927775704004777}.
  \DOIprefix\doi{https://doi.org/10.1016/j.colsurfa.2004.04.079},
  \bibinfo{note}{in honour of the 250th volume of Colloid and Surfaces A and
  the 25th Anniversary of the International Association of Colloid and
  Interface Scientists (IACIS)}.
\bibitem[{Patel et~al.(2017)Patel, Das, Kuipers, Padding, and
  Peters}]{patel2017coupled}
\bibinfo{author}{H.~Patel}, \bibinfo{author}{S.~Das},
  \bibinfo{author}{J.~Kuipers}, \bibinfo{author}{J.~Padding},
  \bibinfo{author}{E.~Peters},
\newblock \bibinfo{title}{{A coupled Volume of Fluid and Immersed Boundary
  Method for simulating 3D multiphase flows with contact line dynamics in
  complex geometries}},
\newblock \bibinfo{journal}{Chemical Engineering Science} \bibinfo{volume}{166}
  (\bibinfo{year}{2017}) \bibinfo{pages}{28--41}.
  \DOIprefix\doi{https://doi.org/10.1016/j.ces.2017.03.012}.
\bibitem[{Gr{\"u}nding et~al.(2020)Gr{\"u}nding, Smuda, Antritter, Fricke,
  Rettenmaier, Kummer, Stephan, Marschall, and Bothe}]{grunding2020comparative}
\bibinfo{author}{D.~Gr{\"u}nding}, \bibinfo{author}{M.~Smuda},
  \bibinfo{author}{T.~Antritter}, \bibinfo{author}{M.~Fricke},
  \bibinfo{author}{D.~Rettenmaier}, \bibinfo{author}{F.~Kummer},
  \bibinfo{author}{P.~Stephan}, \bibinfo{author}{H.~Marschall},
  \bibinfo{author}{D.~Bothe},
\newblock \bibinfo{title}{{A comparative study of transient capillary rise
  using direct numerical simulations}},
\newblock \bibinfo{journal}{Applied Mathematical Modelling}
  \bibinfo{volume}{86} (\bibinfo{year}{2020}) \bibinfo{pages}{142--165}.
  \DOIprefix\doi{https://doi.org/10.1016/j.apm.2020.04.020}.
\bibitem[{Huh and Scriven(1971)}]{huh1971hydrodynamic}
\bibinfo{author}{C.~Huh}, \bibinfo{author}{L.~E. Scriven},
\newblock \bibinfo{title}{{Hydrodynamic model of steady movement of a
  solid/liquid/fluid contact line}},
\newblock \bibinfo{journal}{Journal of Colloid and Interface Science}
  \bibinfo{volume}{35} (\bibinfo{year}{1971}) \bibinfo{pages}{85--101}.
  \DOIprefix\doi{https://doi.org/10.1016/0021-9797(71)90188-3}.
\bibitem[{Qu{\'e}r{\'e} et~al.(1999)Qu{\'e}r{\'e}, Rapha\"{e}l, and
  Ollitrault}]{Quere1999}
\bibinfo{author}{D.~Qu{\'e}r{\'e}}, \bibinfo{author}{{\'E}.~Rapha\"{e}l},
  \bibinfo{author}{J.-Y. Ollitrault},
\newblock \bibinfo{title}{{Rebounds in a Capillary Tube}},
\newblock \bibinfo{journal}{Langmuir} \bibinfo{volume}{15}
  (\bibinfo{year}{1999}) \bibinfo{pages}{3679--3682}.
  \DOIprefix\doi{https://doi.org/10.1021/la9801615}.
\bibitem[{Fries and Dreyer(2009)}]{fries2009dimensionless}
\bibinfo{author}{N.~Fries}, \bibinfo{author}{M.~Dreyer},
\newblock \bibinfo{title}{{Dimensionless scaling methods for capillary rise}},
\newblock \bibinfo{journal}{Journal of Colloid and Interface Science}
  \bibinfo{volume}{338} (\bibinfo{year}{2009}) \bibinfo{pages}{514--518}.
  \DOIprefix\doi{https://doi.org/10.1016/j.jcis.2009.06.036}.
\bibitem[{{OpenFOAM.com}(2023)}]{OpenFOAMv2312}
\bibinfo{author}{{OpenFOAM.com}}, \bibinfo{title}{{OpenFOAM-v2312}},
  \bibinfo{year}{2023}. \URLprefix
  \url{https://develop.openfoam.com/Development/openfoam/-/tree/OpenFOAM-v2312}.
\bibitem[{{Bernhard Gschaider}(2005)}]{pyFoam}
\bibinfo{author}{{Bernhard Gschaider}}, \bibinfo{title}{{Contrib/PyFoam}},
  \bibinfo{year}{2005}.
  \bibinfo{note}{\url{https://openfoamwiki.net/index.php/Contrib/PyFoam}, Last
  accessed on 2020-10-19}.
\bibitem[{Tolle et~al.(2022)Tolle, Gründing, Bothe, and
  Marić}]{TOLLE2022108249}
\bibinfo{author}{T.~Tolle}, \bibinfo{author}{D.~Gründing},
  \bibinfo{author}{D.~Bothe}, \bibinfo{author}{T.~Marić},
\newblock \bibinfo{title}{{triSurfaceImmersion: Computing volume fractions and
  signed distances from triangulated surfaces immersed in unstructured
  meshes}},
\newblock \bibinfo{journal}{Computer Physics Communications}
  \bibinfo{volume}{273} (\bibinfo{year}{2022}) \bibinfo{pages}{108249}.
  \DOIprefix\doi{https://doi.org/10.1016/j.cpc.2021.108249}.
\bibitem[{{Franjo Juretic}(2021)}]{cfMesh}
\bibinfo{author}{{Franjo Juretic}}, \bibinfo{title}{{integration-cfmesh}},
  \bibinfo{year}{2021}.
  \bibinfo{note}{\url{https://develop.openfoam.com/Community/integration-cfmesh/-/commit/f362ee65334e08056abdabab45e588503553e0ef},
  build = {14aeaf8dab-20211220}, Last accessed on 2022-10-19}.
\bibitem[{Jürgen~Riegel and van Havre(2022)}]{freecad}
\bibinfo{author}{W.~M. Jürgen~Riegel}, \bibinfo{author}{Y.~van Havre},
  \bibinfo{title}{{freecad: A 3D parametric modeler}}, \bibinfo{year}{2022}.
  \URLprefix \url{https://www.freecadweb.org/}.
\bibitem[{Youngs(1982)}]{youngs1982time}
\bibinfo{author}{D.~L. Youngs},
\newblock \bibinfo{title}{{Time-dependent multi-material flow with large fluid
  distortion}},
\newblock \bibinfo{journal}{Numerical Methods for Fluid Dynamics}
  (\bibinfo{year}{1982}). \URLprefix
  \url{https://cir.nii.ac.jp/crid/1571417126191472512}.
\bibitem[{Scardovelli and Zaleski(2003)}]{scardovelli2003interface}
\bibinfo{author}{R.~Scardovelli}, \bibinfo{author}{S.~Zaleski},
\newblock \bibinfo{title}{{Interface reconstruction with least-square fit and
  split Eulerian--Lagrangian advection}},
\newblock \bibinfo{journal}{International Journal for Numerical Methods in
  Fluids} \bibinfo{volume}{41} (\bibinfo{year}{2003})
  \bibinfo{pages}{251--274}. \DOIprefix\doi{https://doi.org/10.1002/fld.431}.
\bibitem[{Tolle et~al.(2020)Tolle, Bothe, and Mari{\'c}}]{Tolle2020}
\bibinfo{author}{T.~Tolle}, \bibinfo{author}{D.~Bothe},
  \bibinfo{author}{T.~Mari{\'c}},
\newblock \bibinfo{title}{{SAAMPLE: A segregated accuracy-driven algorithm for
  multiphase pressure-linked equations}},
\newblock \bibinfo{journal}{Computers \& Fluids} \bibinfo{volume}{200}
  (\bibinfo{year}{2020}) \bibinfo{pages}{104450}.
  \DOIprefix\doi{https://doi.org/10.1016/j.compfluid.2020.104450}.
\bibitem[{Renardy et~al.(2001)Renardy, Renardy, and Li}]{renardy2001numerical}
\bibinfo{author}{M.~Renardy}, \bibinfo{author}{Y.~Renardy},
  \bibinfo{author}{J.~Li},
\newblock \bibinfo{title}{{Numerical simulation of moving contact line problems
  using a volume-of-fluid method}},
\newblock \bibinfo{journal}{Journal of Computational Physics}
  \bibinfo{volume}{171} (\bibinfo{year}{2001}) \bibinfo{pages}{243--263}.
  \DOIprefix\doi{https://doi.org/10.1006/jcph.2001.6785}.
\bibitem[{Scheufler and Roenby(2023)}]{scheufler2021twophaseflow}
\bibinfo{author}{H.~Scheufler}, \bibinfo{author}{J.~Roenby},
\newblock \bibinfo{title}{Twophaseflow: A framework for developing two phase
  flow solvers in openfoam},
\newblock \bibinfo{journal}{OpenFOAM® Journal} \bibinfo{volume}{3}
  (\bibinfo{year}{2023}) \bibinfo{pages}{200–224}. \URLprefix
  \url{https://journal.openfoam.com/index.php/ofj/article/view/80}.
\bibitem[{Asghar et~al.(2 02)Asghar, Fricke, Bothe, and Marić}]{fpnotebooks}
\bibinfo{author}{M.~H. Asghar}, \bibinfo{author}{M.~Fricke},
  \bibinfo{author}{D.~Bothe}, \bibinfo{author}{T.~Marić},
  \bibinfo{title}{Numerical wetting benchmarks - advancing the
  plicrdf-isoadvector unstructured volume-of-fluid (vof) method using the
  parabolic fit curvature model- jupyter notebooks, csv files, secondary data,
  parameter variation file}, \bibinfo{year}{2023-02-02}. \URLprefix
  \url{https://tudatalib.ulb.tu-darmstadt.de/handle/tudatalib/3622.5}.
  \DOIprefix\doi{10.48328/tudatalib-983.5}.
\bibitem[{Asghar et~al.(2023)Asghar, Fricke, Bothe, and Marić}]{rdfnotebooks}
\bibinfo{author}{M.~H. Asghar}, \bibinfo{author}{M.~Fricke},
  \bibinfo{author}{D.~Bothe}, \bibinfo{author}{T.~Marić},
  \bibinfo{title}{{Numerical Wetting Benchmarks - Advancing the
  plicRDF-isoAdvector unstructured Volume-of-Fluid (VOF) method using the RDF
  curvature model- Jupyter Notebooks, CSV files, Secondary Data, Parameter
  variation file}}, \bibinfo{year}{2023}. \URLprefix
  \url{https://tudatalib.ulb.tu-darmstadt.de/handle/tudatalib/3730.2}.
  \DOIprefix\doi{10.48328/tudatalib-1069.2}.
\bibitem[{Asghar et~al.(2 02)Asghar, Fricke, Bothe, and Marić}]{hfnotebooks}
\bibinfo{author}{M.~H. Asghar}, \bibinfo{author}{M.~Fricke},
  \bibinfo{author}{D.~Bothe}, \bibinfo{author}{T.~Marić},
  \bibinfo{title}{Numerical wetting benchmarks - advancing the
  plicrdf-isoadvector unstructured volume-of-fluid (vof) method using
  height-function curvature model- jupyter notebooks, csv files, secondary
  data, parameter variation file}, \bibinfo{year}{2023-02-02}. \URLprefix
  \url{https://tudatalib.ulb.tu-darmstadt.de/handle/tudatalib/3729.3}.
  \DOIprefix\doi{10.48328/tudatalib-1068.3}.
\bibitem[{Fricke et~al.(2019)Fricke, K{\"o}hne, and
  Bothe}]{fricke2019kinematic}
\bibinfo{author}{M.~Fricke}, \bibinfo{author}{M.~K{\"o}hne},
  \bibinfo{author}{D.~Bothe},
\newblock \bibinfo{title}{{A kinematic evolution equation for the dynamic
  contact angle and some consequences}},
\newblock \bibinfo{journal}{Physica D: Nonlinear Phenomena}
  \bibinfo{volume}{394} (\bibinfo{year}{2019}) \bibinfo{pages}{26--43}.
  \DOIprefix\doi{https://doi.org/10.1016/j.physd.2019.01.008}.
\bibitem[{Fricke et~al.(2018)Fricke, K{\"o}hne, and
  Bothe}]{fricke2018kinematics}
\bibinfo{author}{M.~Fricke}, \bibinfo{author}{M.~K{\"o}hne},
  \bibinfo{author}{D.~Bothe},
\newblock \bibinfo{title}{{On the kinematics of contact line motion}},
\newblock \bibinfo{journal}{PAMM} \bibinfo{volume}{18} (\bibinfo{year}{2018})
  \bibinfo{pages}{e201800451}.
  \DOIprefix\doi{https://doi.org/10.1002/pamm.201800451}.
\bibitem[{Fricke et~al.(2020)Fricke, Fickel, Hartmann, Gr{\"u}nding, Biesalski,
  and Bothe}]{fricke2020geometry}
\bibinfo{author}{M.~Fricke}, \bibinfo{author}{B.~Fickel},
  \bibinfo{author}{M.~Hartmann}, \bibinfo{author}{D.~Gr{\"u}nding},
  \bibinfo{author}{M.~Biesalski}, \bibinfo{author}{D.~Bothe},
\newblock \bibinfo{title}{{A geometry-based model for spreading drops applied
  to drops on a silicon wafer and a swellable polymer brush film}},
\newblock \bibinfo{journal}{arXiv preprint arXiv:2003.04914}
  (\bibinfo{year}{2020}). \DOIprefix\doi{10.48550/ARXIV.2003.04914}.
\bibitem[{Afkhami et~al.(2009)Afkhami, Zaleski, and Bussmann}]{afkhami2009mesh}
\bibinfo{author}{S.~Afkhami}, \bibinfo{author}{S.~Zaleski},
  \bibinfo{author}{M.~Bussmann},
\newblock \bibinfo{title}{{A mesh-dependent model for applying dynamic contact
  angles to VOF simulations}},
\newblock \bibinfo{journal}{Journal of Computational Physics}
  \bibinfo{volume}{228} (\bibinfo{year}{2009}) \bibinfo{pages}{5370--5389}.
  \DOIprefix\doi{https://doi.org/10.1016/j.jcp.2009.04.027}.
\bibitem[{Khatavkar et~al.(2007)Khatavkar, Anderson, and
  Meijer}]{khatavkar2007capillary}
\bibinfo{author}{V.~Khatavkar}, \bibinfo{author}{P.~Anderson},
  \bibinfo{author}{H.~Meijer},
\newblock \bibinfo{title}{{Capillary spreading of a droplet in the partially
  wetting regime using a diffuse-interface model}},
\newblock \bibinfo{journal}{Journal of Fluid Mechanics} \bibinfo{volume}{572}
  (\bibinfo{year}{2007}) \bibinfo{pages}{367--387}.
  \DOIprefix\doi{https://doi.org/10.1017/S0022112006003533}.
\bibitem[{Villanueva and Amberg(2006)}]{villanueva2006some}
\bibinfo{author}{W.~Villanueva}, \bibinfo{author}{G.~Amberg},
\newblock \bibinfo{title}{{Some generic capillary-driven flows}},
\newblock \bibinfo{journal}{International Journal of Multiphase Flow}
  \bibinfo{volume}{32} (\bibinfo{year}{2006}) \bibinfo{pages}{1072--1086}.
  \DOIprefix\doi{https://doi.org/10.1016/j.ijmultiphaseflow.2006.05.003}.
\bibitem[{{Kitware, Inc, Los Alamos National Laboratory }(2021)}]{ParaView}
\bibinfo{author}{{Kitware, Inc, Los Alamos National Laboratory }},
  \bibinfo{title}{{ParaView-v5.9}}, \bibinfo{year}{2021}.
  \bibinfo{note}{\url{https://www.paraview.org/documentation/}, Last accessed
  on 2022-10-19}.
\bibitem[{Cox(1986)}]{cox1986dynamics}
\bibinfo{author}{R.~Cox},
\newblock \bibinfo{title}{The dynamics of the spreading of liquids on a solid
  surface. part 1. viscous flow},
\newblock \bibinfo{journal}{Journal of fluid mechanics} \bibinfo{volume}{168}
  (\bibinfo{year}{1986}) \bibinfo{pages}{169--194}.
\bibitem[{{Scheufler, Henning and Roenby, Johan}(2024)}]{isoAdvection}
\bibinfo{author}{{Scheufler, Henning and Roenby, Johan}},
  \bibinfo{title}{{isoAdvection scheme}}, \bibinfo{year}{2024}. \URLprefix
  \url{https://www.openfoam.com/documentation/guides/latest/api/isoAdvection_8C_source.html}.
\bibitem[{Duvivier et~al.(2011)Duvivier, Seveno, Rioboo, Blake, and
  De~Coninck}]{DuvivierExp}
\bibinfo{author}{D.~Duvivier}, \bibinfo{author}{D.~Seveno},
  \bibinfo{author}{R.~Rioboo}, \bibinfo{author}{T.~D. Blake},
  \bibinfo{author}{J.~De~Coninck},
\newblock \bibinfo{title}{Experimental evidence of the role of viscosity in the
  molecular kinetic theory of dynamic wetting},
\newblock \bibinfo{journal}{Langmuir} \bibinfo{volume}{27}
  (\bibinfo{year}{2011}) \bibinfo{pages}{13015--13021}. \URLprefix
  \url{https://doi.org/10.1021/la202836q}. \DOIprefix\doi{10.1021/la202836q},
  \bibinfo{note}{pMID: 21919445}.
\bibitem[{Blake et~al.(2015)Blake, Fernandez-Toledano, Doyen, and
  De~Coninck}]{BlakeForcedWetting}
\bibinfo{author}{T.~D. Blake}, \bibinfo{author}{J.-C. Fernandez-Toledano},
  \bibinfo{author}{G.~Doyen}, \bibinfo{author}{J.~De~Coninck},
\newblock \bibinfo{title}{{Forced wetting and hydrodynamic assist}},
\newblock \bibinfo{journal}{Physics of Fluids} \bibinfo{volume}{27}
  (\bibinfo{year}{2015}) \bibinfo{pages}{112101}. \URLprefix
  \url{https://doi.org/10.1063/1.4934703}. \DOIprefix\doi{10.1063/1.4934703}.
\bibitem[{Lippert(2016)}]{anjathesis}
\bibinfo{author}{A.~C. Lippert}, \bibinfo{title}{Direct Numerical Simulations
  of Thermocapillary Driven Motions in Two-phase Flows}, Ph.D. thesis,
  Technische Universit{\"a}t Darmstadt, \bibinfo{address}{Darmstadt},
  \bibinfo{year}{2016}. \URLprefix
  \url{http://tuprints.ulb.tu-darmstadt.de/5817/}.
\bibitem[{Rieber(2004)}]{rieber2004numerische}
\bibinfo{author}{M.~Rieber}, \bibinfo{title}{Numerische Modellierung der
  Dynamik freier Grenzfl{\"a}chen in Zweiphasenstr{\"o}mungen},
  Fortschrittberichte VDI / 7, \bibinfo{publisher}{VDI-Verlag},
  \bibinfo{year}{2004}. \URLprefix
  \url{https://books.google.de/books?id=1tWKAgAACAAJ}.
\bibitem[{Benkenida and Magnaudet(2000)}]{JADIM}
\bibinfo{author}{A.~Benkenida}, \bibinfo{author}{J.~Magnaudet},
\newblock \bibinfo{title}{Une methode de simulation d'ecoulements diphasiques
  sans reconstruction d'interfaces},
\newblock \bibinfo{journal}{Comptes Rendus de l Académie des Sciences - Series
  IIB - Mechanics-Physics-Astronomy} \bibinfo{volume}{328}
  (\bibinfo{year}{2000}). \DOIprefix\doi{10.1016/S1287-4620(00)88412-4}.
\bibitem[{Mari{\'c} et~al.(2018)Mari{\'c}, Marschall, and
  Bothe}]{maric2018enhanced}
\bibinfo{author}{T.~Mari{\'c}}, \bibinfo{author}{H.~Marschall},
  \bibinfo{author}{D.~Bothe},
\newblock \bibinfo{title}{{An enhanced un-split face-vertex flux-based VoF
  method}},
\newblock \bibinfo{journal}{Journal of Computational Physics}
  \bibinfo{volume}{371} (\bibinfo{year}{2018}) \bibinfo{pages}{967--993}.
  \DOIprefix\doi{https://doi.org/10.1016/j.jcp.2018.03.048}.
\bibitem[{Washburn(1921)}]{Washburn1921}
\bibinfo{author}{E.~W. Washburn},
\newblock \bibinfo{title}{{The Dynamics of Capillary Flow}},
\newblock \bibinfo{journal}{Physical Review} \bibinfo{volume}{17}
  (\bibinfo{year}{1921}) \bibinfo{pages}{273--283}.
  \DOIprefix\doi{10.1103/PhysRev.17.273}.
\bibitem[{Bosanquet(1923)}]{Bosanquet1923}
\bibinfo{author}{C.~H. Bosanquet},
\newblock \bibinfo{title}{{On the flow of liquids into capillary tubes}},
\newblock \bibinfo{journal}{The London, Edinburgh, and Dublin Philosophical
  Magazine and Journal of Science} \bibinfo{volume}{45} (\bibinfo{year}{1923})
  \bibinfo{pages}{525--531}.
  \DOIprefix\doi{https://doi.org/10.1080/14786442308634144}.
\bibitem[{Delannoy et~al.(2019)Delannoy, Lafon, Koga, Reyssat, and
  Qu{\'e}r{\'e}}]{delannoy2019dual}
\bibinfo{author}{J.~Delannoy}, \bibinfo{author}{S.~Lafon},
  \bibinfo{author}{Y.~Koga}, \bibinfo{author}{{\'E}.~Reyssat},
  \bibinfo{author}{D.~Qu{\'e}r{\'e}},
\newblock \bibinfo{title}{The dual role of viscosity in capillary rise},
\newblock \bibinfo{journal}{Soft matter} \bibinfo{volume}{15}
  (\bibinfo{year}{2019}) \bibinfo{pages}{2757--2761}.
\bibitem[{Fricke et~al.(2023{\natexlab{a}})Fricke, Raju, Ouro-Koura, Kozymka,
  Coninck, Tukovic, Maric, and Bothe}]{fricke2023bridging}
\bibinfo{author}{M.~Fricke}, \bibinfo{author}{S.~Raju}, \bibinfo{author}{E.~A.
  Ouro-Koura}, \bibinfo{author}{O.~Kozymka}, \bibinfo{author}{J.~D. Coninck},
  \bibinfo{author}{Z.~Tukovic}, \bibinfo{author}{T.~Maric},
  \bibinfo{author}{D.~Bothe}, \bibinfo{title}{Bridging the scales in capillary
  rise dynamics with complexity-reduced models},
  \bibinfo{year}{2023}{\natexlab{a}}.
  \href{http://arxiv.org/abs/2311.11947}{{\tt arXiv:2311.11947}}.
\bibitem[{Fricke et~al.(2023{\natexlab{b}})Fricke, Ouro-Koura, Raju, {von
  Klitzing}, {De Coninck}, and Bothe}]{fricke2023analytical}
\bibinfo{author}{M.~Fricke}, \bibinfo{author}{E.~A. Ouro-Koura},
  \bibinfo{author}{S.~Raju}, \bibinfo{author}{R.~{von Klitzing}},
  \bibinfo{author}{J.~{De Coninck}}, \bibinfo{author}{D.~Bothe},
  \bibinfo{title}{An analytical study of capillary rise dynamics: Critical
  conditions and hidden oscillations}, \bibinfo{year}{2023}{\natexlab{b}}.
  \URLprefix
  \url{https://www.sciencedirect.com/science/article/pii/S016727892300249X}.
  \DOIprefix\doi{https://doi.org/10.1016/j.physd.2023.133895}.
\bibitem[{Gr{\"u}nding(2020)}]{grunding2020enhanced}
\bibinfo{author}{D.~Gr{\"u}nding},
\newblock \bibinfo{title}{{An enhanced model for the capillary rise problem}},
\newblock \bibinfo{journal}{International Journal of Multiphase Flow}
  \bibinfo{volume}{128} (\bibinfo{year}{2020}) \bibinfo{pages}{103210}.
  \DOIprefix\doi{https://doi.org/10.1016/j.ijmultiphaseflow.2020.103210}.
\bibitem[{Tukovic and Jasak(2012)}]{ALE}
\bibinfo{author}{Z.~Tukovic}, \bibinfo{author}{H.~Jasak},
\newblock \bibinfo{title}{A moving mesh finite volume interface tracking method
  for surface tension dominated interfacial fluid flow},
\newblock \bibinfo{journal}{Computers \& Fluids} \bibinfo{volume}{55}
  (\bibinfo{year}{2012}) \bibinfo{pages}{70–84}.
  \DOIprefix\doi{10.1016/j.compfluid.2011.11.003}.
\bibitem[{Gr{\"u}nding(2020)}]{dirkthesis}
\bibinfo{author}{D.~Gr{\"u}nding}, \bibinfo{title}{An Arbitrary
  Lagrangian-Eulerian Method for the Direct Numerical Simulation of Wetting
  Processes}, Ph.D. thesis, Technische Universit{\"a}t Darmstadt,
  \bibinfo{address}{Darmstadt}, \bibinfo{year}{2020}. \URLprefix
  \url{http://tuprints.ulb.tu-darmstadt.de/11442/}.
  \DOIprefix\doi{https://doi.org/10.25534/tuprints-00011442}.
\bibitem[{Kummer(2012)}]{KummerThesis}
\bibinfo{author}{F.~Kummer}, \bibinfo{title}{The BoSSS Discontinuous Galerkin
  solver for incompressible fluid dynamics and an extension to singular
  equations.}, Ph.D. thesis, Technische Universit{\"a}t Darmstadt,
  \bibinfo{address}{Darmstadt}, \bibinfo{year}{2012}. \URLprefix
  \url{http://tuprints.ulb.tu-darmstadt.de/2889/}.
\bibitem[{Kummer(2017)}]{kummer2017extended}
\bibinfo{author}{F.~Kummer},
\newblock \bibinfo{title}{Extended discontinuous galerkin methods for two-phase
  flows: the spatial discretization},
\newblock \bibinfo{journal}{International Journal for Numerical Methods in
  Engineering} \bibinfo{volume}{109} (\bibinfo{year}{2017})
  \bibinfo{pages}{259--289}.
\bibitem[{Huh and Mason(1977)}]{Huh1977}
\bibinfo{author}{C.~Huh}, \bibinfo{author}{S.~G. Mason},
\newblock \bibinfo{title}{{The steady movement of a liquid meniscus in a
  capillary tube}},
\newblock \bibinfo{journal}{Journal of Fluid Mechanics} \bibinfo{volume}{81}
  (\bibinfo{year}{1977}) \bibinfo{pages}{401--419}.
  \DOIprefix\doi{10.1017/S0022112077002134}.
\bibitem[{Shumaly et~al.(2023)Shumaly, Darvish, Li, Saal, Hinduja, Steffen,
  Kukharenko, Butt, and Berger}]{shumaly2023deep}
\bibinfo{author}{S.~Shumaly}, \bibinfo{author}{F.~Darvish},
  \bibinfo{author}{X.~Li}, \bibinfo{author}{A.~Saal},
  \bibinfo{author}{C.~Hinduja}, \bibinfo{author}{W.~Steffen},
  \bibinfo{author}{O.~Kukharenko}, \bibinfo{author}{H.-J. Butt},
  \bibinfo{author}{R.~Berger},
\newblock \bibinfo{title}{Deep learning to analyze sliding drops},
\newblock \bibinfo{journal}{Langmuir} \bibinfo{volume}{39}
  (\bibinfo{year}{2023}) \bibinfo{pages}{1111--1122}.
\bibitem[{Mitra et~al.(2013)Mitra, Sathe, Doroodchi, Utikar, Shah, Pareek,
  Joshi, and Evans}]{mitra2013}
\bibinfo{author}{S.~Mitra}, \bibinfo{author}{M.~J. Sathe},
  \bibinfo{author}{E.~Doroodchi}, \bibinfo{author}{R.~Utikar},
  \bibinfo{author}{M.~K. Shah}, \bibinfo{author}{V.~Pareek},
  \bibinfo{author}{J.~B. Joshi}, \bibinfo{author}{G.~M. Evans},
\newblock \bibinfo{title}{Droplet impact dynamics on a spherical particle},
\newblock \bibinfo{journal}{Chemical Engineering Science} \bibinfo{volume}{100}
  (\bibinfo{year}{2013}) \bibinfo{pages}{105--119}. \URLprefix
  \url{https://www.sciencedirect.com/science/article/pii/S000925091300050X}.
  \DOIprefix\doi{https://doi.org/10.1016/j.ces.2013.01.037},
  \bibinfo{note}{11th International Conference on Gas-Liquid and
  Gas-Liquid-Solid Reactor Engineering}.
\bibitem[{Rieber and Frohn(1999)}]{RIEBER1999455}
\bibinfo{author}{M.~Rieber}, \bibinfo{author}{A.~Frohn},
\newblock \bibinfo{title}{A numerical study on the mechanism of splashing},
\newblock \bibinfo{journal}{International Journal of Heat and Fluid Flow}
  \bibinfo{volume}{20} (\bibinfo{year}{1999}) \bibinfo{pages}{455--461}.
  \URLprefix
  \url{https://www.sciencedirect.com/science/article/pii/S0142727X99000338}.
  \DOIprefix\doi{https://doi.org/10.1016/S0142-727X(99)00033-8}.
\bibitem[{Brackbill et~al.(1992)Brackbill, Kothe, and
  Zemach}]{BRACKBILL1992335}
\bibinfo{author}{J.~Brackbill}, \bibinfo{author}{D.~Kothe},
  \bibinfo{author}{C.~Zemach},
\newblock \bibinfo{title}{A continuum method for modeling surface tension},
\newblock \bibinfo{journal}{Journal of Computational Physics}
  \bibinfo{volume}{100} (\bibinfo{year}{1992}) \bibinfo{pages}{335--354}.
  \URLprefix
  \url{https://www.sciencedirect.com/science/article/pii/002199919290240Y}.
  \DOIprefix\doi{https://doi.org/10.1016/0021-9991(92)90240-Y}.
\bibitem[{Fricke(2021)}]{mathisThesis}
\bibinfo{author}{M.~Fricke}, \bibinfo{title}{Mathematical modeling and
  Volume-of-Fluid based simulation of dynamic wetting}, Ph.D. thesis,
  Technische Universit{\"a}t Darmstadt, \bibinfo{address}{Darmstadt},
  \bibinfo{year}{2021}. \URLprefix
  \url{http://tuprints.ulb.tu-darmstadt.de/14274/}.
  \DOIprefix\doi{https://doi.org/10.12921/tuprints-00014274}.
\bibitem[{Popinet(2009)}]{POPINET20095838}
\bibinfo{author}{S.~Popinet},
\newblock \bibinfo{title}{An accurate adaptive solver for
  surface-tension-driven interfacial flows},
\newblock \bibinfo{journal}{Journal of Computational Physics}
  \bibinfo{volume}{228} (\bibinfo{year}{2009}) \bibinfo{pages}{5838--5866}.
  \URLprefix
  \url{https://www.sciencedirect.com/science/article/pii/S002199910900240X}.
  \DOIprefix\doi{https://doi.org/10.1016/j.jcp.2009.04.042}.
\bibitem[{Afkhami and Bussmann(2008)}]{afkhami2008height}
\bibinfo{author}{S.~Afkhami}, \bibinfo{author}{M.~Bussmann},
\newblock \bibinfo{title}{Height functions for applying contact angles to 2d
  vof simulations},
\newblock \bibinfo{journal}{International journal for numerical methods in
  fluids} \bibinfo{volume}{57} (\bibinfo{year}{2008})
  \bibinfo{pages}{453--472}.
\bibitem[{Hartmann et~al.(2021)Hartmann, Fricke, Weimar, Gründing, Marić,
  Bothe, and Hardt}]{HARTMANN2021103582}
\bibinfo{author}{M.~Hartmann}, \bibinfo{author}{M.~Fricke},
  \bibinfo{author}{L.~Weimar}, \bibinfo{author}{D.~Gründing},
  \bibinfo{author}{T.~Marić}, \bibinfo{author}{D.~Bothe},
  \bibinfo{author}{S.~Hardt},
\newblock \bibinfo{title}{Breakup dynamics of capillary bridges on hydrophobic
  stripes},
\newblock \bibinfo{journal}{International Journal of Multiphase Flow}
  \bibinfo{volume}{140} (\bibinfo{year}{2021}) \bibinfo{pages}{103582}.
  \URLprefix
  \url{https://www.sciencedirect.com/science/article/pii/S0301932221000306}.
  \DOIprefix\doi{https://doi.org/10.1016/j.ijmultiphaseflow.2021.103582}.

\end{thebibliography}


\end{document}